\documentclass{article}
\usepackage{epsfig}
\usepackage{pst-plot}

\newcommand{\be}{\begin{equation}}
\newcommand{\ee}{\end{equation}}
\newcommand{\bea}{\begin{eqnarray}}
\newcommand{\eea}{\end{eqnarray}}

\newcommand{\bra}[1]{\left\langle #1 \right|}
\newcommand{\ket}[1]{\left| #1 \right\rangle}
\begin{document}

\title{Effective interactions and the nuclear shell-model}
\author{D.~J.~Dean$^{1,2}$, T.~Engeland$^{2,3}$, M.~Hjorth-Jensen$^{2,3}$,
M.~P.~Kartamyshev$^{2,3}$, and E.~Osnes$^{2,3}$\\ \\
$^{1}$Physics Division, Oak Ridge National Laboratory, Oak Ridge, TN 37831-6373, U.S.A.\\
$^{2}$Center of Mathematics for Applications, University of Oslo, N-0316 Oslo, Norway\\
$^{3}$Department of Physics, University of Oslo, N-0316 Oslo, Norway}

\maketitle
\begin{abstract}
This review aims at a critical discussion of  the interplay between
effective interactions derived from various many-body approaches
and spectroscopic data extracted from large scale shell-model
studies. To achieve this,
our many-body scheme starts with
the free nucleon-nucleon (NN)
interaction, typically modelled
on various meson exchanges. The NN interaction
is in turn renormalized in order to derive an effective medium
dependent interaction. The latter is in turn used
in shell-model calculations of selected nuclei. We also describe
how to sum up the parquet class of diagrams and present
initial uses of the effective interactions in coupled cluster
many-body theory. 
\end{abstract}

\tableofcontents

\section{Introduction}\label{sec:sec1}

Traditional shell-model studies
have recently received a renewed
interest through large scale shell model calculations
in both no-core calculations \cite{bruce1,bruce2,bruce3,bruce4} for light nuclei, 
the $1s0d$ shell \cite{alex}, the $1p0f$ shell and the $2s1d0g_{7/2}$ shell with
the inclusion of the $0h_{11/2}$ intruder state as well.
It is now therefore fully possible to perform large-scale
shell-model investigations
and study the excitation spectra
for systems with
many basis states. With recent advances in
Monte Carlo methods, see for example 
Refs.~\cite{taka1,r:smmc_pr, vijay,ndrops97,bob1,bob2,bob3}, one is also
able to enlarge the dimensionality
of the systems under study considerably,
and important information on e.g., ground state properties
has thereby been obtained.

An important feature of such large scale calculations
is that it allows one to probe the underlying many-body
physics in a hitherto unprecedented way.
The crucial starting point in all such shell-model
calculations is
the derivation of an effective interaction, be it
either an approach based on a microscopic theory
starting from the free nucleon-nucleon ($NN$) interaction or a more
phenomenologically determined interaction.
In shell-model studies of e.g., the Sn isotopes, one may have
up to 31 valence particles or holes interacting via e.g.,
an effective two-body interaction. The results of such
calculations can therefore yield, when compared with
the available body of experimental data, critical
inputs to the underlying theory of the effective interaction.

Clearly, although the $NN$  interaction is of short
but finite range, with typical interparticle
distances of the order of $1\sim 2$ fm, there are
indications from both studies of few-body systems and
infinite nuclear matter, that at least three-body
interactions, both real and effective ones, may be of
importance.
Thus, with many valence nucleons present, such
large-scale shell-model calculations may
tell us how well e.g., an effective interaction
which only includes two-body terms does in
reproducing properties such as excitation spectra and
binding energies.

This work deals therefore with various ways of
deriving the effective
interaction or effective operator needed
in shell-model calculations, starting from the
free $NN$  interaction.
Normally, the problem of deriving such effective operators and interactions are solved
in a limited space, the so-called model space, which is a subspace of the
full Hilbert space. The effective operator and interaction
theory is then introduced in order to
systematically take into account contributions from the complement
(the excluded space) of the chosen model space. Several formulations
for such expansions of effective operators and interactions exit in the literature, following
time-dependent or time-independent
perturbation theory \cite{so95,brandow67,ko90,hko95,lm85,so84}.
The so-called $\hat{Q}$-box method with the folded-diagram
formulation of Kuo and co-workers \cite{ko90,hko95} has been extensively
applied to systems in nuclear physics, especially for the derivation of
effective interactions for the nuclear shell-model. This method is however limited 
by the fact that the $\hat{Q}$-box includes diagrams to a certain order in perturbation only.
To go beyond third order in perturbation theory is very hard and there is no indications
that the expansion converges. Thus,
formulations like the coupled-cluster
method or exponential ansatz
\cite{lm85,coester58,coester60,bartlett81,comp_chem_rev00,harris92,piotr1,helgaker,arponen97,lk72a,lk72b,zabolitzky74,klz78,ticcm,mh00a,mh00b,mh99,hm99}, which is  much favored
in quantum chemistry,
the summation of the parquet class of diagrams
\cite{dm64,nozieres,babu,jls82,br86,scalapino,ym96,dya97}, although with few applications, and FHNC
theory \cite{br86,adelchi98,apr98}, 
offer the possibility of summing much larger classes of many-body terms. 

Ab initio methods like variational and diffusional Monte Carlo approaches 
\cite{vijay,ndrops97,bob1,bob2,bob3} and no-core-shell model calculations 
\cite{bruce1,bruce2,bruce3,bruce4} for light nuclei
have also been much favored
in nuclear theory.

In this work we will focus on the above-mentioned $\hat{Q}$-box approach combined with the 
nuclear shell-model,
the summation of the so-called parquet diagrams and Coupled cluster theory as three
possible ways of studying the nuclear many-body problem.
The
$\hat{Q}$-box has been introduced in Rayleigh-Schr\"odinger perturbation
theory as the definition of all non-folded diagrams to a given order in
the expansion parameter, in nuclear physics the so-called $G$-matrix.
The $G$-matrix renders the free $NN$  interaction $V$, which
is repulsive at small internucleon distances,  tractable
for a perturbative analysis through the summation of ladders diagrams
to infinite order. Stated differently, the $G$-matrix, through the
solution of the Bethe-Brueckner-Goldstone equation, accounts for the
short-range correlations involving high-lying states.
Folded diagrams are a class of diagrams which arise
due to the removal of the dependence of the exact model-space energy
in the Brillouin-Wigner perturbation expansion. Through the $\hat{Q}$-box
formulation and its derivatives, this set of diagrams can easily be summed
up.

In addition to the evaluation of folded diagrams and the inclusion
of ladder diagrams to infinite order included in the $G$-matrix, there
are other classes of diagrams which can be summed up.
These take into account
the effect of long-range correlations involving low-energy excitations.
A frequently applied formalism is
the Tamm-Dancoff (TDA) or the random-phase
(RPA) approximations. In their traditional formulation one allows
for the summation of all particle-hole excitations, both forward-going
and backward going insertions.
This set of diagrams, as formulated by Kirson \cite{kirson74} and reviewed
in Ref.\ \cite{eo77}, should account for correlations  arising
from collective particle-hole correlations. Another possibility
is to include any number of particle-particle and hole-hole
correlations in the screening of particle-hole correlations.
The inclusion of these correlations is conventionally labelled
particle-particle (pp) RPA. It has been used both in nuclear matter
studies \cite{angels88,rpd89,yhk86,syk87}
and in evaluations of ground state properties of closed-shell
nuclei \cite{hmtk87,emm91,hmm95}.
Ellis, Mavromatis and M\"uther \cite{emm91,hmm95} have
extended the pp RPA to include the particle-hole (ph) RPA, though
screening of two-particle-one-hole (2p1h) and two-hole-one-particle
(2h1p) vertices was not included.
The latter works can be viewed as a step towards the full summation of the
parquet class of diagrams.
The summation of the parquet diagrams entails a self-consistent
summation of both particle-particle and hole-hole ladder diagrams
and particle-hole diagrams. Practical solutions to this many-body
scheme for finite nuclei will be discussed here.

The coupled-cluster method originated in nuclear physics over
forty years ago when Coester and K\"ummel proposed an exponential ansatz
to describe correlations within a nucleus
\cite{coester58,coester60}. This ansatz has been well justified
for many-body problems using a formalism in which the
cluster functions are constructed by cluster operators acting on
a reference determinant \cite{harris92}. Early applications
to finite nuclei were described in Ref.~\cite{klz78}. From that
time to this, a systematic development and implementation
of this interesting many-body theory in nuclear
physics applications has been only sporadic. The view from
computational quantum chemistry is quite different.
In fact, coupled-cluster methods applied to computational chemistry
enjoy tremendous success
\cite{bartlett81,comp_chem_rev00,piotr1,helgaker,arponen97}
over a broad class of
chemistry problems related to chemical and
molecular structure and chemical reactions.
The method is fully
microscopic and is capable of systematic and hierarchical improvements.
Indeed, when one expands the cluster operator in coupled-cluster theory
to all $A$ particles in the system, one exactly produces the fully-correlated
many-body wave function of the system. The only input that the method
requires is the nucleon-nucleon interaction.
The method may also be extended
to higher-order interactions such as the three-nucleon interaction.
Second, the method is size extensive which means that only linked
diagrams appear in the computation of the
energy (the expectation value of the Hamiltonian) and amplitude equations.
Third, coupled-cluster theory is also size
consistent which means that the energy of two non-interacting fragments
computed separately is the same as that computed for both fragments
simultaneously. In chemistry, where the study of reactions
is quite important, this is a crucial property not available
in the interacting shell model (named configuration interaction in
chemistry).

A fourth interesting point involves a comparison of 
coupled-cluster theory and many-body theory. 
The computed energy using the coupled-cluster formalism includes 
a very large class of many-body perturbation theory diagrams.
In standard many-body perturbation theory, one typically sums
all diagrams order by order. The coupled-cluster approach essentially
iterates diagrams so that one may discuss it in terms of an infinite
summation of particular classes of diagrams. The infinite resummation
means that the coupled-cluster theory is nonperturbative. In fact, the 
coupled-cluster energy at the single and double excitation level
contains contributions identical to
those of second order and third order many-body perturbation theory, but
lacks triple excitation contributions necessary to complete fourth-order
many-body perturbation theory; see e.g., the review article of
Bartlett \cite{bartlett81}. Later in this review, we will compare 
second and third-order many-body perturbation theory to coupled-cluster
calculations. 

This work falls in eight sections.
In the next section we present various definitions pertinent
to the determination of effective interactions, with an emphasis
on perturbative methods.
The resummation of the ladder type of
diagrams is then presented in section \ref{sec:sec3}.
In that section we also discuss the summation of so-called
folded diagrams which arise in the evaluation of
valence space effective interactions. Further perturbative
corrections are also discussed. Selected results
for light nuclei in the $1s0d$ and $1p0f$ shells and for
several medium heavy nuclei in the mass region $A=100$ to $A=132$
are presented in the subsequent section.

The summation of the TDA and RPA class of diagrams and the so-called
parquet diagrams is discussed in
section \ref{sec:sec5} whereas section \ref{sec:sec6} is devoted to a discussion
of the coupled cluster method. Section \ref{sec:sec7} presents a critical discussion
of three-body effects in nuclear structure.

We conclude in section \ref{sec:sec8} with a discussion
on the extension of the methods discussed in sections \ref{sec:sec5} and
\ref{sec:sec6} to weakly bound nuclei.

\section{Many-body perturbation theory}
\label{sec:sec2}

In order to derive a microscopic approach to the effective interaction and/or operator
within the framework of perturbation theory, we need to introduce various
notations and definitions pertinent to the methods exposed.
In this section we review how to calculate an effective
operator within the framework of
degenerate Rayleigh-Schr\"{o}dinger
(RS) perturbation theory \cite{ko90,lm85}.

It is common practice in perturbation theory to reduce the infinitely
many degrees of freedom of the Hilbert space to those represented
by a physically motivated subspace, the model space.
In such truncations of the Hilbert space, the notions of a projection
operator $P$ onto the model space and its complement $Q$ are
introduced. The projection operators defining the model and excluded
spaces are defined by
\begin{equation}
        P=\sum_{i=1}^{D} \left|\Phi_i\right\rangle
        \left\langle\Phi_i\right |,
\end{equation}
and
\begin{equation}
        Q=\sum_{i=D+1}^{\infty} \left|\Phi_i\right\rangle
        \left\langle\Phi_i\right |,
\end{equation}
with $D$ being the dimension of the model space, and $PQ=0$, $P^2 =P$,
$Q^2 =Q$ and $P+Q=I$. The wave functions $\left|\Phi_i\right\rangle$
are eigenfunctions
of the unperturbed Hamiltonian $H_0 = T+U$, where $T$ is the kinetic
energy and $U$ and appropriately chosen one-body potential, that of the
harmonic oscillator (h.o.) in most calculations. The full Hamiltonian
is then rewritten as $H=H_0 +H_1$ with $H_1=V-U$, $V$ being e.g.\ the
$NN$    interaction. The eigenvalues
and eigenfunctions of the full Hamiltonian are denoted by
$\left|\Psi_{\alpha}\right\rangle$
and $E_{\alpha}$,
\begin{equation}
                H\left|\Psi_{\alpha}\right\rangle=
                E_{\alpha}\left|\Psi_{\alpha}\right\rangle.
\end{equation}
Rather than solving the full Schr\"{o}dinger equation above, we define
an effective Hamiltonian acting within the model space such
that
\begin{equation}
               PH_{\mathrm{eff}}P\left|\Psi_{\alpha}\right\rangle=
               E_{\alpha}P\left|\Psi_{\alpha}\right\rangle=
              E_{\alpha}\left|\Phi_{\alpha}\right\rangle
\end{equation}
where $\left|\Phi_{\alpha}\right\rangle=P\left|\Psi_{\alpha}\right\rangle$
is the projection of the full wave function
onto the model space, the model space wave function.
In RS perturbation theory, the effective interaction
$H_{\mathrm{eff}}$ can be written out order by order in the
interaction $H_1$ as
\begin{equation}
               PH_{\mathrm{eff}}P=PH_1P +PH_1\frac{Q}{e}H_1 P+
               PH_1\frac{Q}{e}H_1 \frac{Q}{e}H_1 P+\dots,
               \label{eq:effint}
\end{equation}
where terms of third and higher order also
include the aforementioned folded diagrams.
Further, $e=\omega -H_0$,
where $\omega$ is the so-called starting energy, defined as the unperturbed
energy of the interacting particles..
Similarly,
the exact wave
function $\left|\Psi_{\alpha}\right\rangle$
can now be written in terms of the model space wave function as
\begin{equation}
                \left|\Psi_{\alpha}\right\rangle=
                \left|\Phi_{\alpha}\right\rangle+
                \frac{Q}{e}H_1\left|\Phi_{\alpha}\right\rangle
                +\frac{Q}{e}H_1\frac{Q}{e}H_1\left|\Phi_{\alpha}\right\rangle
                +\dots
                \label{eq:wavef}
\end{equation}
The wave operator is often expressed as
\begin{equation}
              \Omega = 1 +\chi,
\end{equation}
where $\chi$ is known as the correlation operator. The correlation
operator generates the component of the wave function in the $Q$-space
and must therefore contain at least one perturbation. Observing
that $P\Omega P = P$, we see that the correlation operator $\chi$
has the properties
\begin{equation}
               P\chi P = 0, \hspace{1cm} Q\Omega P =
              Q\chi P =\chi P. \label{eq:chi1}
\end{equation}
Since  $\left|\Psi_i\right\rangle=\Omega\left|\Psi_i^{M}\right\rangle$
determines the wave operator
only when it operates to the right on the model space, i.e., only the
$\Omega P$  part is defined, the term $\Omega Q$
never appears in the theory,
and we could therefore add the conditions $Q\chi Q =0$ and $P\chi Q =0$
to Eq.\ (\ref{eq:chi1}). This leads to the following choice for $\chi$
\begin{equation}
                   \chi = Q\chi P. \label{eq:chi2}
\end{equation}
This has been the traditional choice in perturbation theory \cite{so95,lm85}.

\begin{figure}[hbtp]
\begin{center}
      \setlength{\unitlength}{1mm}
      \begin{picture}(100,100)
      \put(0,0){\epsfxsize=12cm \epsfbox{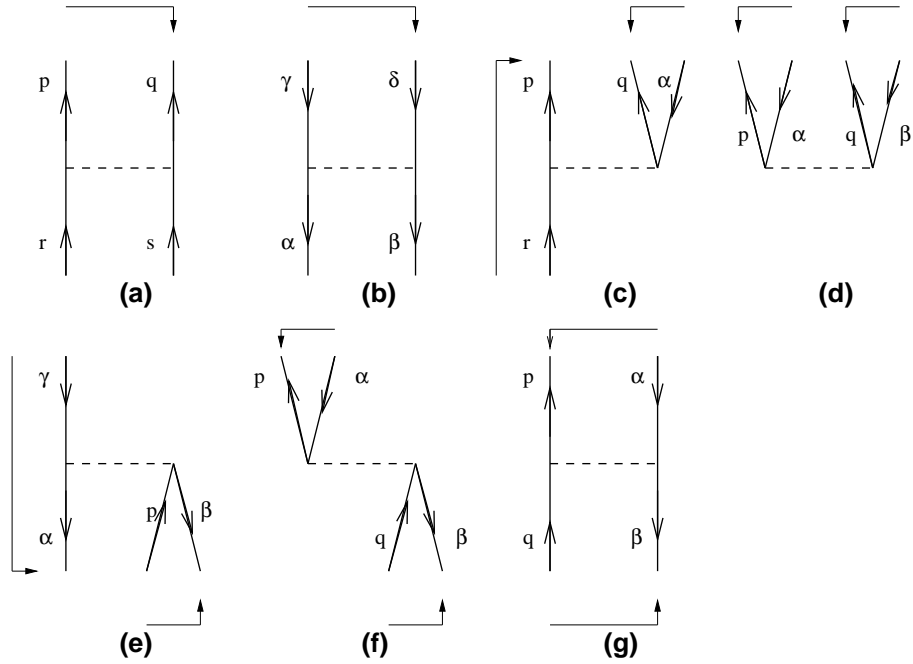}}
      \end{picture}
      \caption{The various vertices to first order in the interaction
               $V$ which contribute to the wave operator
               $\Omega =1+\chi$. Hartree-Fock
               terms are not included. Possible hermitian conjugate
                diagrams are not shown. Indicated are also possible
               angular momentun coupling orders.}
      \label{fig:wavef1}
\end{center}
\end{figure}

The wave operator $\Omega$ can then be ordered in terms of the number
of interactions with the perturbation $H_1$
\begin{equation}
              \Omega = 1 +\Omega^{(1)} + \Omega^{(2)}+\dots ,
\end{equation}
where $\Omega^{(n)}$ means that we have $n$ $H_1$ terms.
Explicitely, the above
equation reads
\begin{eqnarray}
         \Omega\left|\Phi_i\right\rangle=
         &{\displaystyle\left|\Phi_i\right\rangle
         +\sum_{\alpha}\frac{\left|\alpha\right\rangle
         \left\langle\alpha\right|
          V\left|\Phi_i\right\rangle}{\varepsilon_i -\varepsilon_{\alpha}}
         +\sum_{\alpha\beta}\frac{\left|\alpha\right\rangle
        \left\langle\alpha\right| V
         \left|\beta\right\rangle\left\langle\beta\right| V
         \left|\Phi_i\right\rangle }
         {(\varepsilon_i-\varepsilon_{\alpha})
       (\varepsilon_i-\varepsilon_{\beta})} }\\   \label{eq:wavefu}\nonumber
&       {\displaystyle  -\sum_{\alpha j}\frac{\left|\alpha\right\rangle
       \left\langle\alpha\right|
         V\left|\Phi_j\right\rangle
        \left\langle\Phi_j\right| V\left|\Phi_i\right\rangle}
       {(\varepsilon_i-\varepsilon_{\alpha})
      (\varepsilon_i-\varepsilon_{j})} }
       +\dots ,
\end{eqnarray}
where $\varepsilon$ are the unperturbed energies of the $P$-space
and $Q$-space states defined by $H_0$.
The greek letters refer to
$Q$-space states, whereas a latin letter refers to model-space
states.   The second term
in the above equation corresponds to $\Omega^{(1)}$ while the third
and fourth define $\Omega^{(2)}$.
Note that the fourth term diverges
in case we have a degenerate or nearly degenerate model space. It is
actually divergencies like these which are to be removed by the folded
diagram procedure to be discussed in the next section. Terms like these
arise due to the introduction of an energy independent perturbative
expansion. Conventionally, the various contributions to the
perturbative expansion are represented by Feynman-Goldstone diagrams.
In Fig.\ \ref{fig:wavef1} we display the topologically distinct
contributions to first order in the
interaction $V$ to
the wave operator Eq.\ (\ref{eq:wavefu}). These diagrams all
define the correlation operator $\chi$ to first order in $V$.
Diagrams with Hartree-Fock contributions
and single-particle potential terms
are not included. The possible
renormalizations of these diagrams will be discussed
in the next four sections.

\begin{figure}[hbpt]
\begin{center}
      \setlength{\unitlength}{1mm}
      \begin{picture}(100,100)
      \put(0,0){\epsfxsize=12cm \epsfbox{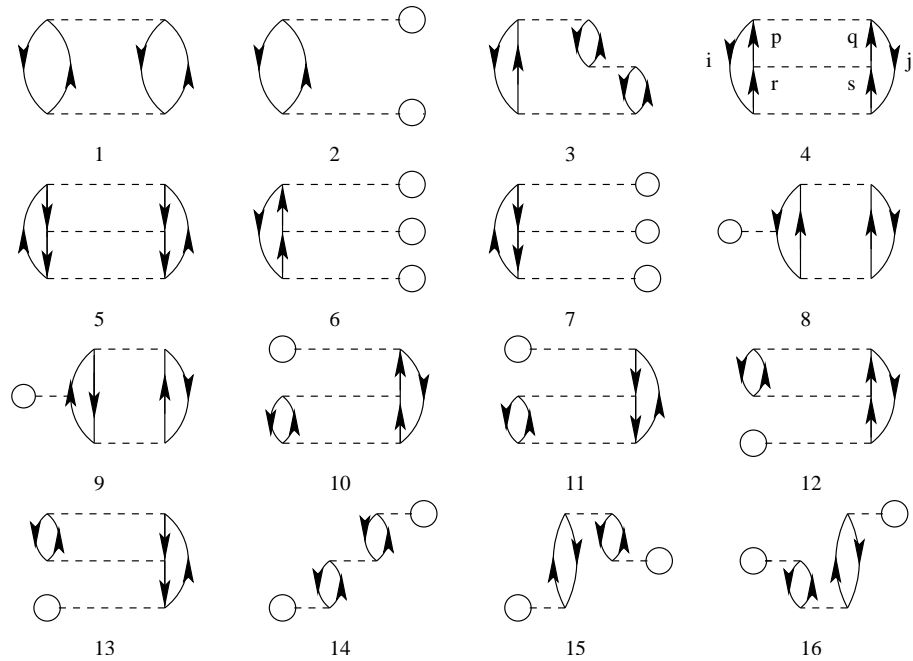}}
      \end{picture}
\caption{Antisymmetrized Goldstone diagrams through third order in perturbation
theory included
in the evaluation of the binding energy. The dashed lines represents the interaction.
Particle and hole states are represented by upward  and downward arrows, respectively.
The first order diagram is omitted. All closed circles stand for a summation
over hole states.}
\end{center}
\label{fig:diagrams}
\end{figure}

The linked-diagram theorem \cite{lm85,br86} can be used to obtain a perturbative
expansion  for the energy
in terms of the perturbation $V$ or  $V=H-H_0$ where $H_0$ represents the unperturbed
part of the Hamiltonian. The expression for the energy $E$ reads
\begin{equation}
  E = \sum_{k=0}^{\infty} \left\langle \Psi_0 \right| H\left[(\omega - H_0)^{-1}H\right]^k
      \left|\Psi_0\right\rangle_{L},
\end{equation}
where $\Psi_0$ is the uncorrelated Slater determinant for the ground state, $\omega$ is
the corresponding
unperturbed  energy and the subscript $L$ stands for linked diagrams only.
In Fig.~\ref{fig:diagrams} we show all antisymmetrized Goldstone diagrams for closed-shell
systems through third order
in perturbation theory (we omit the first order diagram). All closed circles stand for a summation
over hole states. 
The expression for e.g., diagram (4) $\Delta E_4$ in Fig.~\ref{fig:diagrams} reads
in an angular momentum coupled basis (with  $J$ being the total two-body angular
momentum)
\begin{eqnarray}
\Delta E_4 &=& \frac{1}{8}\sum_{\begin{array}{c}ij\leq F\\ pqrs > F\\ J\end{array}}(2J+1)
\left\langle (ij)J\right | V\left | (pq)J 
\right\rangle\frac{1}{\varepsilon_i+\varepsilon_j-\varepsilon_p-\varepsilon_q} \\ \nonumber 
&& \times \left\langle (pq)J\right | V\left | (rs)J
\right\rangle\frac{1}{\varepsilon_i+\varepsilon_j-\varepsilon_r-\varepsilon_s}
\left\langle (rs)J\right |V \left | (ij)J
\right\rangle.
\end{eqnarray}
The symbol $F$ represents the last hole, or Fermi surface.
In the next section we will replace the bare interaction with the so-called $G$-matrix.

\begin{figure}[hbtp]
\begin{center}
      \setlength{\unitlength}{1mm}
      \begin{picture}(100,80)
      \put(0,0){\epsfxsize=10cm \epsfbox{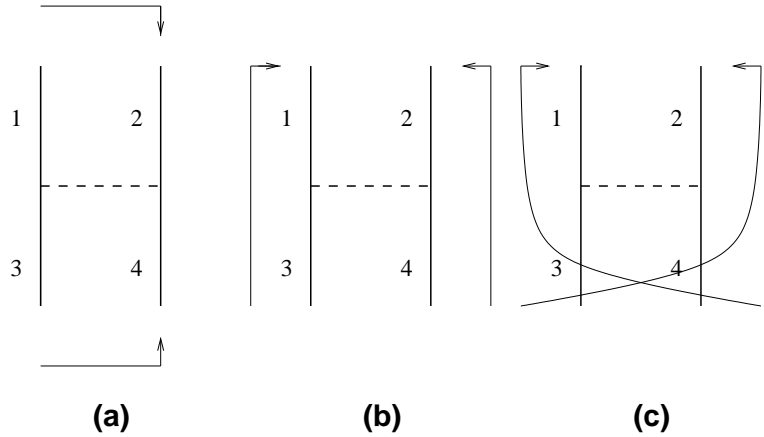}}
      \end{picture}
      \caption{Coupling order for the $[12]$ (a), $[13]$ (b) and
               $[14]$ (c) channels.}
      \label{fig:channelsdef}
\end{center}
\end{figure}

We end this section with the equations for the diagrams
in Fig.\ \ref{fig:wavef1} representing $\chi$ to first order in $V$.
Moreover, in order to introduce the various channels needed to sum the
parquet class of diagrams, we will find it convenient here to
classify these channels in terms of angular momentum
recouplings. Later on, we will also introduce the pertinent
definitions of energy and momentum variables in the various channels.
The nomenclature we will follow in our labelling is that
of Blaizot and Ripka, see Ref.\ \cite{br86} chapter 15.
All matrix elements in the definitions below
are antisymmetrized and unnormalized.
The first channel is the $[12]$ channel, or the $s$-channel
in field theory, and its angular momentum coupling order
is depicted in Fig.\ \ref{fig:channelsdef}.
In this figure
we do not distinguish between particles and holes, all single-particle
labels $1,2,3,4$ can represent either a hole or particle
single-particle state. It is the coupling order which is
of interest here.
The matrix element $V^{[12]}$ in the $[12]$ channel is then
\begin{equation}
       V_{1234J}^{[12]}
       =\left\langle (12)J\right | V
       \left | (34)J\right\rangle,
       \label{eq:12channel}
\end{equation}
meaning that the single-particle state $1(3)$ couples to the state
$2(4)$ to yield a total angular momentum $J$.
This channel is commonly denoted as the particle-particle (pp)
or particle-particle-hole-hole (pphh) channel, meaning that when
we will sum classes of diagrams to infinite order in this channel, the only
intermediate states which are allowed are those of a pphh character,
coupled to a final $J$ in the above order.
In the next section we will explicitely discuss ways to evaluate
the equations for the $[12]$ channel.
This coupling order is also the standard way of representing
effective interactions for shell-model
calculations.
If we now specialize to particles and holes (these matrix
elements were shown in Fig.\ \ref{fig:wavef1}) we obtain for the case
with particle states only\footnote{Note that we only
include angular momentum factors,
other
factors coming from the diagram rules\cite{kstop81},
like number of hole lines,
number of closed loops, energy denominators
etc.\ are omitted here.}, diagram (a),
\begin{equation}
      V_{\mathrm{2p}}=V_{pqrs J}^{[12]}=
       \left\langle (pq)J\right | V\left | (rs)J\right\rangle.
       \label{eq:2pv}
\end{equation}
The corresponding one for holes only, diagran (b), is
\begin{equation}
      V_{\mathrm{2h}}=V_{\alpha\beta\gamma\delta J}^{[12]}=
       \left\langle (\alpha\beta)J\right | V
       \left | (\gamma\delta)J\right\rangle.
       \label{eq:2hv}
\end{equation}
Thus, in the forthcoming discussion, we will always employ as our
basic notation for a matrix element that of the $[12]$ channel,
meaning that matrix elements of the other two channels
can always be rewritten in terms of those in $[12]$ channel
We see this immediately by looking at the expression for the
matrix element in the $[13]$ channel, the $t$-channel in field
theory, see Fig.\
\ref{fig:channelsdef}(b). Here the single-particle state $3(4)$
couples to the single-particle state $1(2)$\footnote{In a Goldstone-Feynman
diagram in an angular momentum representation, the coupling direction
will always be from incoming single-particle states to outgoing
single-particle states.}.
Through simple angular momentum algebra we have
\begin{equation}
      V_{1234J}^{[13]}=
      {\displaystyle \sum_{J'}}(-)^{j_1+j_4+J+J'}\hat{J'}^2
      \left\{
      \begin{array}{ccc}
       j_3&j_1&J\\j_2&j_4&J'
      \end{array}
       \right\}V_{1234J}^{[12]},
       \label{eq:13channel}
\end{equation}
where the symbol with curly brackets represents a $6j$-symbol and
$\hat{J'}=\sqrt{2J'+1}$.
In a similar way we can also express the matrix
element in the $[14]$ channel, the $u$-channel in field theory,
through
\begin{equation}
       V_{1234J}^{[14]}=
      {\displaystyle \sum_{J'}}(-)^{j_1+j_4+J+2j_3}\hat{J'}^2
      \left\{
      \begin{array}{ccc}
       j_4&j_1&J\\j_2&j_3&J'
      \end{array}
       \right\}
       V_{1234J}^{[12]}.
       \label{eq:14channel}
\end{equation}
It is also possible to have the inverse relations or to express
e.g., the $[14]$ channel through the $[13]$ channel as
\begin{equation}
       V_{1234J}^{[14]}=
      {\displaystyle \sum_{J'}}(-)^{2j_1+2j_2+2j_3}\hat{J'}^2
      \left\{
      \begin{array}{ccc}
       j_4&j_1&J\\j_3&j_2&J'
      \end{array}
       \right\}
       V_{1234J}^{[13]}.
       \label{eq:1413channel}
\end{equation}
The matrix elements defined in Eqs.\
(\ref{eq:12channel})-(\ref{eq:1413channel}) and the inverse relations
are the
starting points for various resummation of diagrams.
In the next section we will detail ways of solving equations
in the $[12]$ channel, whereas various approximations for the
$[13]$ channel and $[14]$ channel such as the TDA and RPA
and  vertex and propagator renormalization schemes
will be discussed in section \ref{sec:sec5}. Finally, how to
merge self-consistently  all three channels will also be discussed
in section \ref{sec:sec5}.

We end this section by giving the expressions in an angular momentum
basis for the remaining
diagrams of Fig.\ \ref{fig:wavef1}. The coupling order is indicated
in the same figure.

The 2p1h vertex $V_{\mathrm{2p1h}}$,
diagram (c) in  Fig.\  \ref{fig:wavef1},
is coupled
following the prescription of the $[13]$ channel and reads
\begin{equation}
      V_{\mathrm{2p1h}}=V_{pqr\alpha J}^{[13]}=
      {\displaystyle \sum_{J'}}(-)^{j_{\alpha}+j_p+J+J'}\hat{J'}^2
      \left\{
      \begin{array}{ccc}
       j_r&j_p&J\\j_q&j_{\alpha}&J'
      \end{array}
       \right\}
       V_{pqr\alpha J'}^{[12]}.
       \label{eq:2p1hv}
\end{equation}
The 2p2h ground-state correlation $V_{\mathrm{2p2h}}$, diagram (d),
 which will enter
in the RPA summation discussed in section \ref{sec:sec5} is given by,
the coupling order is that of the $[13]$ channel,
\begin{equation}
      V_{\mathrm{2p2h}}=V_{pq\alpha\beta J}^{[13]}=
      {\displaystyle \sum_{J'}}(-)^{j_{\beta}+j_p+J+J'}\hat{J'}^2
      \left\{
      \begin{array}{ccc}
       j_{\alpha}&j_p&J\\j_q&j_{\beta}&J'
      \end{array}
       \right\}
       V_{pq\alpha\beta J'}^{[12]}.
       \label{eq:2p2hv}
\end{equation}
The 2h1p vertex $V_{\mathrm{2h1p}}$, diagram (e),
still in the representation of
the $[13]$ channel, is defined as
\begin{equation}
      V_{\mathrm{2h1p}}=V_{\alpha\beta\gamma p J}^{[13]}=
      {\displaystyle \sum_{J'}}(-)^{j_{\alpha}+j_p+J+J'}\hat{J'}^2
      \left\{
      \begin{array}{ccc}
       j_{\gamma}&j_{\alpha}&J\\j_{\beta}&j_p&J'
      \end{array}
       \right\}
       V_{\alpha\beta\gamma p J'}^{[12]}.
       \label{eq:2h1pv}
\end{equation}
Note well that the vertices of Eqs.\ (\ref{eq:2p1hv})-(\ref{eq:2h1pv})
and their respective
hermitian conjugates can all be expressed in the $[14]$ channel
or $[12]$ channel as well.
However, it is important to note that the expressions in the various
channels are different, and when solving the equations for the various
channels, the renormalizations will be different. As an example, consider
the two particle-hole vertices $V_{\mathrm{ph}}$
of Fig.\ \ref{fig:wavef1}, i.e., diagrams (f) and (g).
Diagram (g) is just the exchange diagram of (f) when seen in the
$[12]$ channel. However, if (f) is coupled as in the $[13]$ channel,
recoupling this diagram to the $[14]$ channel will not give
two particle-hole
two-body states coupled to a final $J$ but rather
a particle-particle two-body state and a hole-hole  two-body state.
But why bother at all about such petty details? The problem arises when we
are to  sum diagrams in the $[13]$ channel and $[14]$ channel.
In the $[12]$ channel we allow only particle-particle and hole-hole
intermediate states, whereas in the $[13]$ channel and $[14]$ channel
we allow only particle-hole intermediate states, else we may risk
to double-count various contributions.
If we therefore recouple
diagram (f) to the $[14]$ representation, this contribution
does not yield an intermediate particle-hole state
in the $[14]$ channel.
Thus, diagram (f), whose expression is
\begin{equation}
      V_{\mathrm{ph}}=V_{p\beta \alpha q J}^{[13]}=
      {\displaystyle \sum_{J'}}(-)^{j_p+j_q+J+J'}\hat{J'}^2
      \left\{
      \begin{array}{ccc}
       j_{\alpha}&j_p&J\\j_{\beta}&j_q&J'
      \end{array}
       \right\}
       V_{p\beta \alpha q J'}^{[12]},
       \label{eq:ph13}
\end{equation}
yields a particle-hole contribution only in the $[13]$ channel,
whereas the exchange diagram (g), which reads
\begin{equation}
      V_{\mathrm{ph}}=V_{p\beta q\alpha J}^{[14]}=
      {\displaystyle \sum_{J'}}(-)^{2j_q+j_{\alpha}+j_p+J}\hat{J'}^2
      \left\{
      \begin{array}{ccc}
       j_{\alpha}&j_p&J\\j_{\beta}&j_{q}&J'
      \end{array}
       \right\}
       V_{p\beta q\alpha J'}^{[12]},
       \label{eq:ph14}
\end{equation}
results in the corresponding particle-hole contribution in the
$[14]$ channel.
In electron gas theory, the latter expression
is often identified as the starting point for the self-screening
of the exchange term. In the discussion of the TDA series in
section \ref{sec:sec4} we will give the expressions for the screening
corrections based on Eqs.\ (\ref{eq:ph13}) and (\ref{eq:ph14}).

An important aspect to notice in connection with the latter
equations and the discussions in  section \ref{sec:sec5} is that
\begin{equation}
    V_{p\beta q\alpha J}^{[14]}=-V_{p\beta \alpha q J}^{[13]},
\end{equation}
i.e., just the exchange diagram, as it should be.
This is however important to keep in mind, since we later
on will sum explicitely sets of diagrams in the
$[13]$ channel and the $[14]$ channel, implying thereby that
we will obtain screening and vertex corrections
for  both direct and exchange  diagrams.

\section{Diagrams in the $[12]$ channel}
\label{sec:sec3}

\begin{figure}[hbtp]
\begin{center}
      \setlength{\unitlength}{1mm}
      \begin{picture}(100,100)
      \put(0,0){\epsfxsize=10cm \epsfbox{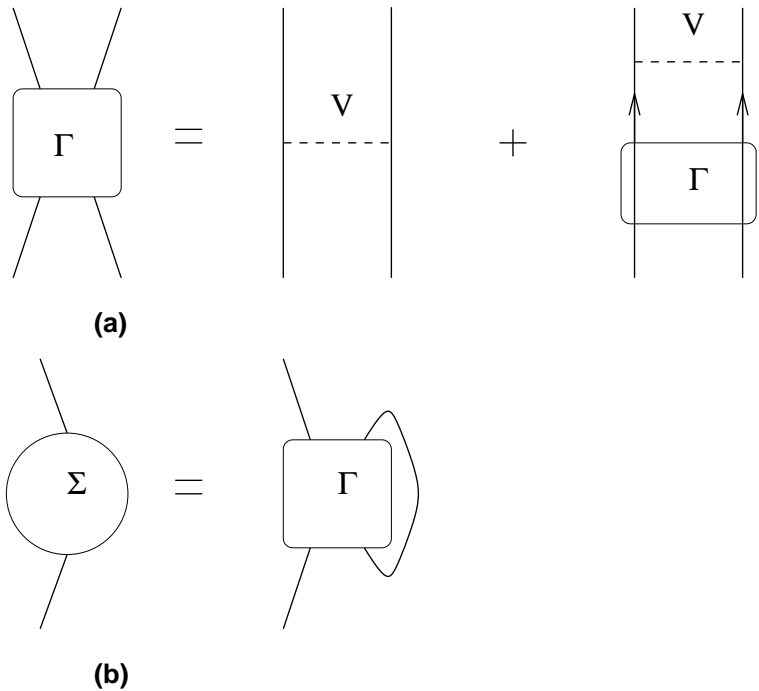}}
      \end{picture}
      \caption{(a) represents the two-body vertex $\Gamma$ function while (b)
               represents the self-energy $\Sigma$.}
      \label{fig:selfcons12}
\end{center}
\end{figure}

In order to write down the equation for the renormalized
interaction $\Gamma^{[12]}$ in the
$[12]$ channel we need first to present some further definitions.
We will also assume that the reader has some familiarity with the theory
of Green's function. In our presentation below we will
borrow from the monograph of Blaizot and Ripka \cite{br86}, see also the recent review 
articles of Dickhoff and Barbieri \cite{db04} and M\"uther and Polls \cite{mp00}.
The vertex $\Gamma^{[12]}$ is in lowest order identical with the
interaction $V^{[12]}$ and obeys also the same symmetry relations
as $V$, i.e.,
\begin{equation}
     \Gamma^{[12]}_{1234J}=\Gamma^{[12]}_{2143J}=-\Gamma^{[12]}_{2134J}=
     \Gamma^{[12]}_{1243J}.
     \label{eq:symproperties}
\end{equation}
We also need to define energy variables. Since we are going to
replace the interaction $V$ with the $G$-matrix, or certain
approximations to it,  defined below in all
of our practical calculations, the momentum variables are already
accounted for in $G$. The basis will be that of harmonic oscillator
wave functions, and the labels $1234$ will hence refer to oscillator
quantum numbers, which in turn can be related to the momentum
variables. The labels $1234$, in addition to representing
single-particle quantum numbers, define also the energy of the single-particle
states. With a harmonic oscillator basis, the starting point for the
single-particle energies $\varepsilon_{1,2,3,4}$ are the unperturbed
oscillator energies. When iterating the equations for $\Gamma^{[12]}$,
self-consistent single-particle energies can be introduced.
The total energy in the $[12]$ channel $s$ is
\begin{equation}
    s=\varepsilon_1+\varepsilon_2=\varepsilon_3+\varepsilon_4.
    \label{eq:energy12}
\end{equation}
The equation for the vertex $\Gamma^{[12]}$ is,
in a compact matrix notation, given by \cite{br86}
\begin{equation}
     \Gamma^{[12]}=V^{[12]}+V^{[12]}(gg)\Gamma^{[12]},
     \label{eq:schematic12}
\end{equation}
where $g$ is the one-body Green's function representing
the intermediate states.
The diagrammatic expression for this equation is
given in Fig.\ \ref{fig:selfcons12}.
The expression of Eq.\ (\ref{eq:schematic12}) is known as the Feynman-Galitskii
equation. This equation is normally solved iteratively.
In the first iteration
the irreducible
vertex $V^{[12]}$ is then often chosen as the bare $NN$
interaction. This interaction is then typically assumed to be energy
independent and we can drop the $s$ dependence of $V^{[12]}$. Moreover,
the single-particle propagators are chosen as the
unperturbed ones. The first iteration of
Eq.\ (\ref{eq:schematic12}) can then be rewritten as
\begin{equation}
      \Gamma^{[12]}_{1234J}(s) =
      V^{[12]}_{1234J}+\frac{1}{2}
      \sum_{56}
      V^{[12]}_{1256J}\hat{{\cal G}}^{[12]}
      \Gamma^{[12]}_{5634J}(s).
      \label{eq:first12}
\end{equation}
We have defined 
the unperturbed particle-particle and hole-hole propagators
\begin{equation}
    \hat{{\cal G}}^{[12]}=
    \frac{Q^{[12]}_{\mathrm{pp}}}{s-\varepsilon_5-\varepsilon_6+\imath \eta}-
    \frac{Q^{[12]}_{\mathrm{hh}}}{s-\varepsilon_5-\varepsilon_6-\imath \eta},
    \label{eq:paulioperator12}
\end{equation}
which results from the integration over the energy variable
in the product of the two single-particle
propagators in Eq.\ (\ref{eq:schematic12}).
In our discussions we will not deal with dressed propagators, 
neither for the one-particle nor the two-particle propagators. 
For this and related topics such as the spreading of single-particle strength we refer to
the work of Dickhoff and Barbieri \cite{db04}. In our discussion of the 
coupled cluster method however, such features can be extracted at the end of the calculations 
through various expectation values. These expectation values will however depend on the actual
size of the model space used. In our case this applies to the number of harmonic oscillator
shells. 

The factor $1/2$ in Eq.~(\ref{eq:first12}) follows from
one of the standard Goldstone-Feynman diagram
rules \cite{kstop81}, which state
that a factor $1/2$ should be associated with each pair of lines
which starts at the same interaction vertex and ends at the same
interaction vertex. This rule follows from the fact 
that we sum freely over the intermediate single-particle states
$56$.
The reader should note that the intermediate states $56$
can represent a two-particle state or a two-hole state.
In Eq.\ (\ref{eq:paulioperator12}) we have assumed unperturbed single-particle
energies.
In our iterations we will approximate the single-particle energies
with their real part only. Thus,
the two-particle propagator
$\hat{{\cal G}}^{[12]}$
with renormalized single-particle energies has the same
form as the unperturbed one.
The operators $Q^{[12]}_{\mathrm{pp}}$ and $Q^{[12]}_{\mathrm{hh}}$
ensure that the intermediate states are of two-particle
or two-hole character.
In order to obtain a self-consistent scheme, Eq.\ (\ref{eq:first12})
has also to be accompanied with the
equation for the single-particle propagators $g$
given by Dyson's equation
\begin{equation}
    g=g_0+g_0\Sigma g,
    \label{eq:dyson12}
\end{equation}
with $g_0$ being the unperturbed single-particle
propagator and $\Sigma$ the self-energy. We will however defer a discussion
of these quantities to section \ref{sec:sec5}. Here it will suffice to state
that  the self-energy is related to the vertex
$\Gamma^{[12]}$ as
\begin{equation}
      \Sigma \sim g\Gamma.
      \label{eq:sigma12}
\end{equation}
The similarity sign is meant to indicate that, although being formally
correct, great care has to be exercised in order not to double-count
contributions to the self-energy \cite{jls82}.
The set of equations for the vertex function and the self-energy
is shown pictorially in Fig.\ \ref{fig:selfcons12}.
Assume now that we have performed the first iteration. The question which now
arises  is whether the obtained vertex $\Gamma^{[12]}$ from the solution of
Eq.\ (\ref{eq:first12}) should replace the bare vertex $V^{[12]}$
in the next iteration. Before answering this question, let us give some examples
of diagrams which can be generated from the first iteration.
\begin{figure}[hbtp]
\begin{center}
      \setlength{\unitlength}{1mm}
      \begin{picture}(100,80)
      \put(0,0){\epsfxsize=10cm \epsfbox{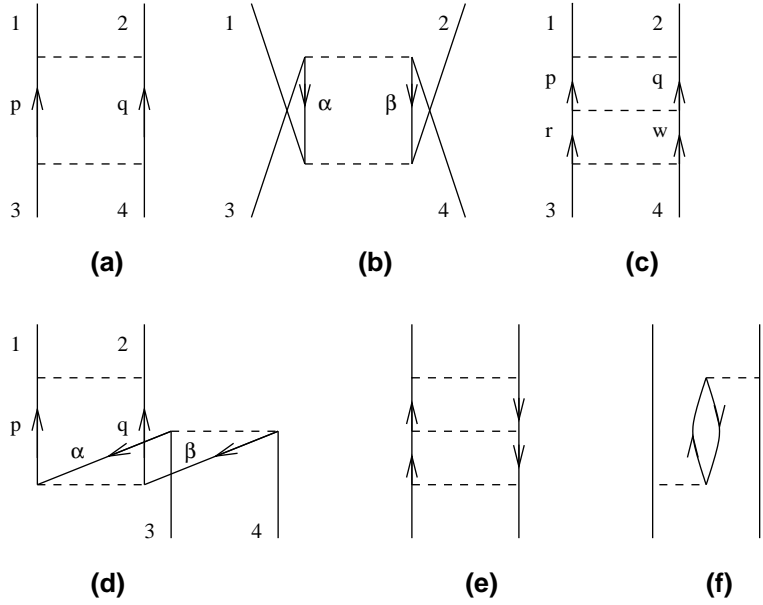}}
      \end{picture}
      \caption{Diagrams (a)-(d) give examples of
               diagrams which are summed up by
               the use of Eq.\ (\protect{\ref{eq:schematic12}}).
               Diagrams (e) and (f) are examples of core-polarization
               terms which are not generated by the $[12]$ channel.}
      \label{fig:gamma12}
\end{center}
\end{figure}
These contributions
are given by e.g., diagrams (a)-(d) in Fig.\ \ref{fig:gamma12}.
Diagrams (a) and (b) are examples of contributions to second order
in perturbation theory, while diagrams (c) and (d) are higher
order terms. Diagrams (e) and (f) are higher-order core-polarization terms,
which can e.g., be generated through the solution of the equations for
the $[13]$ and $[14]$ channels discussed in the next section.
If we were to include diagrams (a)-(d) in the definition of the bare
vertex in our next iteration,
we are prone to double-count, since such contributions are generated
once again. Diagrams which contain particle-hole intermediate state are however
not generated by the solution of Eq.\ (\ref{eq:first12}).
We need therefore to define the vertex $V^{[12]}$ used in every iteration
to be the sum of diagrams which are irreducible in the $[12]$ channel.
With irreducible we will mean all diagrams which cannot be reduced to
a piece containing the particle states $12$  entering or leaving
the same interaction vertex
and another   part containing the states $34$ at the same interaction
vertex by cutting two internal lines.
Clearly, if we cut diagrams (a) and (b) we are just left with two bare
interaction vertices. Similarly, cutting two lines of an intermediate
state in diagrams (c) and (d) leaves us with two second-order terms
of the type (a) and (b) and one bare interaction.
Diagrams (e) and (f) are however examples of diagrams which are
irreducible in the $[12]$ channel. Diagram (e) is
irreducible in the $[13]$ channel, but not in the $[14]$ channel.
Similarly, diagram (g) is reducible in the $[13]$ channel and
irreducible in the $[14]$ channel.
This means that, unless we solve equations similar to Eq.\ (\ref{eq:first12})
in the $[13]$ channel and $[14]$ channels as well, changes
from further iterations
of Eq.\ (\ref{eq:first12}) will only come from
the single-particle terms defined
by Dyson's equation in Eq.\ (\ref{eq:dyson12}).

In the remaining part of this section, we will try to delineate ways
of solving the above equations, and discuss possible approximations,
their merits and faults. First of all, we will reduce
the propagator of  Eq.\ (\ref{eq:paulioperator12}) to only include
particle-particle intermediate states. This will lead us to the familiar
$G$-matrix in nuclear many-body theory.
Based on the $G$-matrix, we will construct effective interactions
through perturbative summations. Applications of such
effective interacions to selected nuclei will then be discussed.
Thereafter, we will try to
account for hole-hole contributions and 
self-consistent determinations of the single-particle
energies through the solution of Dyson's equation.

\subsection{\it The $G$-matrix}

In nuclear structure and nuclear matter calculations one has to
face the problem that any realistic $NN$ interaction $V$
exhibits a strong short-range repulsion, which in turn makes
a perturbative treatment of the nuclear many-body problem
prohibitive. If the interaction has a so-called hard core,
the matrix elements of such an interaction $\bra{\psi}V\ket{\psi}$
evaluated
for an uncorrelated two-body wave function $\psi (r)$ diverge,
since the uncorrelated wave function is different from zero also for
relative distances $r$ smaller than the hard-core radius. Similarly,
even if one uses interactions with softer cores, the matrix elements of the
interaction become very large at short distances.
The above problem was however overcome by
introducing the reaction matrix $G^{[12]}$ (displayed by  the summation
of ladder type of diagrams in Fig.\ \ref{fig:gamma12}
with particle-particle intermediate states only),
accounting thereby for short-range two-nucleon correlations.
The $G^{[12]}$-matrix represents just a subset to the solution
of the equations for the interaction $\Gamma^{[12]}$ in the $[12]$ channel,
we have clearly neglected the possibility of having intermediate states
which are of the hole-hole type.
The matrix elements of the
interaction $V^{[12]}$ then become
\begin{equation}
\bra{\psi}G^{[12]}\ket{\psi} =\bra{\psi}V^{[12]}\ket{\Psi}
\end{equation}
where $\Psi$ is now the correlated wave function
containing selected correlation from the excluded space.
By accounting for these
correlations in the two-body wave functon $\Psi$, the matrix elements of
the interaction become finite, even for a hard-core interaction $V$. Moreover,
as will be discussed below, compared with the uncorrelated
wave function, the correlated wave function enhances the
matrix elements of $V$ at distances for which the interaction is
attractive.
The type of correlations which typically are included in the evaluation
of the $G^{[12]}$-matrix are those of the two-particle type.
If we label the operator $Q$ in this case by $Q^{[12]}_{\mathrm{pp}}$, we can write
the integral equation for the $G$-matrix as
\begin{equation}
   G^{[12]}(s)=V^{[12]}+V^{[12]}\frac{Q^{[12]}_{\mathrm{pp}}}
           {s -H_0+\imath \eta}G^{[12]}(s),
   \label{eq:g1}
\end{equation}
implicitely assuming that $\lim \eta \rightarrow \infty$.
The variable $s$ represents normally the unperturbed
energy of the incoming two-particle state. We will suppress $\imath \eta$ in the
following equations. Moreover, since one is often interested only in the
$G^{[12]}$-matrix
for negative starting energies, the $G^{[12]}$-matrix commonly used in studies
of effective interactions has no divergencies. Note also, that compared with 
Eq.~(\ref{eq:first12}), we express the $G$-matrix in terms of operators. The explicit form,
with e.g., the sum over imtermediate states is implicitely included here.
We can also write
\begin{equation}
      G^{[12]}(s )=V^{[12]}+V^{[12]}Q^{[12]}_{\mathrm{pp}}
      \frac{1}{s -Q^{[12]}_{\mathrm{pp}}H_0Q^{[12]}_{\mathrm{pp}}}
      Q^{[12]}_{\mathrm{pp}}
      G^{[12]}(s ).
      \label{eq:g2}
\end{equation}
The former equation applies if the Pauli operator $Q^{[12]}_{\mathrm{pp}}$  commutes
with the unperturbed Hamiltonian $H_0$, whereas the latter is
needed if $[H_0,Q^{[12]}_{\mathrm{pp}}]\neq 0$.
Similarly, the correlated wave function $\Psi$
is given as
\begin{equation}
    \ket{\Psi}=\ket{\psi}+
    \frac{Q^{[12]}_{\mathrm{pp}}}{s - H_0}G^{[12]}\ket{\psi},
    \label{eq:wave}
\end{equation}
or
\begin{equation}
   \ket{\Psi}=\ket{\psi}+Q^{[12]}_{\mathrm{pp}}\frac{1}
    {s - Q^{[12]}_{\mathrm{pp}}H_0Q^{[12]}_{\mathrm{pp}}}
    Q^{[12]}_{\mathrm{pp}}G^{[12]}\ket{\psi}.
\end{equation}

In order to evaluate the $G^{[12]}$-matrix for finite nuclei,
we define first a useful identity following Bethe, Brandow and
Petschek \cite{bbp63}. Suppose we have two
different $G$-matrices\footnote{For notational economy,
we drop the superscript $^{[12]}$. Furthermore,
in the subsequent discussion in
this subsection it is understood that all operators $Q$
refer to particle-particle intermediate states only. The subscript
$\mathrm{pp}$ is also dropped.}, defined by
\begin{equation}
    G_1=V_1+V_1\frac{Q_1}{e_1}G_1,
\end{equation}
and
\begin{equation}
    G_2=V_2+V_2\frac{Q_2}{e_2}G_2,
\end{equation}
where $Q_1/e_1$ and $Q_2/e_2$ are the propagators of
either Eq.\ (\ref{eq:g1}) or Eq.\ (\ref{eq:g2}). $G_1$ and $G_2$
are two different $G$-matrices having two different interactions
and/or different propagators. We aim at an identity
which will enable us to calculate $G_1$ in terms of $G_2$,
or vice versa.
Defining the wave operators
\begin{equation}
    \Omega_1=1+\frac{Q_1}{e_1}G_1,
\end{equation}
and
\begin{equation}
    \Omega_2=1+\frac{Q_2}{e_2}G_2,
\end{equation}
we can rewrite the above $G$-matrices as
\begin{equation}
    G_1=V_1\Omega_1,
    \label{eq:omega1}
\end{equation}
and
\begin{equation}
    G_2=V_2\Omega_2.
    \label{eq:omega2}
\end{equation}
Using these relations, we rewrite $G_1$ as
\begin{eqnarray}
   G_1=&G_1 -{\displaystyle
         G_2^{\dagger}\left(\Omega_1-1-\frac{Q_1}{e_1}G_1\right)
        +\left(\Omega_2^{\dagger}-1-G_2^{\dagger}\frac{Q_2}{e_2}\right)G_1}
         \nonumber \\
       =&{\displaystyle G_2^{\dagger} +G_2^{\dagger}\left(\frac{Q_1}{e_1}-
        \frac{Q_2}{e_2}\right)G_1
        +\Omega_2^{\dagger}G_1 -G_2^{\dagger}\Omega_1},
\end{eqnarray}
and using Eqs.\ (\ref{eq:omega1}) and (\ref{eq:omega2}) we obtain
the identity
\begin{equation}
        G_1=G_2^{\dagger} +G_2^{\dagger}
        \left(\frac{Q_1}{e_1}-\frac{Q_2}{e_2}\right)G_1
        +\Omega_2^{\dagger}(V_1-V_2)\Omega_1.
        \label{eq:gidentity}
\end{equation}
The second term on the rhs.\ is called the propagator-correction term;
it vanishes if $G_1$ and $G_2$ have the same propagators. The third term
is often referred to as the potential-correction term, and it disappears
if $G_1$ and $G_2$  have the same potentials.
\begin{figure}[hbtp]
\begin{center}
    \setlength{\unitlength}{1mm}
    \begin{picture}(100,80)
    \put(0,0){\epsfxsize=8cm \epsfbox{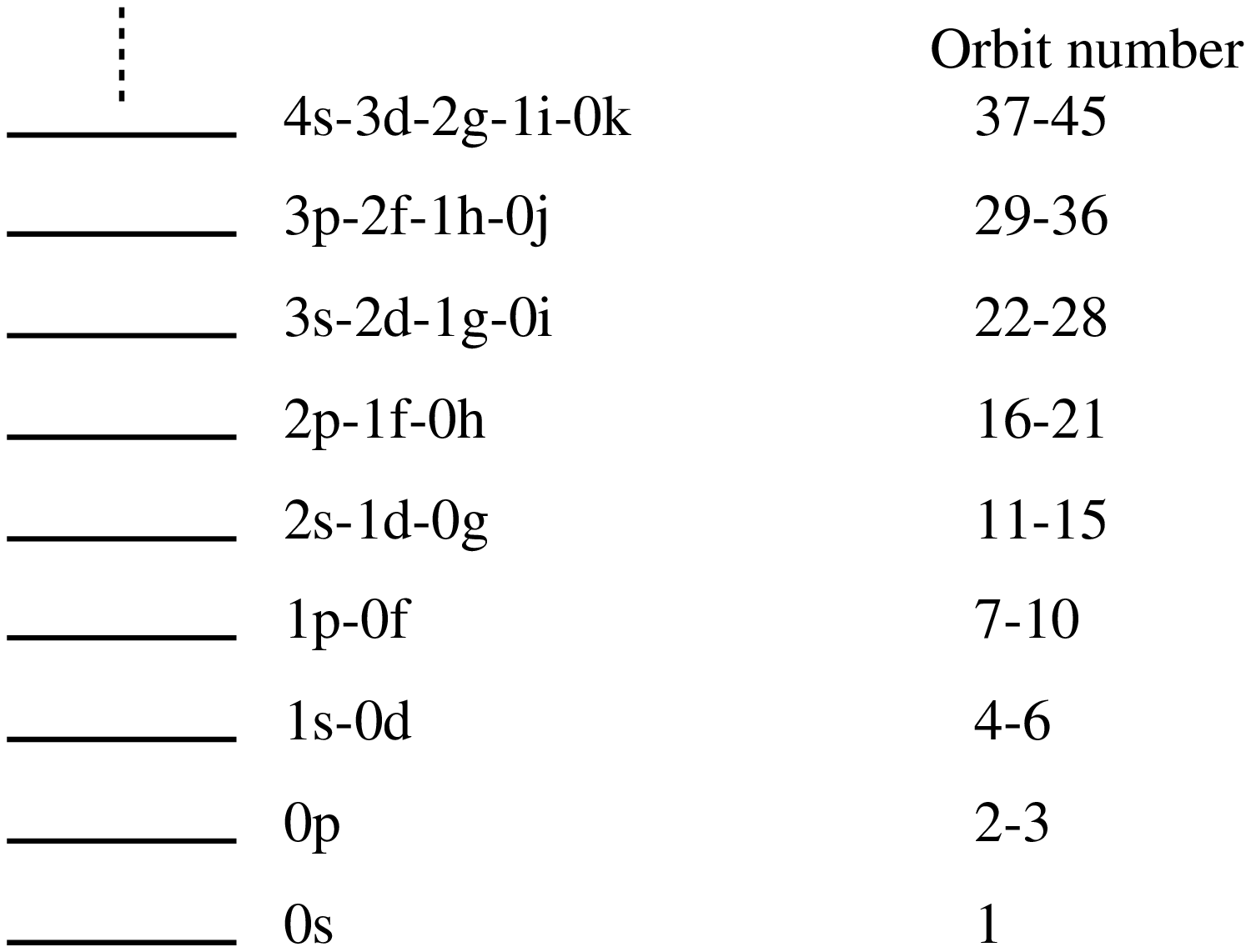}}
    \end{picture}
\caption{Classification of harmonic oscillator single-particle
orbits.}
\label{fig:orbits}
\end{center}
\end{figure}
The reader may now ask what is the advantage of the above identity. If
we assume that by some physical reasoning we are able to calculate
$G_2$ and that the expression for $G_2$ can be calculated
easily, and further that $G_2$ is a good approximation
to the original $G$-matrix, then we can use the above identity to
perform a perturbative calculation of $G_1$ in terms of $G_2$.

Before we proceed in detailing the calculation of the $G$-matrix
appropriate for finite nuclei, certain approximations need to be explained.

\begin{figure}[hbtp]
\begin{center}
      \setlength{\unitlength}{1mm}
      \begin{picture}(100,80)
      \put(0,0){\epsfxsize=10cm \epsfbox{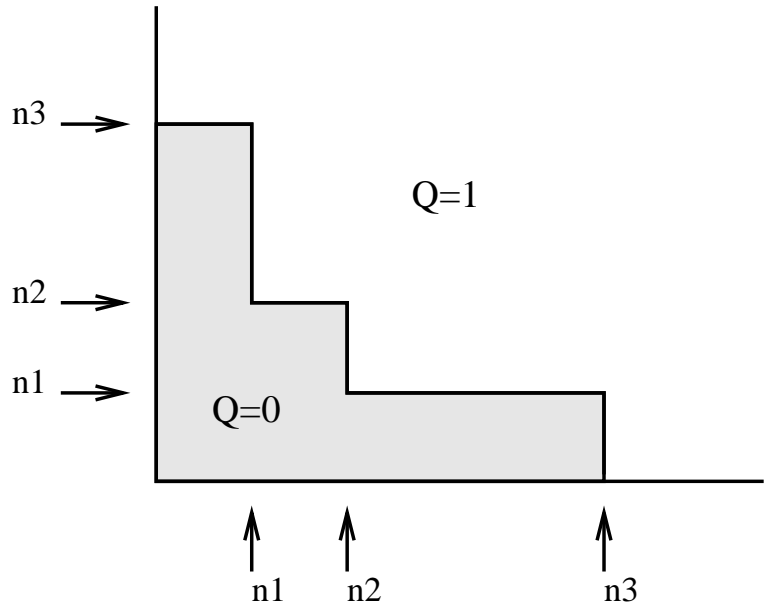}}
      \end{picture}
\caption{Definition of the $P$ (shaded area) and $Q$ operators
appropriate for the definition of the $G$-matrix and the effective
interaction. Outside the shaded area limited by the boundaries $n_1$,
$n_2$ and $n_3$ $P=0$ and $Q=1$.}
\label{fig:qoperat}
\end{center}
\end{figure}

\begin{figure}[hbtp]
\begin{center}
      \setlength{\unitlength}{1mm}
      \begin{picture}(100,60)
       \put(0,0){\epsfxsize=10cm \epsfbox{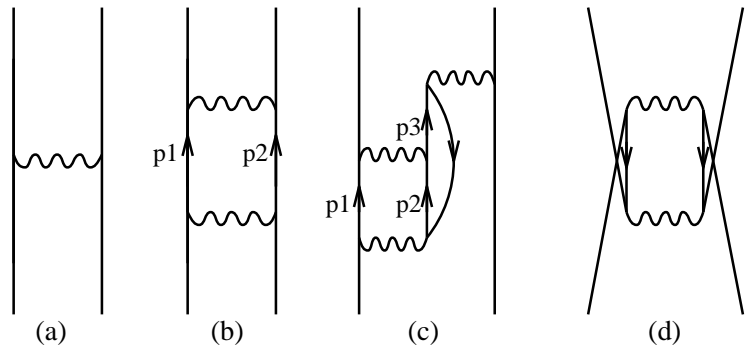}}
      \end{picture}
      \caption{Examples of diagrams which may define the effective valence space
         interaction. The wavy line is the $G$-matrix.}
\label{fig:qboxexam1}
\end{center}
\end{figure}

As discussed above, the philosophy behind perturbation theory is
to reduce the intractable full Hilbert space problem to one which
can be solved within a physically motivated model space, defined by the
operator $P$. The excluded degrees of freedom are represented by the
projection operator $Q$. The definition of these operators is connected
with the nuclear system and the perturbative expansions discussed
in section \ref{sec:sec2}. Consider the evaluation of the effective interaction
needed in calculations of the low-lying states of $^{18}$O. From
experimental data and theoretical calculations the belief is that
several properties of this nucleus can be described by a model
space consisting of a closed $^{16}$O core (consisting of the filled
$0s$- and $0p$-shells) and two valence neutrons
in the $1s0d$-shell. In Fig.\ \ref{fig:orbits} we exhibit this division
in terms of h.o.~sp orbits.
The active sp states in the $1s0d$-shell are then given by the  $0d_{5/2}$,
$0d_{3/2}$ and $1s_{1/2}$ orbits, labels $4-6$ in Fig.\ \ref{fig:orbits}.
The remaining states enter the definition of
$Q$. Once we have defined $P$ and $Q$ we proceed in constructing the $G$-matrix
and the corresponding perturbative expansion in terms of the $G$-matrix.
There are however several ways of choosing $Q$. A common procedure is to
specify the boundaries of $Q$ by three numbers, $n_1$, $n_2$ and $n_3$, explained
in Fig.\ \ref{fig:qoperat}.
For $^{18}$O we would choose $(n_1=3,n_2=6,n_3=\infty)$.
Our choice of
$P$-space implies that the single-particle states outside the model space
start from the
$1p0f$-shell (numbers 7--10 in Fig.\ \ref{fig:orbits}), and orbits 1, 2
and 3 are hole states. Stated differently, this means that $Q$
is constructed so as to prevent scattering into intermediate
two-particle states
with one particle in the $0s$- or $0p$-shells or both particles
in the $1s0d$-shell. This definition of the $Q$-space influences the determination
of the effective shell-model interaction. Consider the diagrams displayed
in Fig.\ \ref{fig:qboxexam1}.
Diagram (a) of this figure is just the $G$-matrix and is allowed in the definition
of the $\hat{Q}$-box. With our choice $(n_1=3,n_2=6,n_3=\infty)$, diagram (b) is not
allowed since the intermediate state consists of passive particle
states  and is already included in the evaluation of the $G$-matrix. Similarly,
diagram (c) is also not allowed whereas diagram (d) is allowed. Now an important
subtlety arises. If we evaluate the $G$-matrix with the boundaries
$(n_1=3,n_2=10,n_3=\infty)$, and define the $P$-space of {\em
the effective interaction}
by including orbits 4 to 6 only, then diagrams (b) and (c)
are allowed if $7\leq p_1 , p_2 \leq 10$
In this way we allow for
intermediate two-particle states as well with orbits outside the
model-space of the effective interaction. The reader should notice the above
differences, i.e.\ that the $Q$-space defining the $G$-matrix and
$H_{\mathrm{eff}}$
may differ.
In order to calculate the $G$-matrix we will henceforth employ a
so-called double-partitioned scheme.
Let us be more specific and detail this double-partitioned procedure.
We define first a reference $G$-matrix $\tilde{G}$
in terms of plane wave intermediate states only, meaning that $H_0$ is
replaced by the kinetic energy operator $T$ only
while $G$ has harmonic oscillator intermediate states (this is one
possible choice for $U$). We divide the exclusion operator
into two parts, one which represents the low-lying states $Q_l$ and
one which accounts for high-lying states $Q_h$, viz.\
\[
    Q=Q_l+Q_h=Q_l+\tilde{Q}.
\]
If we consider $^{18}$O as our pilot nucleus, we may define $Q_l$ to consist
of the sp orbits of the $1p0f$-shell, orbits $7-10$ in Fig.\ \ref{fig:orbits},
described by h.o.\ states. $Q_h$ represents then the remaining orthogonalized
intermediate states.
Using the identity of Bethe, Brandow and Petschek \cite{bbp63} of
Eq.\ (\ref{eq:gidentity}) we
can first set up $\tilde{G}$ as
\begin{equation}
     \tilde{G}=V+V\frac{\tilde{Q}}{s -T}\tilde{G},
\label{eq:gfinite}
\end{equation}
and  express $G$ in terms of $\tilde{G}$ as
\begin{equation}
        G=\tilde{G} +\tilde{G}
        \left(\frac{Q_l}{s -H_0}\right)G,
        \label{eq:gidfinite}
\end{equation}
and we have assumed that $\tilde{G}$ is hermitian and that $[Q_l,H_0]=0$.
Thus, we first calculate
a ``reference'' $G$-matrix ($\tilde{G}$ in our case), and then insert this
in the expression for the full $G$-matrix. The novelty here is that
we are able to calculate $\tilde{G}$ exactly through operator relations
to be discussed below. In passing we note that $G$ depends significantly
on the choice of $H_0$, though the low-lying intermediate states
are believed to be fairly well represented by h.o.\ states.
Also, the authors of ref.\ \cite{kkko76} demonstrate that low-lying
intermediate states are not so important in $G$-matrix calculations,
being consistent with the short-range nature of the $NN$ interaction.
Since we let $Q_l$ to be defined by the orbits of the $1p0f$-shell,
and the energy difference between two particles in the
$sd$-shell and $pf$ shell is of the order $-14$ MeV, we can treat
$G$ as a perturbation expansion in $\tilde{G}$.
Eq.\ (\ref{eq:gidfinite}) can then be written as
\begin{equation}
        G=\tilde{G} +\tilde{G}
        \left(\frac{Q_l}{s -H_0}\right)\tilde{G}
        +\tilde{G}
        \left(\frac{Q_l}{s -H_0}\right)\tilde{G}
        \left(\frac{Q_l}{s -H_0}\right)\tilde{G} +\dots
\end{equation}
The only intermediate states are those defined by the $1p0f$-shell.
The second term on the rhs.\ is nothing but the second-order
particle-particle ladder. The third term is then the third-order ladder
diagram in terms of
$\tilde{G}$. As shown by the authors of ref.\ \cite{kkko76}, the inclusion
of the second-order particle-particle diagram in the evaluation
of the $\hat{Q}$-box, represents a good approximation.
The unsettled problem is however how to define
the boundary between
$Q_l$ and $Q_h$.

Now we will discuss how to compute $\tilde{G}$.
One can solve the equation for the $G$-matrix
for finite nuclei by employing
a formally
exact technique for handling $\tilde{Q}$
discussed in e.g., Ref.\ \cite{kkko76}.
Using the matrix identity
\begin{equation}
  \tilde{Q}\frac{1}{\tilde{Q}A\tilde{Q}}
  \tilde{Q}=\frac{1}{A}-
   \frac{1}{A}\tilde{P}\frac{1}{\tilde{P}A^{-1}\tilde{P}}\tilde{P}\frac{1}{A},
   \label{eq:matrix_relation_q}
\end{equation}
with $A=s -T$, to rewrite Eq.\ (\ref{eq:gfinite}) as\footnote{We will omit the
label $\tilde{G}$ for the $G$-matrix for finite nuclei, however it is
understood that the $G$-matrix for finite nuclei is calculated according
to Eq.\ (\ref{eq:gfinite}) This means that we have to
include the particle-particle ladder diagrams in the
$\hat{Q}$-box. }
\begin{equation}
   G = G_{F} +\Delta G,\label{eq:gmod}
\end{equation}
where $G_{F}$ is the free $G$-matrix defined as
\begin{equation}
   G_{F}=V+V\frac{1}{s - T}G_{F}. \label{eq:freeg}
\end{equation}
The term $\Delta G$ is a correction term defined entirely within the
model space $\tilde{P}$ and given by
\begin{equation}
   \Delta G =-V\frac{1}{A}\tilde{P}
   \frac{1}{\tilde{P}A^{-1}\tilde{P}}\tilde{P}\frac{1}{A}V.
\end{equation}
Employing the definition for the free $G$-matrix of Eq.\ (\ref{eq:freeg}),
one can rewrite the latter equation as
\begin{equation}
  \Delta G =-G_{F}\frac{1}{e}\tilde{P}
  \frac{1}{\tilde{P}(e^{-1}+e^{-1}G_{F}e^{-1})
  \tilde{P}}\tilde{P}\frac{1}{e}G_F,
\end{equation}
with $e=s -T$.
We see then that the $G$-matrix for finite nuclei
is expressed as the sum of two
terms; the first term is the free $G$-matrix with no Pauli corrections
included, while the second term accounts for medium modifications
due to the Pauli principle. The second term can easily
be obtained by some simple matrix operations involving
the model-space matrix $\tilde{P}$ only.
However, the second term is a function of the variable
$n_3$. The convergence in terms of $n_3$ was discussed ad extenso
in Ref.\ \cite{hko95} and we refer the reader to that work.
The equation for the free matrix $G_F$ is solved in momentum space in the
relative and centre of mass system and thereafter transformed to the
relevant expression in terms of harmonic ocillator single-particle
wavefunctions in the laboratory system. This yields final
matrix elements of the type
\begin{equation}
  \bra{(ab)J}G\ket{(cd)J}
\end{equation}
where $G$ is the given by the sum $G = G_{F} +\Delta G$.
The label $a$ represents here all the single particle quantum numbers
$n_{a}l_{a}j_{a}$.

\subsection{\it Modified Hilbert space for resummation of large sets of diagrams}

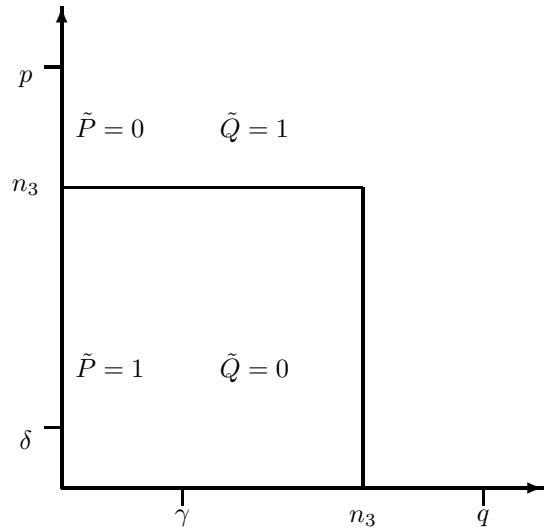
\begin{figure}[htbp]
\begin{center}
\setlength{\unitlength}{0.8cm}
\begin{picture}(9,10)
\thicklines
   \put(1,0.5){\makebox(0,0)[bl]{
              \put(0,1){\vector(1,0){8}}
              \put(0,1){\vector(0,1){8}}
              \put(7,0.5){\makebox(0,0){${q}$}}
              \put(7,0.7){\line(0,1){0.3}}
              \put(2,0.5){\makebox(0,0){${\gamma}$}}
              \put(2,0.7){\line(0,1){0.3}}
              \put(-0.6,6){\makebox(0,0){$n_3$}}
              \put(5,0.5){\makebox(0,0){$n_3$}}
              \put(2,3){\makebox(0,0){$\tilde{P}=1\hspace{1cm}\tilde{Q}=0$}}
              \put(2,7){\makebox(0,0){$\tilde{P}=0\hspace{1cm}\tilde{Q}=1$}}
              \put(-0.6,7.8){\makebox(0,0){${p}$}}
              \put(-0.3,8){\line(1,0){0.3}}
              \put(-0.6,1.8){\makebox(0,0){${\delta}$}}
              \put(-0.3,2){\line(1,0){0.3}}
              \put(0,6){\line(1,0){5}}
              \put(5,1){\line(0,1){5}}
         }}
\end{picture}
\caption{Definition of the exclusion operator used to compute the $G$-matrix for large
spaces.\label{fig:paulioperator}}
\end{center}
\end{figure}

In Fig.~\ref{fig:qoperat} the numbers $n_1$, $n_2$ and $n_3$ where used to 
define hole and particle states with respect to a given nucleus. We could also define a
huge model space which does not reflect a particular core and thereby nucleus.
One possible way of defining such a  no-core model space is obtained by setting
$n_1=n_2=n_3$, with $n_3$ representing a large number, at least some eight-ten major
oscillator shells.   
The single-particle states labeled by $n_3$ 
represent then the last orbit of the model space $\tilde{P}$, 
following the numbering indicated in 
Fig.~\ref{fig:orbits}. 
This so-called no-core model space 
is indicated in Fig.~\ref{fig:paulioperator} and will be used in our definitions of 
model spaces for the resummations of parquet diagrams and many-body terms in 
coupled cluster theory. 
In Fig.~\ref{fig:paulioperator} the two-body state
$\left| (pq)JT_Z\right \rangle$ 
does not belong to the model space and is included
in the computation of
the $G$-matrix.
Similarly,
$\left| (p\gamma)JT_Z\right \rangle$
and
$\left| (\delta q)JT_Z\right \rangle$
also enter the definition of $\tilde{Q}$ whereas
$\left| (\delta\gamma)JT_Z\right \rangle$
is not included in the computation of $G$.
This means that correlations not defined in the $G$-matrix need
to be computed by other non-perturbative
resummations or many-body schemes.
This is where the coupled-cluster scheme and the parquet approaches enter.

\begin{figure}[htbp]
\begin{center}
\setlength{\unitlength}{0.8cm}
\begin{picture}(9,10)
\thicklines
   \put(1,0.5){\makebox(0,0)[bl]{
              \put(0,1){\vector(1,0){8}}
              \put(0,1){\vector(0,1){8}}
              \put(7,0.5){\makebox(0,0){b}}
              \put(-0.6,6){\makebox(0,0){$n_p$}}
              \put(5,0.5){\makebox(0,0){$n_p$}}
              \put(-0.6,3){\makebox(0,0){$n_{\alpha}$}}
              \put(2,0.5){\makebox(0,0){$n_{\alpha}$}}
              \put(-0.6,8){\makebox(0,0){a}}
              \put(0,6){\line(1,0){5}}
              \put(5,1){\line(0,1){5}}
              \put(0,3){\line(1,0){2}}
              \put(2,1){\line(0,1){2}}
         }}
\end{picture}
\caption{Definition of particle and hole states
for coupled-cluster, parquet diagrams  and perturbative many-body calculations
in large spaces.
The orbit represented by 
 $n_{\alpha}$ stands for the last  hole state whereas $n_{p}$
represents the last particle orbit included in the $G$-matrix model space.
The hole states define the
Fermi energy.\label{fig:finalp}}
\end{center}
\end{figure}
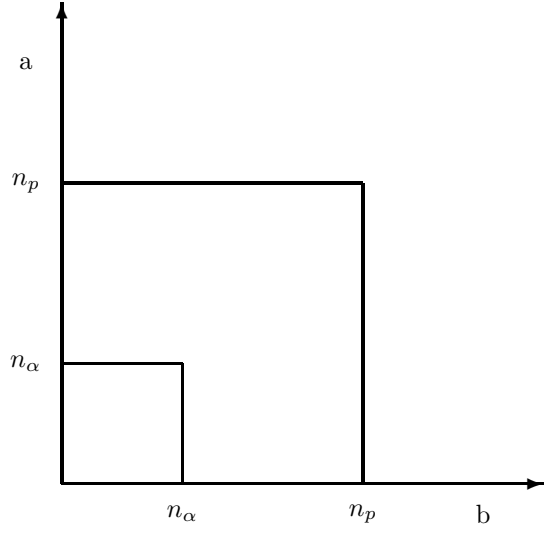

With the $G$-matrix model space $\tilde{P}$ of Fig.~\ref{fig:paulioperator} we can now define
an appropriate space for many-body perturbation theory, parquet diagrams or 
coupled-cluster calculations where correlations
not included in the $G$-matrix are to be generated. This model space is defined
in Fig.~\ref{fig:finalp}, where the label $n_{p}$ represents the same
single-particle orbit as $n_{3}$ in  Fig.~\ref{fig:paulioperator}.

The $G$-matrix computed according to Fig.~\ref{fig:paulioperator}
does not reflect a specific nucleus and
thereby single-particle orbits which define the uncorrelated
Slater determinant.  For a nucleus like
$^{4}$He the $0s_{1/2}$ orbit is fully occupied and defines thereby single-hole states.
These are labeled by $n_{\alpha}$ in Fig.~\ref{fig:finalp}.
For $^{16}$O the corresponding hole states are represented by the orbits
$0s_{1/2}$,  $0p_{3/2}$ and  $0p_{1/2}$. With this caveat we can then generate
correlations not included in the $G$-matrix and hopefully perform resummations of larger 
classes of diagrams.

\begin{figure}
\begin{center}
\includegraphics[angle=270, scale=0.5]{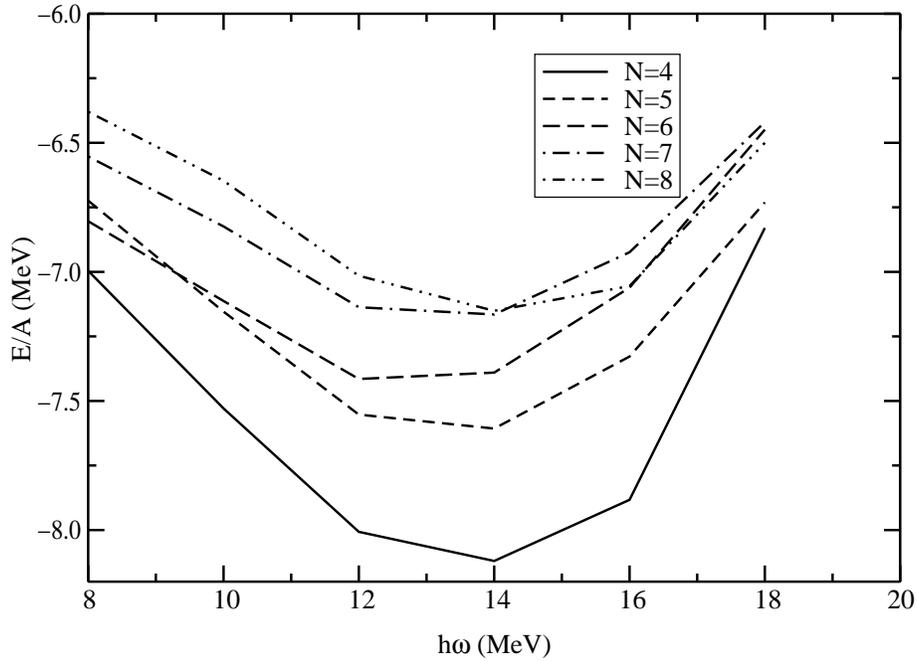}
\caption{Binding energy per particle $E/A$ 
from third-order perturbation theory for $^{16}$O as function 
of the number of major harmonic oscillator shells $N$ 
and the oscillator energy $\hbar\omega$. For $N=8$ we have the optimal value of
$E/A=-7.12$ MeV at $\hbar\omega = 13.6$ MeV. The experimental value is
$E/A=-7.98$ MeV.}
\label{fig:mbptox}
\end{center}
\end{figure}

To demonstrate the dependence upon the size of the enlarged model-space
defined by $n_p$, 
We present here results from third-order in perturbation theory for the 
binding energy of $^{16}$O as function of the size of the model
space and the chosen oscillator energy $\hbar\omega$.
These results are shown in Fig.~\ref{fig:mbptox}. 

There are several features to be noted. First of all one notices that the results 
seem to stabilize between seven and eight  
major shells. 
The fact that the energies seem to converge at this level of truncation is a welcome
feature which can be exploited in the coupled-cluster and parquet diagram 
calculations. These calculations,
see below, are much more challenging from a computational point of view since we in principle
generate a much larger class of diagrams. 
coupled-cluster calculations. 

Secondly, although the minimum shifts a little 
as function of the oscillator energy as we increase the
oscillator space, we notice
that as the number of major shells is increased, the 
dependence of the binding energy 
upon the oscillator parameter weakens. A similar 
feature is seen in the coupled-cluster calculations
below.  
For $^{16}$O the minimum for seven shells 
takes place at $E/A=-7.16$ MeV for $\hbar\omega = 12.9$ MeV 
and for eight shells we have   $E/A=-7.12$ MeV at $\hbar\omega = 13.6$ MeV. For six shells
we obtain  $E/A=-7.42$ MeV at $\hbar\omega = 12.6$ MeV. 
The curvature for larger values of $\hbar\omega$ decreases with increasing number of shells $N$.
At $\hbar\omega =18$~MeV we have for $^{16}$O and $N=8$ that $d(E/A)/d\omega=0.22$, for 
$N=7$ we obtain $d(E/A)/d\omega=0.28$
and $N=6$ we have $d(E/A)/d\omega=0.35$. 
The reader should also note that 
in the limit $\hbar\omega \rightarrow 0$ we have
$E\rightarrow 0$.  

We do not expect to reproduce the experimental binding energies. This calculation
does not include the Coulomb interaction or realistic three-body interactions.
The latter will be discussed in section \ref{sec:sec7}.

\subsection{\it Folded diagrams and the effective valence space interaction}

Here we discuss further classes of diagrams
which can be included in the evaluation of effective interactions
for the shell model.
Here we will focus on the summations of so-called folded
diagrams.

One way of obtaining the wave operator
$\Omega$ is through the generalized Bloch
equation given by Lindgren and Morrison \cite{lm85}
\begin{equation}
[\Omega, H_0]P=QH_1\Omega P-\chi PH_1\Omega P,
\label{eq:lind}
\end{equation}
which offers a suitable way of generating the RS perturbation expansion.
Writing Eq.\ (\ref{eq:lind}) in terms of $\Omega^{(n)}$ we have
\begin{equation}
[\Omega^{(1)}, H_0]P=QH_1P,
\end{equation}
\begin{equation}
[\Omega^{(2)}, H_0]P=QH_1\Omega^{(1)} P- \Omega^{(1)} PH_1P,
\end{equation}
and so forth,  which can be generalized to
\begin{equation}
[\Omega^{(n)}, H_0]P=QH_1\Omega^{(n-1)} P- \sum_{m=1}^{n-1}
\Omega^{(n-m)} PH_1\Omega^{(m-1)}P.
\end{equation}
The effective interaction to a given order can then be obtained from
$\Omega^{(n)}$, see \cite{lm85}.
Another possibility is obviously the coupled-cluster method discussed
below.

Here we will assume that we can start with a given
approximation to $\Omega$, and through an iterative scheme generate
higher order terms. Such schemes will in general differ from the
order-by-order scheme of Eq.\ (\ref{eq:lind}).  Two such iterative
schemes were derived  by Lee and Suzuki \cite{ls80}. We will focus
on the folded diagram method of Kuo and co-workers \cite{ko90}.

\begin{figure}[hbtp]
\begin{center}
      \setlength{\unitlength}{1mm}
      \begin{picture}(100,80)
      \put(0,0){\epsfxsize=10cm \epsfbox{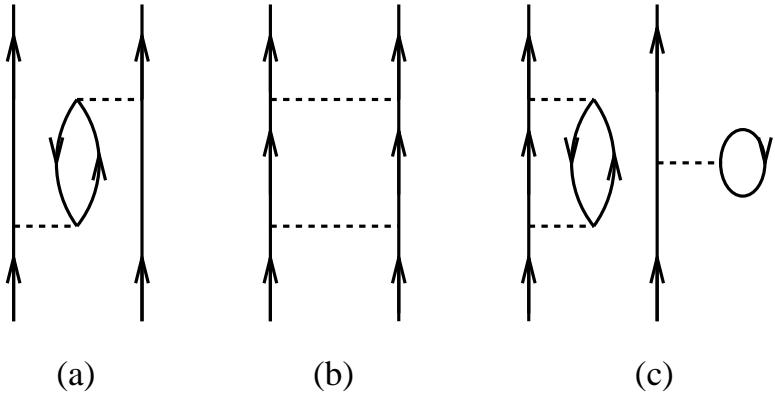}}
      \end{picture}
\caption{Different types of valence-linked diagrams. Diagram (a)
is irreducible and connected, (b) is reducible, while (c) is irreducible
and disconnected.}
\label{fig:diagsexam}
\end{center}
\end{figure}

Having defined the wave operator $\Omega = 1 +\chi$ (note that $
\Omega^{-1}=1-\chi$) with
$\chi$ given by Eq.\ (\ref{eq:chi2}) we can obtain
\begin{equation}
     QHP-\chi HP +QH\chi - \chi H\chi = 0. \label{eq:basic}
\end{equation}
This is the basic equation to which a solution to $\chi$ is
to be sought.
If we choose to  work with a degenerate model space we define
\[
   PH_0 P = s P,
\]
where $s$ is the unperturbed model space eigenvalue
(or starting energy) in the degenerate case,
such that Eq.\ (\ref{eq:basic}) reads in a slightly modified form
($H=H_0 + H_1$)
\[
    (s -QH_0 Q -QH_1 Q)\chi = QH_1 P -\chi PH_1 P -\chi PH_1 Q\chi,
\]
which yields the following equation for $\chi$
\begin{equation}
    \chi = \frac{1}{s - QHQ}QH_1 P -\frac{1}{s -QHQ}\chi\left(PH_1 P +
    PH_1 Q\chi P\right).\label{eq:chi3}
\end{equation}
Observing that  the $P$-space effective Hamiltonian is given as
\[
     H_{\mathrm{eff}}= PHP+PH\chi=PH_0 P + V_{\mathrm{eff}}(\chi),
\]
with $V_{\mathrm{eff}}(\chi)= PH_1 P + PH_1Q\chi P$, Eq. (\ref{eq:chi3}) becomes
\begin{equation}
     \chi = \frac{1}{s - QHQ}QH_1 P -\frac{1}{s -QHQ}
     \chi V_{\mathrm{eff}}(\chi ).
     \label{eq:chi4}
\end{equation}
Now we find it convenient to introduce the so-called $\hat{Q}$-box,
defined as
\begin{equation}
     \hat{Q}(s)=PH_1 P + PH_1 Q\frac{1}{s - QHQ}
      QH_1 P.\label{eq:qbox}
\end{equation}
The $\hat{Q}$-box is made up of non-folded diagrams which are irreducible
and valence linked. A diagram is said to be irreducible if between each pair
of vertices there is at least one hole state or a particle state outside
the model space. In a valence-linked diagram the interactions are linked
(via fermion lines) to at least one valence line. Note that a valence-linked
diagram can be either connected (consisting of a single piece) or
disconnected. In the final expansion including folded diagrams as well, the
disconnected diagrams are found to cancel out \cite{ko90}.
This corresponds to the cancellation of unlinked diagrams
of the Goldstone expansion.
We illustrate
these definitions by the diagrams shown in Fig.\
\ref{fig:diagsexam}.
Diagram (a) is irreducible, valence linked and connected,
while (b) is reducible since
the intermediate particle states belong to the model space.
Diagram (c) is irreducible, valence linked and disconnected.
It is worth noting that general form of the $\hat{Q}$-box
is the same as that of the $G$-matrix, or the equations
of the $[12]$ channel or those of the $[13]$ and $[14]$
channels to be discussed in section \ref{sec:sec4}.
In Ref.\ \cite{hko95}, the $\hat{Q}$-box was defined
to be the sum all diagrams to third order in the $G$-matrix.

Multiplying both sides of Eq.\ (\ref{eq:chi4}) with $PH_1$ and
adding $PH_1 P$ to both sides we get
\[
    PH_1 P + PH_1 \chi =
    PH_1 P + PH_1 Q\frac{1}{s - QHQ}QH_1 P -
    PH_1 \frac{1}{s -QHQ}\chi V_{\mathrm{eff}}(\chi ),
\]
which gives
\begin{equation}
     V_{\mathrm{eff}}(\chi )=\hat{Q}(s)-
     PH_1 \frac{1}{s -QHQ}\chi V_{\mathrm{eff}}(\chi ).
     \label{eq:veff}
\end{equation}

There are several ways to solve Eq.\ (\ref{eq:veff}). The idea is
to set up an iteration scheme where we determine $\chi_n$ and
thus $V_{\mathrm{eff}}(\chi_n )$ from
$\chi_{n-1}$ and $V_{\mathrm{eff}}(\chi_{n-1})$.
For the mere sake of simplicity we write
$V_{\mathrm{eff}}^{(n)}=V_{\mathrm{eff}}(\chi_{n})$.

Let us write Eq.\ (\ref{eq:veff}) as
\[
   V_{\mathrm{eff}}^{(n)}=\hat{Q}(s)-
   PH_1 \frac{1}{s -QHQ}\chi_n V_{\mathrm{eff}}^{(n-1)}.
\]
The solution to this equation can be shown to be \cite{ls80}
\begin{equation}
    V_{\mathrm{eff}}^{(n)}=\hat{Q}+{\displaystyle\sum_{m=1}^{\infty}}
    \frac{1}{m!}\frac{d^m\hat{Q}}{ds^m}\left\{
    V_{\mathrm{eff}}^{(n-1)}\right\}^m .
    \label{eq:fd}
\end{equation}
Observe also that the
effective interaction is $V_{\mathrm{eff}}^{(n)}$
is evaluated at a given model space energy
$s$. If
$V_{\mathrm{eff}}^{(n)}=V_{\mathrm{eff}}^{(n-1)}$, the iteration is said to
converge. In the limiting case $n\rightarrow \infty$, the
solution $V_{\mathrm{eff}}^{(\infty)}$ agrees with the formal solution of
Brandow
\cite{brandow67} and Des Cloizeaux \cite{des}
\begin{equation}
    V_{\mathrm{eff}}^{(\infty)}=\sum_{m=0}^{\infty}\frac{1}{m!}
    \frac{d^{m}\hat{Q}}{ds^{m}}\left\{
    V_{\mathrm{eff}}^{(\infty)}\right\}^{m}.\label{eq:pert}
\end{equation}

Alternatively, we can generate the contribution
from $n$ folds by the following way. In
an $n$-folded $\hat{Q}$-box there are of course $n+1$ $\hat{Q}$-boxes. The
general expression for an $n$-folded $\hat{Q}$-box is then
\begin{equation}
        \hat{Q}   -\hat{Q}\int\hat{Q}
    +\hat{Q}\int\hat{Q}\int\hat{Q} -\dots=
    {\displaystyle\sum_{m_1m_2\dots m_n}}
    \frac{1}{m_1!}\frac{d^{m_1}\hat{Q}}{ds^{m_1}}P
    \frac{1}{m_2!}\frac{d^{m_2}\hat{Q}}{ds^{m_2}}P
    \dots
    \frac{1}{m_n!}\frac{d^{m_n}\hat{Q}}{ds^{m_n}}P\hat{Q},
\label{eq:fdfinal}
\end{equation}
where we have the constraints
\[
  m_1+m_2+\dots m_n=n,
\]
\[
m_1\geq 1,
\]
\[
m_2, m_3, \dots m_n \geq 0,
\]
and
\[
m_k \leq n-k+1.
\]
The last restriction follows from the fact that there are only
$n-k+1$ $\hat{Q}$-boxes to the right of $k^{\mathrm{th}}$ $\hat{Q}$-box.
Thus, it can at most be differentiated $n-k+1$ times.
We have inserted the model-space projection operator in the above
expression, in order to emphasize that folded diagrams have as intermediate
states between successive $\hat{Q}$-boxes
only model-space states. Therefore, the sum in Eq.\ (\ref{eq:fdfinal})
includes a sum over all model-space states with the same quantum
numbers such as isospin and total angular momentum. It is understood
that the $\hat{Q}$-box  and its derivatives are evaluated at the
same starting energy, which should correspond to the unperturbed
energy of the model-space state.
It is then straightforward to recast Eq.\ (\ref{eq:fdfinal})
into the form of Eq.\ (\ref{eq:fd}).

Note that although $\hat{Q}$ and its derivatives contain disconnected
diagrams, such diagrams cancel exactly in each order \cite{ko90}, thus
yielding a fully connected expansion in Eq.\ (\ref{eq:fd}).
However, in order to achieve this, disconnected diagrams have
to be included in the definition of the $\hat{Q}$-box. An example
is given by diagram (c) Fig.\ \ref{fig:diagsexam}.
Such a diagram will generate a contribution to
the first fold $\frac{d\hat{Q}}{ds}\hat{Q}=-\hat{Q}\int\hat{Q} $
which cancels exactly diagram (c) when all time-ordered
contributions to this diagram are accounted for, see Ref.\ \cite{ko90}
for more details.
It is moreover important to note in connection with the above
expansion, that a term like
$F_1= \hat{Q}_1 \hat{Q}$ actually means $P\hat{Q}_1 P\hat{Q}P$ since
the $\hat{Q}$-box is defined in the model space only. Here we have defined
$\hat{Q}_{m}=\frac{1}{m!}\frac{d^{m}\hat{Q}}
{ds^{m}}$.
Due to this structure, only so-called folded diagrams
contain $P$-space intermediate states.

The folded diagram expansion discussed above yields however
a non-hermitian effective interaction. This happens even at the
level of the $G$-matrix.
A hermitian effective interaction has recently been derived by
Suzuki and co-workers \cite{so95,so84,kehlsok93} through the
following steps\footnote{The reader who wishes more details can consult
Refs.\ \cite{so95,kehlsok93}.}.
To obtain a hermitian effective interaction,
let us define
a model-space eigenstate
$\left | b_{\lambda}\right\rangle$ with eigenvalue $\lambda$ as
\begin{equation}
     \left | b_{\lambda}\right\rangle=\sum_{\alpha =1}^{D}
     b_{\alpha}^{(\lambda )}\left | \psi_{\alpha}\right\rangle
\end{equation}
and the biorthogonal wave function
\begin{equation}
     \left | \overline{b}_{\lambda}\right\rangle=\sum_{\alpha=1}^{D}
      \overline{b}_{\alpha}^{(\lambda )}
     \left | \overline{\psi}_{\alpha}\right\rangle,
\end{equation}
such that
\begin{equation}
     {\left\langle \overline{b}_{\lambda} | b_{\mu} \right\rangle}=
     \delta_{\lambda\mu}.
\end{equation}
The model-space eigenvalue problem can be written
in terms of the above non-hermitian effective interaction unperturbed
wave functions
\begin{equation}
     {\displaystyle
     \sum_{\gamma =1}^{D}b_l^{(\lambda )}\left\langle \psi_{\sigma}\right |
     H_0+V+VQ\chi\left | \psi_{\gamma}\right\rangle}  =
     E_{\lambda}b_{\sigma}^{(\lambda )}.
\end{equation}
The exact wave function expressed in terms of the correlation
operator is
\begin{equation}
      \left | \Psi_{\lambda}\right\rangle=
      ({\bf 1}+\chi)\left | \psi_{\lambda}\right\rangle.
\end{equation}
The part $\chi\left | \psi_{\lambda}\right\rangle$ can be expressed in terms
of the time-development operator or
using the time-independent formalism as
\begin{equation}
  \chi\left | \psi_{\lambda}\right\rangle=
  \frac{Q}{E_{\lambda}-QHQ}QVP\left | \psi_{\lambda}\right\rangle,
  \label{eq:newchi}
\end{equation}
where $Q$ is the exclusion operator. Note that this equation is given
in terms of the Brillouin-Wigner perturbation expansion, since
we have the exact energy $E_{\lambda}$ in the denominator.

Using the normalization condition for the true wave function
we obtain
\begin{equation}
  \left\langle \Psi_{\gamma} |\Psi_{\lambda}\right\rangle
  N_{\lambda}\delta_{\lambda\gamma}=
  \left\langle \psi_{\gamma}\right |({\bf 1} +
   \chi^{\dagger}\chi\left | \psi_{\lambda}\right\rangle,
  \label{eq:newnorm}
\end{equation}
where we have used the fact that
$\left\langle \psi_{\gamma}\right |
\chi\left | \psi_{\lambda}\right\rangle=0$. Recalling that the
time-development operator is hermitian we have that $\chi^{\dagger}\chi$ is
also hermitian. We can then define an orthogonal basis $d$ whose eigenvalue
relation is
\begin{equation}
   \sum_{\alpha}\left\langle \psi_{\beta}\right | \chi^{\dagger}
   \chi\left | \psi_{\alpha}
   \right\rangle d_{\alpha}^{\lambda}=\mu^{2}_{\lambda}
   d_{\beta}^{\lambda},
   \label{eq:newbasis1}
\end{equation}
with eigenvalues greater than $0$.
Using the definition in Eq.\ (\ref{eq:newchi}), we note
that the diagonal element of
\begin{equation}
   \left\langle \psi_{\lambda}\right | \chi^{\dagger}
   \chi\left | \psi_{\lambda}\right\rangle=
   \left\langle \psi_{\lambda}\right | PVQ\frac{1}{(E_{\lambda}-QHQ)^2}QVP
   \left | \psi_{\lambda}\right\rangle,
   \label{eq:chichi}
\end{equation}
which is nothing but the derivative of the $\hat{Q}$-box, with an additional
minus sign. Thus, noting that if $\gamma\neq\lambda$
\begin{equation}
  \left\langle \Psi_{\gamma} |\Psi_{\lambda}\right\rangle=0=
    \left\langle \psi_{\gamma} |\psi_{\lambda}\right\rangle
   +\left\langle \psi_{\gamma}\right |
    \chi^{\dagger}\chi\left | \psi_{\lambda}\right\rangle,
\end{equation}
we can write $\chi^{\dagger}\chi$
in operator form as
\begin{equation}
   \chi^{\dagger}\chi =-\sum_{\alpha}\left |
    \overline{\psi}_{\alpha}\right\rangle
   \left\langle \psi_{\alpha}\right |
    \hat{Q}_1(E_{\alpha})\left | \psi_{\alpha}\right\rangle
   \left\langle \overline{\psi}_{\alpha} \right |
   -\sum_{\alpha\neq\beta}\left | \overline{\psi}_{\alpha}\right\rangle
    \left\langle \psi_{\alpha} |\psi_{\beta}\right\rangle
   \left | \overline{\psi}_{\beta}\right\rangle.
\end{equation}
Using the new basis in Eq.\ (\ref{eq:newbasis1}), we see that
Eq.\ (\ref{eq:newnorm}) allows us to define another
orthogonal basis $h$
\begin{equation}
  h_{\alpha}^{\lambda}=\sqrt{\mu_{\alpha}^2+1}\sum_{\beta}
  d_{\beta}^{\alpha}b_{\beta}^{\lambda}\frac{1}{\sqrt{N_{\lambda}}}
  =\frac{1}{\sqrt{\mu_{\alpha}^2+1}}\sum_{\beta}
  d_{\beta}^{\alpha}\overline{b}_{\beta}^{\lambda}\sqrt{N_{\lambda}},
  \label{eq:hbasis}
\end{equation}
where we have used the orthogonality properties of the vectors
involved. The vector $h$ was used by the authors of Ref.\ \cite{kehlsok93}
to obtain a hermitian effective interaction as
\begin{equation}
   \left\langle \psi_{\alpha}\right |
   V_{\mathrm{eff}}^{\mathrm{(her)}}\left | \psi_{\beta}\right\rangle=
   \frac{\sqrt{\mu_{\alpha}^2+1}\left\langle \psi_{\alpha} \right |
   V_{\mathrm{eff}}^{\mathrm{(nher)}}\left | \psi_{\beta}\right\rangle
   +\sqrt{\mu_{\beta}^2+1}\left\langle \psi_{\alpha}\right |
   V_{\mathrm{eff}}^{\dagger\mathrm{(nher)}}\left |
   \psi_{\beta} \right\rangle}
   {\sqrt{\mu_{\alpha}^2+1}+\sqrt{\mu_{\beta}^2+1} },
   \label{eq:hermitian}
\end{equation}
where (her) and (nher) stand for hermitian and non-hermitian
respectively.
This equation is rather simple to compute, since
we can use the folded-diagram method
to obtain the non-hermitian part.
To obtain the total hermitian effective interaction, we have to
add the $H_0$ term. The above equation is manifestly hermitian.
Other discussion of the hermiticity problem can be found
in Refs.\ \cite{lm85,arponen97}.
The remaining question is how to evaluate the $\hat{Q}$-box.
Obviously, we are not in the position where we can evaluate it
exactly, i.e., to include all possible many-body terms.
Rather, we have to truncate somewhere.
Several possible approaches exist, but all have in common that
there is no clear way which tells us where to stop.
However, as argued by the authors of Ref.\ \cite{jls82}, there is a
minimal class of diagrams which need to be included
in order to fulfil necessary conditions. This class
of diagrams includes both diagrams which account for short-range
correlations such as the $G$-matrix and long-range
correlations such as those accounted for by various
core-polarization terms.
The importance of such diagrams has been extensively documented
in the literature and examples can be found
in Refs.\ \cite{hko95,eo77}. In Ref.\ \cite{hko95} we included
all core-polarization contributions to third-order
in the $G$-matrix, in addition to including other diagrams
which account for short-range correlations as well.

In our all results presented in section \ref{sec:sec4} we employ a 
third-order $\hat{Q}$-box using the CD-Bonn interaction model described in
Refs.~\cite{cdbonn,cdbonn2000}. Moreover, the procedure for obtaining a
hermitian procedure discussed in Eq.~(\ref{eq:hermitian}) is employed in
subsection \ref{subsec:sdpfshells}. In our coupled-cluster calculations we employ the 
recent chiral model Idaho-A of Entem and Machleidt, see for example Ref.~\cite{machleidt02}.

\subsection{\it Center of mass corrections}

Momentum conservation requires 
that a many-body wave function must factorize
as $\Psi(\bf{r}) = \phi(R)\Psi(\bf{r}_{\rm rel})$ where 
$R$ is the center-of-mass (CoM) coordinate and $\bf{r}_{\rm rel}$ the 
relative coordinates. If we choose to expand our wave functions in
the harmonic oscillator basis, then we are able to exactly separate the
center-of-mass motion from the problem provided that we work in a model
space that includes all $n\hbar\Omega$ excitations. 
In our coupled-cluster calculations we have a $Q$ operator that allows 
for all possible two-particle interactions within a given 
set of oscillator shells.  This means that we are 
$n\hbar\Omega$ incomplete in a given 
calculation so that our method of separation of the center-of-mass 
motion becomes approximate. For example, for $^4$He in four major oscillator
shells, we can excite all particles to $n=12\hbar\Omega$ excitations, 
but we can only excite one particle to $n=3\hbar\Omega$ excitations. 
Thus, care must be taken when correcting for center-of-mass contamination
in our calculations. We have taken a variational approach based on the 
the work of Whitehead {\em et al.}, \cite{whit77}. The idea is 
to add $\beta_{\rm CoM}H_{\rm CoM}$ to the Hamiltonian, but 
with $\beta_{\rm CoM}$ remaining fairly
small. This minimizes the effects of the center-of-mass contamination on
low-lying state properties, and partially pushes unwanted states out of the
spectrum. If we were to use a large $\beta_{\rm CoM}$, we would find 
spurious states entering into the calculated low-lying spectrum due to 
the incompleteness of our model space. 

The CoM Hamiltonian is then
\begin{equation}
H_{\rm CoM} = \frac{{\bf P}^2}{2MA} 
+\frac{1}{2}mA\Omega^2{\bf R}^2 - \frac{3}{2}\hbar\Omega \;,
\end{equation}
where ${\bf P}=\sum_{i=1,A}{\bf p}_i$ and $R=(\sum_{i=1,A}{\bf r}_i)/A$. 
The term $H_{\rm CoM}$ can be rewritten as a one-body harmonic potential, and a
two-body term that depends on both the relative and center-of-mass 
coordinates of the two interacting particles. The matrix elements for
the two body terms may be found in Ref.~\cite{lawson}. Operationally, 
we add $H_{\rm CoM}$ to our Hamiltonian 
\begin{equation}
H' = H+\beta_{\rm CoM}H_{\rm CoM}\;,
\end{equation}
where we choose $\beta_{\rm CoM}$ so that the expectation value of 
$H_{\rm CoM}$ is zero \cite{drhklz99}. This 
insures that our center-of-mass contamination 
within the many-body wave function is minimized. We also find that this
procedure yields reasonable spectra (in a space of four major
oscillator shells) for ${^4}$He \cite{papenbrock03}. 
This approach is used in our coupled-cluster calculations in section \ref{sec:sec6}
and in the shell-model Monte Carlo calculations of properties of $1s0d-1p0f$-shell
nuclei, see section \ref{sec:sec4} and Ref.~\cite{drhklz99}. In our shell-model
calculations we employ the standard procedure of Lawson, see Ref.~\cite{lawson}.

There are at least two other widely used ways of dealing with the CoM corrections.
The philosophy is
\begin{itemize} 
   \item One starts with a {\em translationally} invariant Hamiltonian,
         kinetic energy plus e.g., two-body interaction.
         From that one the Harmonic Oscillator field is introduced.
         This is then split up in a pure CoM term and a two-body term
         which depends on the radial distance.
   \item One starts with the the harmonic oscillator as the one-body piece
         plus a two-body interaction. The one-body piece can be rewritten 
         in terms of a pure CoM term and a two-body term which now
         depends on the relative distance and momentum. 
\end{itemize}

In deriving the CoM corrections, the following expressions are helpful.
The CoM momentum is
\begin{equation}
   P=\sum_{i=1}^A\vec{p}_i,
\end{equation}
and we have that
\begin{equation}
\sum_{i=1}^A\vec{p}_i^2 =
\frac{1}{A}\left[\vec{P}^2+\sum_{i<j}(\vec{p}_i-\vec{p}_j)^2\right]
\end{equation}
meaning that
\begin{equation}
\left[\sum_{i=1}^A\frac{\vec{p}_i^2}{2m} -\frac{\vec{P}^2}{2mA}\right]
=\frac{1}{2mA}\sum_{i<j}(\vec{p}_i-\vec{p}_j)^2.
\end{equation}
The last expression is explicitely translationally invariant.

In a similar fashion we can define the CoM coordinate
\begin{equation}
    \vec{R}=\frac{1}{A}\sum_{i=1}^{A}\vec{r}_i,
\end{equation}
which yields
\begin{equation}
\sum_{i=1}^A\vec{r}_i^2 =
\frac{1}{A}\left[\vec{R}^2+\sum_{i<j}(\vec{r}_i-\vec{r}_j)^2\right].
\end{equation}

If we then introduce the harmonic oscillator one-body Hamiltonian
\begin{equation}
     H_0= \sum_{i=1}^A\left(\frac{\vec{p}_i^2}{2m}+
          \frac{1}{2}m\Omega^2\vec{r}_i^2\right),
\end{equation}
with $\Omega$ the oscillator frequency,
we can rewrite the latter as 
\begin{equation}
     H_{\mathrm{HO}}= \frac{\vec{P}^2}{2mA}+\frac{mA\Omega^2\vec{R}^2}{2}
           +\frac{1}{2mA}\sum_{i<j}(\vec{p}_i-\vec{p}_j)^2
           +\frac{m\Omega^2}{2A}\sum_{i<j}(\vec{r}_i-\vec{r}_j)^2,
    \label{eq:obho}
\end{equation}
or just
\begin{equation}
H_{\mathrm{HO}}= H_{\mathrm{CoM}}+\frac{1}{2mA}\sum_{i<j}(\vec{p}_i-\vec{p}_j)^2
           +\frac{m\Omega^2}{2A}\sum_{i<j}(\vec{r}_i-\vec{r}_j)^2,
\end{equation}
with 
\begin{equation}
     H_{\mathrm{CoM}}= \frac{\vec{P}^2}{2mA}+\frac{mA\Omega^2\vec{R}^2}{2}.
\end{equation}

In shell model studies the translationally invariant one- and two-body 
Hamiltonian reads
for an A-nucleon system,
\begin{equation}\label{eq:ham}
H=\left[\sum_{i=1}^A\frac{\vec{p}_i^2}{2m} -\frac{\vec{P}^2}{2mA}\right] +\sum_{i<j}^A V_{ij} \; ,
\end{equation}
where $m$ is the nucleon mass and $V_{ij}$ the nucleon-nucleon interaction,
is  modified by including the harmonic oscillator potential through the 
following
\begin{equation}
\sum_{i=1}^A\frac{1}{2}m\Omega^2\vec{r}_i^2-
\frac{m\Omega^2}{2A}\left[\vec{R}^2+\sum_{i<j}(\vec{r}_i-\vec{r}_j)^2\right]=0.
\end{equation}
We can rewrite the Hamiltonian of Eq.~(\ref{eq:ham}) as

\begin{equation}\label{hamomega}
H^\Omega=\sum_{i=1}^A \left[ \frac{\vec{p}_i^2}{2m}
+\frac{1}{2}m\Omega^2 \vec{r}^2_i
\right] + \sum_{i<j}^A \left[ V_{ij}-\frac{m\Omega^2}{2A}
(\vec{r}_i-\vec{r}_j)^2
\right] -H_{\mathrm{CoM}}.
\end{equation}

Shell-model calculations are carried out in a model space defined
by a projector $P$. 
The complementary space to the model space
is defined by the projector $Q=1-P$. Consequently, for the $P$-space
part of the shell-model Hamiltonian we get
\begin{equation}\label{phamomega}
H^\Omega_P=\sum_{i=1}^A P\left[ \frac{\vec{p}_i^2}{2m}
+\frac{1}{2}m\Omega^2 \vec{r}^2_i
\right]P + \sum_{i<j}^A P\left[ V_{ij}-\frac{m\Omega^2}{2A}
(\vec{r}_i-\vec{r}_j)^2
\right]_{\rm eff} P -PH_{\mathrm{CoM}}P.
\end{equation}
The effective interaction appearing in Eq.~(\ref{phamomega}) is in general
an A-body interaction and if it is determined without any approximations,
the model-space Hamiltonian provides an identical description 
of a subset of states as the full-space Hamiltonian (\ref{hamomega}).
The intrinsic properties of the many-body system still do not depend
on $\Omega$. From among the eigenstates of the Hamiltonian 
(\ref{phamomega}),
it is necessary to choose only those corresponding to the same 
CoM energy. This can be achieved by projecting 
the CoM eigenstates
with energies greater than $\frac{3}{2}\hbar\Omega$ upwards in the
energy spectrum.

The effective interaction should be determined from $H^\Omega$ 
(\ref{hamomega}). Calculation of the exact A-body effective
interaction is, however, as difficult as finding the full space solution.
Usually, the effective interaction is approximated by a 
two-body effective interaction determined from a two-nucleon
problem. The relevant two-nucleon Hamiltonian 
obtained from (\ref{hamomega}) is then
\begin{equation}\label{hamomega2}
H^\Omega_2\equiv H^\Omega_{02}+V_2^\Omega=
\frac{\vec{p}_1^2+\vec{p}_2^2}{2m}
+\frac{1}{2}m\Omega^2 (\vec{r}^2_1+\vec{r}^2_2)
+ V(\vec{r}_1-\vec{r}_2)-\frac{m\Omega^2}{2A}(\vec{r}_1-\vec{r}_2)^2 \; .
\end{equation}

With this Hamiltonian we can then compute a
starting-energy-dependent effective interaction 
or $G$-matrix corresponding to a two-nucleon model
space defined by the projector $P_2$. This $G$-matrix reads
\begin{equation}\label{G}
G(\omega) = V_2^\Omega 
+ V_2^\Omega Q_2 \frac{1}{\omega-Q_2H_2^\Omega Q_2} 
Q_2 V_2^\Omega \; ,
\end{equation}
where $Q_2=1-P_2$ and $V_2^\Omega$ is the interaction given by the last 
two terms on the righthand side of Eq.~(\ref{hamomega2}).
Using a similarity transformation introduced by Suzuki and Okamoto \cite{so95},
Navr\'atil {\em et al.} used the above Hamiltonian to define an effective two-body
interaction for no-core calculations, see for example Ref.~\cite{petr_erich2002}. 
The CoM corrections can easily be implemented within such a scheme.
For a standard $G$-matrix this is more difficult since such calculations are typically
done in momentum space. We have however applied the CoM correction to the 
bare potential in a perturbative way in Ref.~\cite{dean03}. 

If we however start with a harmonic oscillator basis we
have
\begin{equation}
H^\Omega=\sum_{i=1}^A \left[ \frac{\vec{p}_i^2}{2m}
+\frac{1}{2}m\Omega^2 \vec{r}^2_i
\right] + \sum_{i<j}^A V_{ij}.
\label{newho}
\end{equation}

If we then use Eq.~(\ref{eq:obho}), we obtain the following expression
for our Hamiltonian
\begin{equation}
H^\omega=\frac{\vec{P}^2}{2mA}+\frac{mA\Omega^2\vec{R}^2}{2}
           +\frac{1}{2mA}\sum_{i<j}(\vec{p}_i-\vec{p}_j)^2
           +\frac{m\Omega^2}{2A}\sum_{i<j}(\vec{r}_i-\vec{r}_j)^2
           + \sum_{i<j}^A V_{ij},
\end{equation}
or 
\begin{equation}
H^\omega=H_{\mathrm{CoM}}
           +\frac{1}{2mA}\sum_{i<j}(\vec{p}_i-\vec{p}_j)^2
           +\frac{m\Omega^2}{2A}\sum_{i<j}(\vec{r}_i-\vec{r}_j)^2
           + \sum_{i<j}^A V_{ij},
\end{equation}
which differs from Eq.~(\ref{phamomega}) by the sign of the CoM contribution
and the two-body part which now reads
\begin{equation}
V_2^\omega=V(\vec{r}_1-\vec{r}_2)-\frac{m\Omega^2}{2A}(\vec{r}_1-\vec{r}_2)^2
-\frac{1}{2mA}(\vec{p}_i-\vec{p}_j)^2,
\end{equation}
which should be contrasted to 
\[
   V_2^\Omega=V(\vec{r}_1-\vec{r}_2)-\frac{m\Omega^2}{2A}(\vec{r}_1-\vec{r}_2)^2.
\]

A thorough discussion of various approaches to the CoM problem within the context of the 
coupled cluster method and summation of parquet diagrams will be presented in future 
works \cite{dean04}.

\section{Shell-model studies of selected nuclei}\label{sec:sec4}

The shell model problem requires the solution of a real symmetric
$n \times n$ matrix eigenvalue equation
\be
H\ket{vec_k} = E_k \ket{vec_k},
\label{e1}
\ee
with $k = 1,\ldots, K$. The eigenvalues $E_k$ are understood to be
numbered in increasing order. In a typical shell model problem
we are interested in only the lowest eigenstates of Eq.~(\ref{e1}),
so $K$ may be of the order of 10 to 50.
The total dimension $n$ of the eigenvalue matrix $H$ is large,
for the Sn isotopes of interest up to $n \approx 2 \times 10^{7}$.
Consider for example the case of a shell model calculation
outside the $Z = 50\;\,\; N = 50$ core with the single-particle orbits
$1d_{5/2}$, $0g_{7/2}$, $1d_{3/2}$, $2s_{1/2}$ and $0h_{11/2}$ defining the shell-model
space. Table \ref{tab:dims} shows examples of dimensionalities encountered for 
different nuclei.
\begin{table}
\caption{Shell-model dimensionalities for selected nuclei around $A\sim 100$.\label{tab:dims}}
\begin{center}
\begin{tabular}{|lrlr|} \hline
System & Dimension & System & Dimension\\ \hline
$^{104}$Sn  & $\approx 1.5 \cdot 10^3$&  $^{112}$Sn  & $\approx 6.2 \cdot 10^6$\\
$^{108}$Sn  & $\approx 3.2 \cdot 10^5$&  $^{116}$Sn  & $\approx 1.6 \cdot 10^7$\\ 
$^{104}$Sb  & $\approx 6.5 \cdot 10^3$ &  $^{112}$Sb  & $\approx 1.1 \cdot10^8$ \\
$^{108}$Sb  & $\approx 3.2 \cdot 10^6$ &  $^{116}$Sb  & $\approx 1.9 \cdot 10^9$ \\ \hline
\end{tabular}
\end{center}
\end{table}
Such large matrix problems are increasingly used in science and
enginering and practical numerical algorithms for determining their
properties are continually being developed.
These methods are closely related to the development of modern
computer technology, both  hardware and software.
What was impossible to solve a few years ago may now be within
reach and we should also be prepared for an increased  future
development. To indicate the present possibility,
in a work by J.~Olsen {\sl et al.}~\cite{ols90} in a quantum
chemistry configuration interaction calculation, Eq.~(\ref{e1})
was solved in a basis with $n = 10^{9}$. In nuclear physics there are presently
several groups dedicated in the development of 
shell-model codes which can deal with systems with similar large dimensionalities.
Notably are 
the Iowa-Livermore-Tucson group \cite{petr_erich2002}, the Michigan group with e.g., the OXBASH code
\cite{alex,mihai1,mihai2},
the Oak Ridge group \cite{andrius}, the Oslo group
\cite{oslo1} and 
the Strasbourg group \cite{etienne1}, just to mention a few. 
All these codes are nowadays capable of 
dealing with nuclear systems with dimensionalities of $n \sim 10^{9}$ basic states.
Extensive full $fp$-shell model and partial $sd-fp$ 
calculations have been carried out by the Strasbourg group
\cite{etienne2,etienne3,etienne4,etienne6}. 
Similarly, the Iowa-Livermore-Tucson group has dealt with
no-core shell-model calculations of light nuclei \cite{bruce1,bruce2,bruce3,bruce4}, 
with a very good agreement with the 
ab-initio Green's function Monte Carlo calculations of the Argonne group \cite{bob1,bob2,bob3}. 

Even larger shell-model dimensionalities can be dealt with e.g., Monte Carlo shell-model
approaches developed by Otsuka and collaborators \cite{taka1,taka2,taka3} and the shell-model
Monte Carlo approach of Koonin {\em et al.}~\cite{r:smmc_pr}.

Below we will focus on results for medium heavy nuclei in the mass regions from
$A=100$ to $A=140$ based on the Oslo shell-model code \cite{oslo1}.

Different computational approaches to solve Eq.(\ref{e1}) can
be distinguished
based on the size of $n$.
For $n$ small, i.e. $10^2 < n < 10^3$ and with  the number of
matrix elements of $H$
less than $10^6$ such  problems can be accomodated within the direct
access memory of a modern work station and can be diagonalized by
standard matrix routines.
In a second domain  with  $ n > 10^3$ but small enough that $H$
has no more than $10^8$ elements. This will require $\approx 1.5$~Gbyte
of storage.Then all matrix elements may be stored in memory
or alternatively on a standard disk.
In these cases the complete diagonalization of
$H$ will not be of physical interest and efficient iteration
procedures have been developed to find the lowest energy eigenvalues
and eigenvectors.

Based on the present computer methods we have developed a code
which is under continuous improvement
to solve the eigenvalue problem given in Eq.~(\ref{e1}).
The basic requirement
is to be able to handle problems with $n > 10^6$. In the following
we discuss some of the important elements which enter the algorithm.

We separate the discussion into three parts:
\begin{itemize}
\item The m--scheme representation of the basic states.
\item The Lanczos iteration algorithm.
\item The Davidson--Liu iteration technique.
\end{itemize}
\subsection{\it The m--scheme representation.}
We write the eigenstates in Eq.~(\ref{e1})  as  linear combinations
of Slater determinants. In a second quantization representation
a Slater determinant (SD) is given by
\be
\ket{SD_{\nu}(N)} = \prod_{(jm)\in {\nu}} a_{jm}^{\dagger}\ket{0},
\ee
and the complete set is generated by distributing the $N$ particles
in all possible ways throughout the basic one--particle states constituting
the P--space. This is a very efficient
representation. A single $\ket{SD}$ requires only one  computer word
(32 or 64 bits) and  in memory a $\ket{SD}$ with $N$ particles
is given by
\be
\ket{SD} \longrightarrow (\underbrace{00111101010 \cdots}_{N 1's}),
\ee
where each 0 and 1 corresponds to an m--orbit in the valence
P--space. Occupied orbits  have a 1 and empty orbits a 0.
 Furthermore, all important calculations  can
be handled in Boolean algebra which is very efficient on modern computers.
The action  of operators of the form $a_{\alpha}^{\dagger} a_{\beta}$ or
$a_{\alpha}^{\dagger} a_{\beta}^{\dagger} a_{\gamma} a_{\delta}$
acting on an $\ket{SD}$ is easy to perform.

The $m$-scheme allows also for a straightforward definition of many-body operators 
such as one--, two-- and three--particle operators
     \begin{equation}
     a_{\alpha}^{\dagger} a_{\beta},
     \end{equation}
\begin{equation}
     a_{\alpha_1}^{\dagger}a_{\alpha_2}^{\dagger} a_{\beta_1} a_{\beta_2} ,
\end{equation}
\begin{equation}
     a_{\alpha_1}^{\dagger}a_{\alpha_2}^{\dagger}a_{\alpha_3}^{\dagger}
            a_{\beta_1} a_{\beta_2} a_{\beta_3},
\end{equation}
respectively, 
or 
generalized seniority operators. The seniority operators can be very useful in preparing a
starting vector for the Lanczos iteration process. This option is fully 
implemented in our codes. 

The generalized seniority operators \cite{talmi} can then be written as 
\begin{equation}
    S^{\dagger}= \sum_{j}
    \frac{1}{\sqrt{2j+1}}C_{j}\sum_{m \geq 0} (-1)^{j-m} a^{\dagger}_{jm}
    a_{j-m}^{\dagger}
    \label{eq:zerosen}
\end{equation}
for seniority zero,
\begin{equation}
     D_{J M}^{\dagger}
         = \sum_{j \leq j', m,m'} (1+\delta_{j,j'})^{-1/2}
           \beta_{j,j'} \langle j m j' m'
           \left | J M \right \rangle a^{\dagger}_{jm}a^{\dagger}_{j' m'}
\label{eq:twosen}
\end{equation}
for seniority two. The coefficients $C_{j}$ and $\beta_{jj'}$ can be  obtained from the a chosen 
two-particle system such as the $^{130}$Sn
ground state and the excited states, respectively.
We can also define a
seniority four operator
\begin{eqnarray}
G(n_1,j_1, n_2,j_2; J,M)
      &=& \left \{D_{n_1,j_1}^{\dagger} D_{n_2,j_2}^{\dagger}
               \right \}_{J.M = 0}  \nonumber\\
      &=& \sum_{\nu_1 \ldots \nu_4} g_{\nu_1 \ldots \nu_4}^{JM}
           a_{\nu_1}^{\dagger}a_{\nu_2}^{\dagger}a_{\nu_3}^{\dagger}
                        a_{\nu_4}^{\dagger} 
  \end{eqnarray}
and  a seniority six operator
\begin{eqnarray}
I(n_1,j_1,(n_2,j_2,n_3,j_3)j_{23}; J,M)
        &=& \left \{ D_{n_1,j_1} G(n_2,j_2, n_3,j_3; j_{23})
              \right \}_{J,M = 0} \nonumber\\
        &=& \sum_{\nu_1 \ldots \nu_6} g_{\nu_1 \dots \nu_6}^{JM}
           a_{\nu_1}^{\dagger}a_{\nu_2}^{\dagger}a_{\nu_3}^{\dagger}
           a_{\nu_4}^{\dagger} a_{\nu_5}^{\dagger} a_{\nu_6}^{\dagger}
\end{eqnarray}

Finally, our shell-model code allows also for the inclusion
of effective and real three-body interactions. Results from such calculations
will be discussed in section \ref{sec:sec7}.

\subsection{\it The Lanczos iteration process.}
At present our basic approach to finding solutions to Eq.~(\ref{e1})
is the Lanczos algorithm. This method was already applied to nuclear
physics problems by Whitehead {\sl et al.} in 1977.
In a review article \cite{whit77} they describe the technique in detail.
For the present discussion we outline the basic elements
of the method.
\begin{enumerate}
\item We choose  an initial Lanczos vector $\ket{lanc_0}$ as the zeroth order
approximation to the lowest eigenvector in Eq.~(\ref{e1}). Our experience
is that any  reasonable choice  is acceptable as long as the
vector does not have special properties such as good angular momentum.
That would usually terminate the iteration process at too early a
stage.
\item The next step involves generating a new  vector
through the process $|new_{p+1}> = H |lanc_p>$.
Throughout this process we construct the energy matrix elements
of $H$ in this Lanczos basis. First, the diagonal matrix elements of $H$
are then obtained by

\be
\bra{lanc_p} H \ket{lanc_p} = \bra{lanc_p} \left . new_{p+1} \right \rangle,
\label{lanc1}
\ee

\item The new vector $\ket{new_{p+1}}$ is then orthogonalized to all
previously calculated Lanczos vectors
\be
\ket{new_{p+1}^{'}} = \ket{new_{p+1}} - \ket{lanc_p} \cdot
	                \bra{lanc_p} \left . new_{p+1} \right \rangle		 - \sum_{q = 0}^{p-1} \ket{lanc_q} \cdot
	          \bra{lanc_q} \left . new_{p+1} \right \rangle,
\ee
and finally normalized

\be
\ket{lanc_{p+1}} = \frac{1}{\sqrt{\bra{new_{p+1}^{'}}
                      \left . new_{p+1}^{'} \right \rangle}}
						\ket{new_{p+1}^{'}},
\ee
to produce a new Lanczos vector.
\item The off--diagonal matrix elements of $H$ are calculated by
\be
\bra{lanc_{p+1}} H \ket{lanc_p} = \bra{new_{p+1}^{'}}
                                \left . new _{p+1}^{'}\right \rangle,
\label{off1}
\ee

and all others are zero.
\item After n iterations we have an energy matrix of the form
\be
\left \{
\begin{array}{ccccc}
H_{0,0} & H_{0,1} & 0       & \cdots   & 0  \\
H_{0,1} & H_{1,1} & H_{1,2} & \cdots   & 0  \\
0       & H_{2,1} & H_{2,2} & \cdots   & 0  \\
\vdots  & \vdots  & \vdots  & \vdots   & H_{p-1,p}  \\
0       & 0       & 0       & H_{p,p-1}   & H_{p,p}\\
\end{array}
\right \}
\label{matr1}
\ee
as the p'th approximation to the eigenvalue problem in Eq.~(\ref{e1}).
The number p is a reasonably small number and we can diagonalize
the matrix by standard methods to obtain eigenvalues and eigenvectors
which are linear combinations of the Lanczos vectors.
\item This process is repeated until a suitable convergence
criterium has been reached.
\end{enumerate}
In this method each Lanczos vector is a linear combination
of the basic $\ket{SD}$ with dimension $n$. For
$n \approx 10^6-10^8$, as in our case of interest.
Here is  one of the important difficulties associated
with the Lanczos method. Large disk storage is needed when
the number of Lanczos vector exceeds $\approx 100$.
Another difficulty is found in the calculation of
$|new_{p+1}> = H |lanc_p>$ when $n > 10^6$.

One important objection found in the computer literature \cite{dav89}
to the Lanczos method is its slow convergence. This
is also our experience so far and means
that a large number of Lanczos vectors have to be calculated
and stored in order to obtain convergence.
One improvement which we have implemented is to
terminate the Lanczos process earlier, diagonalize the energy
matrix and choose some
of the lowest thus obtained eigenvectors as a starting point
for a new Lanczos iteration.
This modifies the energy matrix in Eq.~(\ref{matr1}) slightly,
since the matrix will not be tri--diagonal any more.
It reduces the disk storage requirement, but the convergence
problem is left. A possible way out here is the Davidson--Liu
method.

\subsection{\it The Davidson--Liu method.}
We outline the basic elements of this technique and refer to
the literature \cite{dav89} for details.
The first important improvement to the Lanczos process given in
Eqs.~(\ref{lanc1},\ref{matr1}) is to start with
several orthogonalized  and normalized initial vectors
and the second difference is the  way new additional vectors
are chosen. The method can be viewed
as an improvement of the Lanczos technique and can be described
in the following steps:
\begin{enumerate}
\item Choose a (small) set  of start vectors
$\ket{x_k^{(0)}},\;\; k = 1,\ldots, K$.
\item $K$ new vectors are generated through the process
$\ket{new_k^{(0)}} = H \ket{x_k^{(0)}}$.
\item The matrix elements of $H$ are again obtained by

\be
\bra{x_p^{(0)}} H \ket{x_q^{(0)}}
= \bra{x_p^{(0)}} \left . new_q^{(0)}\right \rangle,
\label{dav1}
\ee
and the matrix $H$ is no longer tri--diagonal.
$H$ is diagonalized within the set of states
$\ket{x_k^{(0)}},\;\; k = 1, \ldots, K$
and $K$ eigenvalues and corresponding eigenstates are obtained
as the zero'th approximation to the lowest true eigenstates
of Eq.~(\ref{e1}) in the form

\be
\ket{x_p^{(1)}} = \sum_{q = 1}^{K} a_{q, p} \ket{x_q^{(0)}}
                = \sum_{\nu = 1}^{n} C_{\nu,p}^{(1)} \ket{SD_{\nu}}.
\label{dav2}
\ee

The vectors $\ket{x_p^{(1)}}$ now constitute the basis for a new iteration.
\item New correction vectors are calculated through the steps
\begin{eqnarray}
\ket{r_p} &=&(H - \varepsilon_p^{(n)}) \ket{x_p^{(n)}}
	 =\ket{h_p^{(n)}} - \varepsilon_p^{(n)} \ket{x_p^{(n)}},
			                       \nonumber\\
\ket{\delta_p} &=& (H_{diag}-\varepsilon_p^{(n)})^{-1} \ket{r_p}.
\label{dav3}
\end{eqnarray}
Then new additional vectors are obtained by orthogonalizing
to all previous vectors, and finally normalized

\be
\ket{x_{k'+p}^{(n)}}
	= \sqrt{\frac{1}{norm}} \{ \ket{\delta_p}
		  -\sum_{q=1}^K \ket{x_q^{(n)}} \cdot
	                    \bra{x_q^{(n)}} \delta_p \rangle 
  -\sum_{q<p}^K \ket{x_{k'+q}^{(n)}} \cdot \bra{x_{K+q}^{(n)}}
		                  \delta_p \rangle  \},
\ee
witk $k' = 1, \ldots, K' \leq K$. Thus up to $K$ new vectors
may be generated through one iteration.
Then the diagonalization process in Eqs.~(\ref{dav1}--\ref{dav2}) is
repeated and a new iteration to the true eigenvectors is obtained.
\item Again this process may be repeated until some convergence
criterium has been reached.
\end{enumerate}

\subsection{\it Break-Up of the Doubly-Magic $^{100}$Sn Core}

Doubly-magic nuclei and their immediate neighbors are of great interest
as they provide direct information on the basic shell structure that is
ultimately responsible for most nuclear properties. Lying at the proton
drip line and being
the heaviest particle-stable, self-conjugate nucleus, 
$^{100}$Sn is particularly relevant in this context.
An important property of this nucleus is the degree of rigidity of its 
spherical shape which is reflected in the excitation energy 
of the lowest 2$^+$ state and in the associated B(E2;2$^+\rightarrow$0$^+$)
transition rate. 
The main component of the wave function of 
this level in a microscopic description is presumably
an isoscalar mixture of proton and 
neutron $2d_{5/2} 1g_{9/2}^{-1}$ excitations across the $N$$=$$Z$$=$$50$ shell 
gaps. This state is at present 
not known experimentally and its observation may well require 
the availability of intense exotic beams. 
Some guidance about its excitation energy can perhaps come from other
doubly-magic nuclei. In the $N$$=$$Z=$$28$ doubly-magic nucleus,
the first 2$^+$ state is located rather low, at 2.7 MeV. In contrast,
the 2$^+$ levels in $^{132}$Sn and $^{208}$Pb are much higher in 
excitation energy, 4.0 and 4.1 MeV, respectively, and in the 
latter nucleus the size of the shell gaps can also be appreciated from
the fact that this state is not even the lowest excitation, but instead
lies above a 3$^-$ (octupole) vibrational state.

In order to estimate the position of the 2$^+$ state in $^{100}$Sn,  
both proton and neutron shell gaps have to be known.
The energy splittings of the relevant single 
particle orbitals in the other heavy doubly magic nuclei are comparable. The 
neutron $2p_{3/2}$ and $1f_{7/2}$ orbitals are 6.4 MeV apart in $^{56}$Ni, 
while the splittings between the $2f_{7/2}$ and $1h_{11/2}$ levels 
in $^{132}$Sn and $2g_{9/2}$ 
and $1i_{13/2}$ states in $^{208}$Pb are 4.9 and 5.1 MeV, respectively.  
Here, the splitting between the $2d_{5/2}$ and $1g_{9/2}$ neutron orbits 
will be shown to be of the order of 6 MeV as in 
the $^{56}$Ni case. However, a sizable proton-neutron
interaction could, as in $^{56}$Ni, decrease the 
excitation energy and increase the transition rate for the lowest 2$^+$ 
state in $^{100}$Sn. Such an interaction is expected to be especially strong 
in $N$$=$$Z$ nuclei, where protons and neutrons occupy the same 
single-particle orbitals, which results in a large spatial
overlap of their wave functions.
\begin{figure}
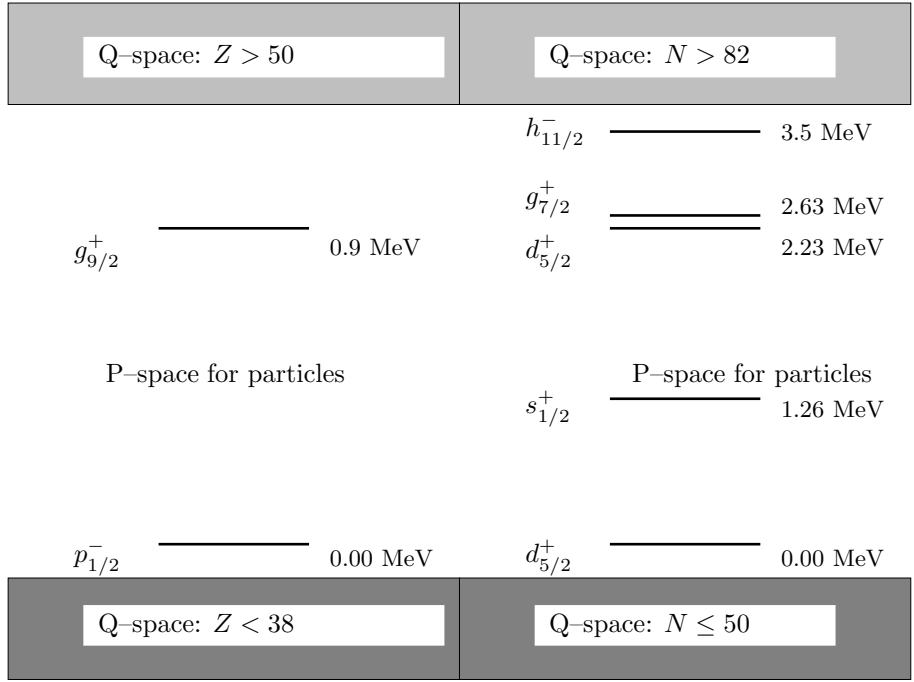

\setlength{\unitlength}{1cm}
\begin{center}

\Cartesian(1cm,0.9cm)
\pspicture(0,1)(12,12)
\psframe[linewidth=0.0pt,fillstyle=solid,fillcolor=gray](6,1.5)(12,3)
\psframe*[linecolor=white](7,2)(11,2.6)
\uput[0](6.9,2.3){ Q--space: $N \leq 50$}
%
%	single-particle spectrum
%
\psline[linewidth=1pt](8,3.5)(10,3.5)
\uput[0](6.7,3.3){$d_{5/2}^{+}$}
\uput[0](10.1,3.3){\small 0.00 MeV}
\psline[linewidth=1pt](8,5.65)(10,5.65)
\uput[0](6.7,5.5){$s_{1/2}^{+}$}
\uput[0](10.1,5.5){\small 1.26 MeV}
\psline[linewidth=1pt](8,8.17)(10,8.17)
\uput[0](6.7,7.8){$d_{5/2}^{+}$}
\uput[0](10.1,7.9){\small 2.23 MeV}
\psline[linewidth=1pt](8,8.36)(10,8.36)
\uput[0](6.7,8.6){$g_{7/2}^{+}$}
\uput[0](10.1,8.5){\small 2.63 MeV}
\psline[linewidth=1pt](8,9.6)(10,9.6)
\uput[0](6.7,9.6){$h_{11/2}^{-}$}
\uput[0](10.1,9.6){\small 3.5 MeV}
\uput[0](8.0,6.0){ P--space for particles}

\psframe[linewidth=0.0pt,fillstyle=solid,fillcolor=lightgray](6,10)(12,11.5)
\psframe*[linecolor=white](7,10.4)(11,11)
\uput[0](6.9,10.7){ Q--space: $N > 82$}
\psframe[linewidth=0.0pt,fillstyle=solid,fillcolor=gray](0,1.5)(6,3)
\psframe*[linecolor=white](1,2)(5.8,2.6)
\uput[0](0.9,2.3){ Q--space: $Z < 38 $}
%
%	single-particle spectrum
%
\psline[linewidth=1pt](2,3.5)(4,3.5)
\uput[0](0.7,3.3){$p_{1/2}^{-}$}
\uput[0](4.1,3.3){\small 0.00 MeV}
\psline[linewidth=1pt](2,8.17)(4,8.17)
\uput[0](0.7,7.8){$g_{9/2}^{+}$}
\uput[0](4.1,7.9){\small  0.9 MeV}
\uput[0](1.0,6.0){ P--space for particles}

%%%%%%%%%%%%%%
\psframe[linewidth=0.0pt,fillstyle=solid,fillcolor=lightgray](0,10)(6,11.5)
\psframe*[linecolor=white](1,10.4)(5.8,11)
\uput[0](0.9,10.7){ Q--space: $Z > 50$}
\endpspicture
\caption{Possible shell-model space for nuclei around $A\sim 100$ using 
$^{88}$Sr as closed-shell core with the neutron orbitals $2s_{1/2}$, 
$1d_{5/2}$, $1d_{3/2}$, $0g_{7/2}$ and $0h_{11/2}$ and 
proton single-particle orbitals  $0g_{9/2}$ and $1p_{1/2}$ defining the model space.
Their respective single-particle energies are displayed as well. \label{fig:sr88ms}}
\end{center}
\end{figure}
Nuclei near $^{100}$Sn were studied using the 
$^{58}$Ni+$^{50}$Cr reaction.
A $^{58}$Ni beam of 225 MeV was provided by the ATLAS
superconducting linear accelerator at Argonne National Laboratory.
The effective two-body interactions were applied in a shell-model
space spanning $2s_{1/2}$, 
$1d_{5/2}$, $1d_{3/2}$, $0g_{7/2}$ and $0h_{11/2}$ neutron, and 
$0g_{9/2}$ and $1p_{1/2}$ proton single-particle orbitals \cite{anne}. 
The model space is shown in Fig.~\ref{fig:sr88ms}.
In this model space, $^{99}$Cd has one neutron and 10 protons outside the core.
The results of the calculation, denoted as SMH, are compared with the
experimental levels in Fig.~\ref{fig:cd99}.
The calculation favors a $J^{\pi}$=5/2$^+$ assignment for the ground state, in 
agreement with the systematics of odd-A, $N$=51 isotones.  
   \begin{figure}
   \setlength{\unitlength}{1mm}
   \begin{picture}(150,160)
   \put(0,0){\epsfxsize=16cm \epsfbox{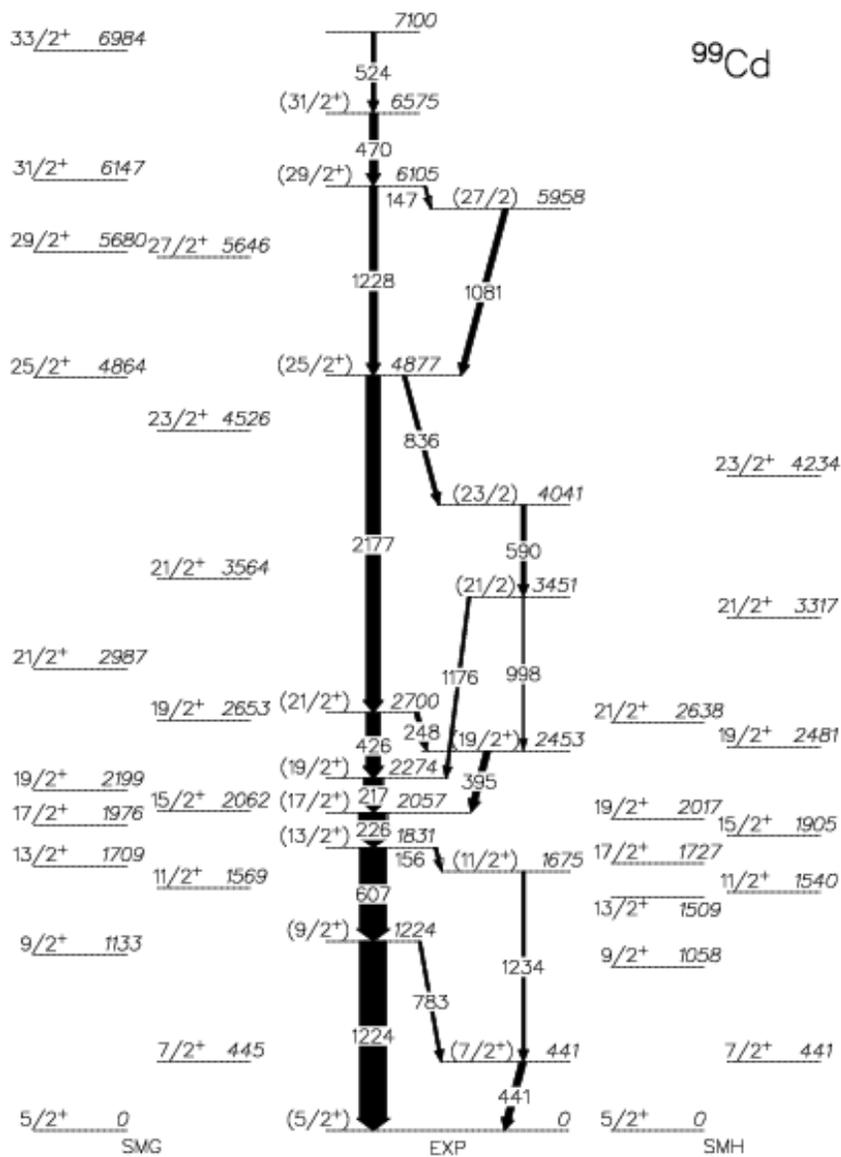}}
   \end{picture}
\caption{
Experimental (EXP) and calculated (SMH, SMG) level schemes for $^{99}$Cd. 
The widths of the arrows are proportional to the intensities of
the $\gamma$ rays observed in the experiment. 
As discussed in the text, the $J=25/2^+$ and higher-lying states
require excitations across the $^{100}$Sn core. 
\label{fig:cd99}}
   \end{figure}
The wave functions of 
the states with $J^{\pi}$=5/2$^+$, 9/2$^+$, 13/2$^+$, 17/2$^+$, 19/2$^+_1$ and 
21/2$^+_1$ all have main configurations where the valence neutron 
is in the $d_{5/2}$ orbit, 
while the 7/2$^+$, 11/2$^+$, 15/2$^+$, 19/2$^+_2$, 21/2$^+_2$ and 
23/2$^+$ levels are associated with the predominant occupation of
the $g_{7/2}$ orbit. For all these states, the two proton 
holes always remain assigned to the $g_{9/2}$ orbit. 
The lowest 7/2$^+$ states in N=51 isotones from $^{91}$Zr to $^{97}$Pd 
are well reproduced by using the experimental $g_{7/2}$ single-particle
energy from zirconium \cite{anne}, but in $^{99}$Cd this state was 
calculated 250 keV too low. Therefore, the 
single-particle energy of the neutron $g_{7/2}$ orbit, 
relative to the $^{88}$Sr, core was increased from 2.63 to 2.89~MeV 
in order to reproduce the experimental excitation energy of the 7/2$^+$ state. 
This yielded excellent agreement between calculations and experiment up to the 
$J^{\pi}$=23/2$^+$ state, see for example Fig.~\ref{fig:cd99}, the
highest spin that can be generated in this model space for positive-parity 
states in $^{99}$Cd. The description of
higher-spin states requires the excitation of one or more 
$g_{9/2}$ particles across the $N,Z$=50 shell gaps,
an excitation similar to that
responsible for the first 2$^+$ and higher-lying states
in $^{100}$Sn.

To study the high-spin states, another shell-model calculation was performed
using this time $^{78}$Sr as a closed shell core. 
The results are denoted as SMG in Fig.~\ref{fig:cd99}.
The same model space as in the previous calculation was employed except for 
the $h_{11/2}$ neutron orbit which was replaced by the $g_{9/2}$ one. 
(Since this $h_{11/2}$ state lies at a relatively high excitation energy, 
its contribution to the positive-parity states is small.)
Due to limitations in computing time, only up to two neutrons were allowed to 
leave the $g_{9/2}$ orbit. (Note that the $g_{9/2}$ orbit lies below the $N$=50
shell gap, while all other neutron orbitals in this model space lie above 
this gap.) Opening of the $N$=50 shell only is justified 
since a neutron hole in the 
$g_{9/2}$ orbit, together with an existing proton hole, produces a very 
attractive 
9$^+$ proton-neutron coupling. Similarly, the excited neutron, together with 
the one already occupying the same orbit, 
produces a 0$^+$ neutron pair with a strong 
attractive interaction. Opening the $Z$=50 shell would not result in such
a strong attraction.
Single-particle energies with respect to the $^{78}$Sr core are not known and 
were 
kept the same as in Ref.~\cite{anne} for the $^{88}$Sr core. The energy of the 
$g_{9/2}$ orbit was placed 5.0 MeV below that of the $d_{5/2}$ one. 
The wave functions of the states below the 25/2$^+$ level are very 
similar to those obtained in the SMH calculation
(which assumed that the $g_{9/2}$ neutron orbit is completely filled),
except for the two 19/2$^+$ levels, where the occupation numbers are
reversed for the $d_{5/2}$ and $g_{7/2}$ orbits.  
The 25/2$^+$ level and the higher-lying states have 9 neutrons in the $g_{9/2}$ 
orbit with the remaining neutron pair almost evenly distributed over the 
$d_{5/2}$ and $g_{7/2}$ orbits.  These states represent the break-up 
of the doubly-magic $^{100}$Sn core.
Thus, the excitation 
energy of the states with $J^\pi$$\geq$25/2$^+$ is sensitive to the 
position of the $g_{9/2}$ orbit. By fitting this single particle energy 
to 5.0 MeV we, 
therefore, indirectly deduced the size of the $N$=50 shell gap to be 6.5 MeV
as defined by 2BE($^{100}$Sn)-BE($^{99}$Sn)-BE($^{101}$Sn), where BE stands
for binding energy. This value agrees well with earlier predictions from
Hartree-Fock calculations by Leander et al.~\cite{leander}, as well as
a single-particle energy parametrization by Duflo and Zuker~\cite{dz1999} 
and an extrapolation by Grawe et al.~\cite{grawe}. To investigate the
effect of the truncation of the model space on the deduced size of the
$N$=50 shell gap we calculated the excitation energy of the 
$J^\pi$=25/2$^+$ state in $^{99}$Cd using two different model spaces.
In the first one, we allowed up to four neutrons to leave the $g_{9/2}$
orbit and the second truncation scheme restricted the valence particles
to only two in each active orbit. Different truncations of the model space
required adjustments of the $g_{9/2}$ single-particle energy to reproduce
the experimental excitation energy of the 25/2$^+$ level, but when this
was achieved the size of the $N$=50 shell gap remained within 0.5 MeV of
the above value. This illustrates a relative insensitivity of our 
inferred result to the truncation procedure used in the shell model
calculation.

\begin{figure}
   \setlength{\unitlength}{1mm}
   \begin{picture}(150,160)
   \put(0,0){\epsfxsize=16cm \epsfbox{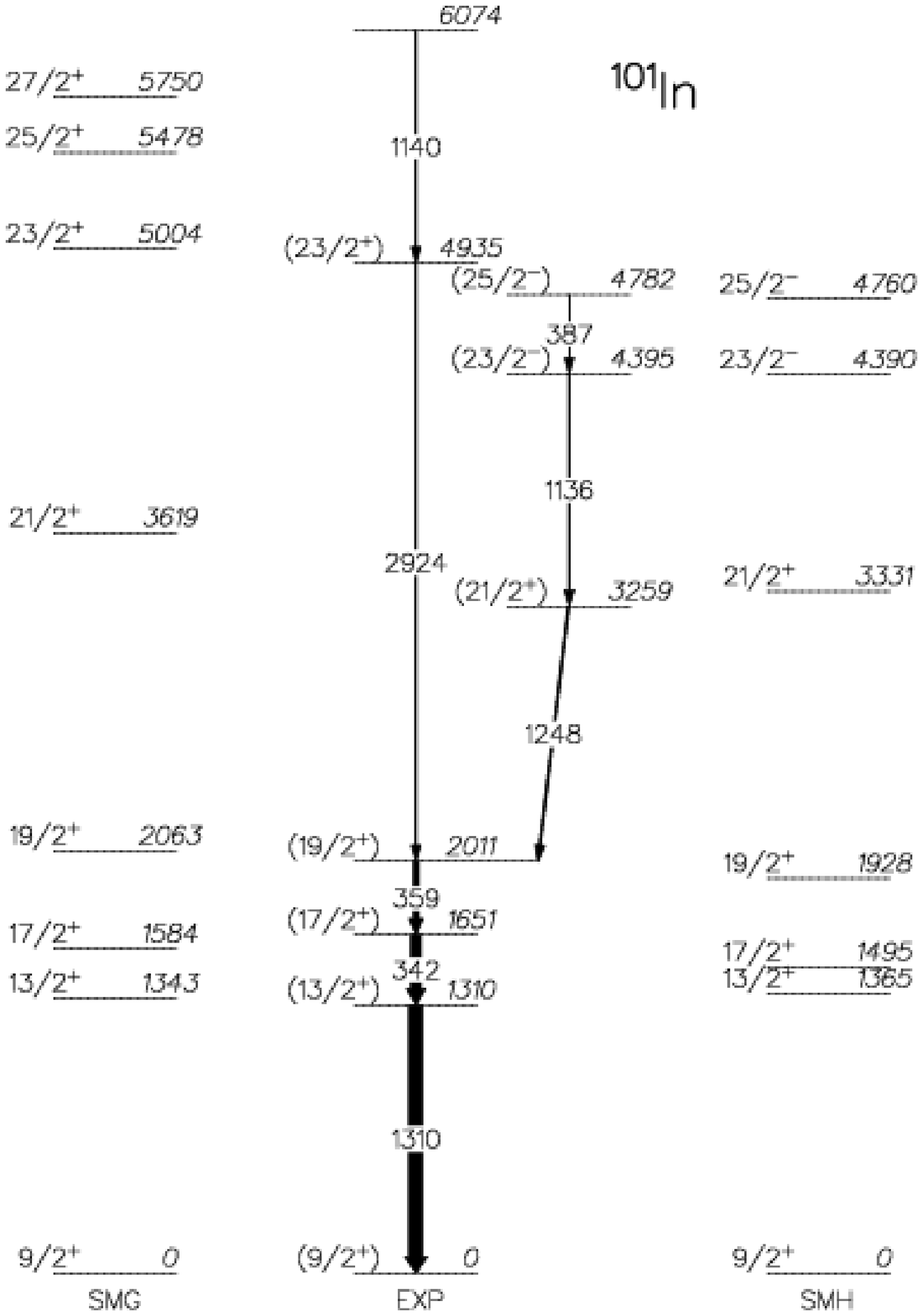}}
   \end{picture}
\caption{
Experimental (EXP) and calculated (SMH, SMG) level schemes for $^{101}$In. 
The widths of the arrows are proportional to the intensities of
the $\gamma$ rays observed in the experiment. The ($J=23/2^+$) state 
requires excitations across the $^{100}$Sn core.
\label{fig:in101}}
   \end{figure}

Calculations using the same single particle energies and
matrix elements were also performed
for $^{101}$In, which has one proton hole and two neutrons outside the 
$N$$=$$Z$$=$$50$ doubly-closed shells. Again,
the results, labeled as SMH and SMG, are in very good agreement with
the experimental level scheme in Fig.~\ref{fig:in101}.
In these calculations, the proton hole remains in the
$g_{9/2}$ orbital and a
$J^{\pi}$=9/2$^+$ assignment follows for the ground state.  
This is in agreement with the systematics of odd-A In isotopes. 
The valence neutron pair is distributed over the $d_{5/2}$ and $g_{7/2}$ orbits 
for the ground state, while it occupies
predominantly the $d_{5/2}$ state for the 13/2$^+$ level. 
One neutron is promoted to the $g_{7/2}$ and $h_{11/2}$ orbitals
to generate the 17/2$^+$, 19/2$^+$, 21/2$^+$ positive-parity,
and 23/2$^-$, 25/2$^-$ negative-parity states, respectively.
The large separation between the 19/2$^+$ and 
21/2$^+$ states, which are associated with the same dominant configuration, 
is attributed to the strong repulsion between the aligned $g_{9/2}$ 
proton hole and the $g_{7/2}$ neutron. 
Note that the $\pi g_{9/2}\nu h_{11/2}$ matrix elements used
to calculate the negative-parity states in 
$^{102}$In described in Ref.~\cite{in102} were adopted here as well. 
The maximum spin that can be reached with the 
$\pi g_{9/2} \nu d_{5/2} g_{7/2}$ configuration is 21/2.  
Excitation of one or more particles across the $N,Z$=50
shell gaps is again needed to account for 
the positive-parity states of higher spin.
Thus, the 23/2$^+$ level is the lowest-lying core-excited state in $^{101}$In. 
The calculated energy gap of 2941 keV between the 23/2$^+$
and  19/2$^+$ states is in excellent agreement with the
experimental value of 2924 keV.
Experimentally known lowest lying core excited states were also calculated 
in the nuclei $^{98}$Ag \cite{matjaz} and $^{96}$Pd \cite{alber} 
using the SMG model space. Their 
excitation energies were reproduced to better than 100 keV. It is
more difficult to identify core excited states in nuclei near $^{90}$Zr,
because in those nuclei high-spin states can be easily reached by
promoting a pair of protons from the low-spin $p_{1/2}$ orbit into the
empty $g_{9/2}$ orbit.

The interactions used in these SMH and SMG calculations 
were also applied
to a shell model calculation of the lowest-lying levels in $^{101}$Sn.
Both SMH and SMG calculations favor $J^{\pi}$=5/2$^+$ quantum numbers
for the ground state, while 
the 7/2$^+$ state lies only $\sim$100 keV above. This is in excellent agreement 
with the extrapolated energy deduced from the systematics
of odd-$A$ Sn isotopes down to $^{103}$Sn \cite{sn103}.

It is also worth pointing out that the level schemes of $^{99}$Cd and $^{101}$In 
closely mirror those of their analogs  
in the $^{56}$Ni region; i.e., $^{55}$Fe~\cite{fe55} and 
$^{57}$Co~\cite{co57}.
In particular, the lowest-lying core-excited states have almost identical 
excitation energies.
This observation, together with the deduced $N$=50 shell gap,
again points to a large similarity between the two heaviest self-conjugated 
doubly-magic nuclei and one may conclude that
the lowest 2$^+$ states in $^{56}$Ni and $^{100}$Sn would have closely
related excitation energies if the sizes of the $Z$=28 and $Z$=50 shell
gaps were also similar.

\subsection{\it $^{100}$Sn Core Excitations in $^{102}$In}
\begin{figure}
   \setlength{\unitlength}{1mm}
   \begin{picture}(150,160)
   \put(0,0){\epsfxsize=16cm \epsfbox{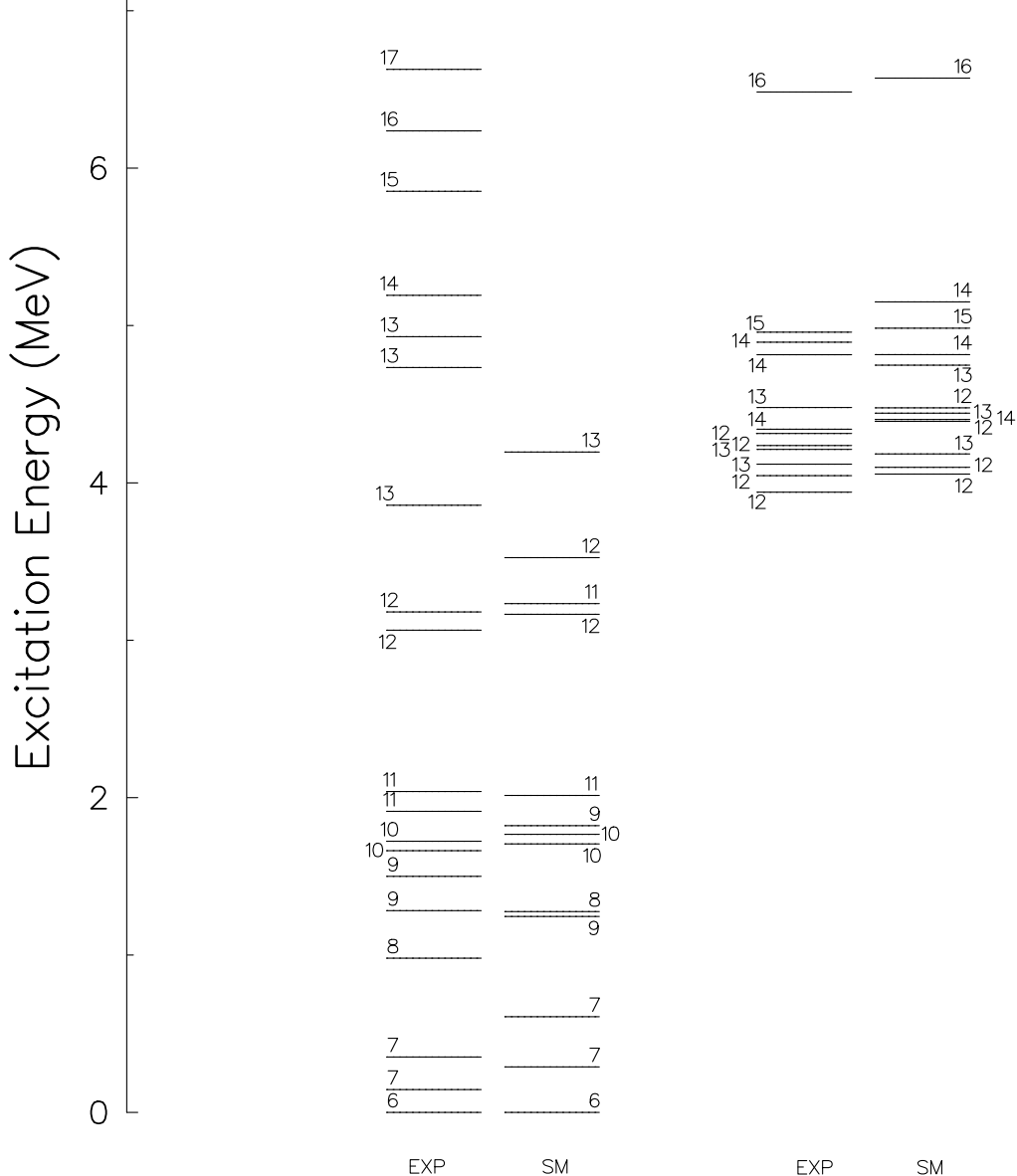}}
   \end{picture}
   \caption{Comparison of experimental and calculated levels in
    $^{102}$In.\label{fig:in102fig}}
   \end{figure}

In the above model space $^{102}$In has three valence neutrons and 11 
valence protons, which is equivalent to one proton hole in the 
doubly magic $^{100}$Sn core. In all calculated levels shown in
Fig.~\ref{fig:in102fig} this proton hole remains in the $g_{9/2}$ orbit. Therefore,
the only proton contribution to the calculated level scheme of 
$^{102}$In is through proton-neutron pairing. 
The calculation favors $J^{\pi}$=6$^+$ for 
the ground state. The three valence neutrons are mainly in the
$d_{5/2}$ orbit in this state, while about 40\% of the wave function 
amplitude comes from the contribution of the $g_{7/2}$ orbit. All
other orbits have an insignificant contribution to the wave function
of the ground state. Surprisingly, all levels up to the $J^\pi$=10$^+_2$
level have a very similar wave function configuration in the calculation.
The 10$^+_2$ level is the lowest observed level for which the 
$d_{5/2}$ and $g_{7/2}$ orbits switch their occupation numbers. 
The larger wave function difference results in lower mixing between 
the 10$^+_1$ and 10$^+_2$ levels. This in turn leads to two close
lying 10$^+$ states, correctly predicted by the calculation.
However, all other non-yrast states are calculated too high and
we expect that the $g_{7/2}$ orbit has a larger than calculated
contribution to their wave functions. This is especially true for
the 11$^+_2$ state that is calculated more than 1 MeV too high.
We may conclude that the interaction used in the calculation
gives a too strong attraction between the $g_{9/2}$ protons
and the $g_{7/2}$ neutrons. 
  
The very nice one to one correspondence between experimental 
and calculated positive parity states ends with the 13$^+_1$
state, which is the highest spin that can be reached by coupling
one proton hole in the $g_{9/2}$ orbit with three neutrons 
in $d_{5/2}$ and $g_{7/2}$ orbits. The wave function of this
state is therefore $\pi$($g_{9/2}$)$^{-1}$$\nu$$d_{5/2}$($g_{7/2}$)$^{2}$.

The experimental positive parity levels above the 3858 keV level
do not have calculated counterparts. The only possibility to
reach such spins and parities within the model space used in
the calculation is to promote a neutron pair to the $h_{11/2}$ orbit. 
However, even after changing the $\pi$$g_{9/2}$$\nu$$h_{11/2}$ 
effective interaction to reproduce the negative parity states
the $J^{\pi}$=13$^+_2$ to 17$^+$ levels were calculated much higher in
energy than the experimental ones. These states must, therefore,
be due to excitations across the doubly closed $N=Z=50$ shell.
Most likely they are due to the excitation of a neutron from the
$g_{9/2}$ orbit just below $N=50$ shell gap to the $d_{5/2}$
orbit just above the gap, since this produces the very 
attractive ($\pi$$g_{9/2}^{-1}$$\nu$$g_{9/2}^{-1}$)$^{9+}$
coupling combined with four neutron particle states like
the $J^{\pi}$=6$^+$ to 8$^+$ and 10$^+$ states in $^{104}$Sn.

\subsection{\it Tin isotopes for $100 \le A \le 132$}

\begin{figure}
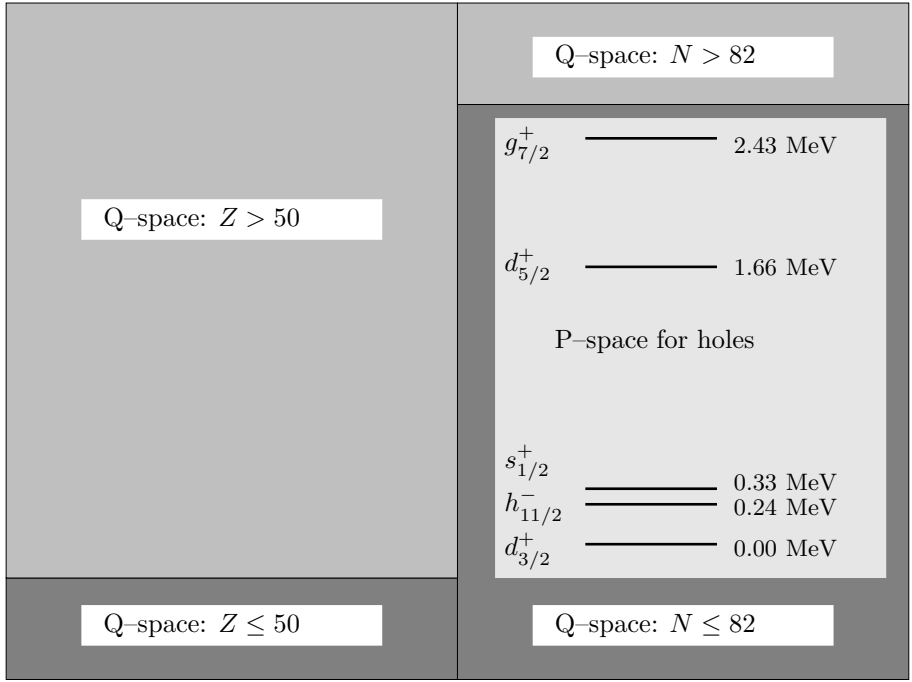

\setlength{\unitlength}{1cm}
\begin{center}

\Cartesian(1cm,0.9cm)
\pspicture(0,1)(12,12)
\newgray{whitegray}{.9}

\psframe[linewidth=0.0pt,fillstyle=solid,fillcolor=gray](0,1.5)(6,3)
\psframe*[linecolor=white](1,2)(5,2.6)
\uput[0](1.0,2.3){ Q--space: $Z \leq 50$}
\psframe[linewidth=0.0pt,fillstyle=solid,fillcolor=gray](6,1.5)(12,10)
\psframe*[linecolor=whitegray](6.5,3)(11.7,9.8)
\psframe*[linecolor=white](7,2)(11,2.6)
\uput[0](7.0,2.3){ Q--space: $N \leq 82$}
%
%	single-particle spectrum
%
\psline[linewidth=1pt](7.7,3.5)(9.45,3.5)
\uput[0](6.45,3.40){$d_{3/2}^{+}$}
\uput[0](9.5,3.45){\small 0.00 MeV}
\psline[linewidth=1pt](7.7,4.09)(9.45,4.09)
\uput[0](6.45,4.05){$h_{11/2}^{-}$}
\uput[0](9.5,4.05){\small 0.24 MeV}
\psline[linewidth=1pt](7.7,4.32)(9.45,4.32)
\uput[0](6.45,4.70){$s_{1/2}^{+}$}
\uput[0](9.5,4.42){\small 0.33 MeV}
\psline[linewidth=1pt](7.7,7.6)(9.45,7.6)
\uput[0](6.45,7.6){$d_{5/2}^{+}$}
\uput[0](9.5,7.6){\small 1.66 MeV}
\psline[linewidth=1pt](7.7,9.5)(9.45,9.5)
\uput[0](6.45,9.4){$g_{7/2}^{+}$}
\uput[0](9.5,9.4){\small 2.43 MeV}
\uput[0](7,6.5){ P--space for holes }

%%%%%%%%%%%%%%
\psframe[linewidth=0.0pt,fillstyle=solid,fillcolor=lightgray](0,3)(6,11.5)
\psframe*[linecolor=white](1,8)(5,8.6)
\uput[0](1.0,8.3){ Q--space: $Z > 50$}

\psframe[linewidth=0.0pt,fillstyle=solid,fillcolor=lightgray](6,10)(12,11.5)
\psframe*[linecolor=white](7,10.4)(11,11)
\uput[0](7.0,10.7){ Q--space: $N > 82$}
\endpspicture
\caption{Shell-model space for tin isotopes for $100 \le A \le  132$. The valence neutrons are hole states.\label{fig:sn132sm}}
\end{center}
\end{figure}

In Fig.~\ref{fig:sn132sm} we present the model space and pertinent single-particle energies
for a shell-model space for neutron valence holes using $^{132}$Sn as closed shell core.
In Table \ref{tab:sn132spectra} we display  
selected states. 
As can be seen, 
the well-known
near constant $0^{+} - 2^{+}$ spacing is well reproduced.
all the way down to
$^{116}$Sn.
Also the additional calculated states are in very good agreement
with experiment. However more detailed analysis of
the results close to $^{116}$Sn
indicates that our effective two-particle interaction
has difficulties in reproducing
the shell closure which is believed to occur in this region.
The increase of the
the $0^{+} - 2^{+}$ splitting is not as sharp as found
experimentally, even if the
phenomenon is rather weak in the case of Sn.
We have observed a similar feature around $^{48}$Ca \cite{hko95}
which is generally agreed
to be a good closed shell nucleus. There the deviation
between theory and experiment
is severe. Preliminary analysis indicates that our effective interaction
may be slightly too attractive when the two particles
occupy different single-particle orbits.
This may be related to the radial wave functions which in our calculation are
chosen to be harmonic oscillator functions. Further analysis of these spectra
can be found in Refs.~\cite{ehho98,haavard2004}.
\begin{table}[t]
\caption{Exitation spectra for the heavy Sn isotopes. \label{tab:sn132spectra}}
\vspace{0.2cm}
\begin{center}
\footnotesize
\begin{tabular}{|cccccccc|}
\hline
&&&&&&&\\[-5pt]
\multicolumn{4}{|c}{$^{130}$Sn}&\multicolumn{4}{c|}{$^{128}$Sn}\\
$J^{\pi}$&Exp.&$J^{\pi}$&Theory&$J^{\pi}$&Exp.&$J^{\pi}$&Theory\\
\hline
$(2^{+})$ & $1.22$ & $2^{+}$   & $1.46$ &$(2^{+})$ & $1.17$ & $2^{+}$ & $1.28$\\
$(4^{+})$ & $2.00$ & $4^{+}$   & $2.39$ &$(4^{+})$ & $2.00$ & $4^{+}$ & $2.18$\\
$(6^{+})$ & $2.26$ & $6^{+}$   & $2.64$ &$(6^{+})$ & $2.38$ & $6^{+}$ & $2.53$\\\hline
\multicolumn{4}{|c}{$^{126}$Sn}&\multicolumn{4}{c|}{$^{124}$Sn}\\
$J^{\pi}$&Exp.&$J^{\pi}$&Theory&$J^{\pi}$&Exp.&$J^{\pi}$&Theory\\\hline
$2^{+}$ & $1.14$ & $2^{+}$   & $1.21$ &$2^{+}$ & $1.13$ & $2^{+}$ & $1.17$\\
$4^{+}$ & $2.05$ & $4^{+}$   & $2.21$ &$4^{+}$ & $2.10$ & $4^{+}$ & $2.26$\\
$     $ &        & $6^{+}$   & $2.61$ &        &        & $6^{+}$ & $2.70$\\\hline
\multicolumn{4}{|c}{$^{122}$Sn}&\multicolumn{4}{c|}{$^{120}$Sn}\\
$J^{\pi}$&Exp.&$J^{\pi}$&Theory&$J^{\pi}$&Exp.&$J^{\pi}$&Theory\\
\hline
$2^{+}$   & $1.14$ & $2^{+}$   & $1.15$ & $2^{+}$  & $1.17$ & $2^{+}$ & $1.14$\\
$4^{+}$   & $2.14$ & $4^{+}$   & $2.30$ & $4^{+}$  & $2.19$ & $4^{+}$ & $2.30$\\
$6^{+}$   & $2.56$ & $6^{+}$   & $2.78$ &          &        & $6^{+}$ & $2.86$\\\hline
\multicolumn{4}{|c}{$^{118}$Sn}&\multicolumn{4}{c|}{$^{116}$Sn}\\
$J^{\pi}$&Exp.&$J^{\pi}$&Theory&$J^{\pi}$&Exp.&$J^{\pi}$&Theory\\
\hline
$2^{+}$   & $1.22$ & $2^{+}$   & $1.15$ & $2^{+}$  & $1.30$ & $2^{+}$ & $1.17$\\[3pt]\hline
\end{tabular}
\end{center}
\end{table}

It is also instructive to study the degree of pairing in these systems.
We calculate the squared overlaps between the constructed generalized seniority
states in Eqs.~(\ref{eq:zerosen}) and (\ref{eq:twosen}) and our shell model states
through
\begin{equation}
\begin{array}{|lccccl|}
(v=0) & & & & &
|\langle ^{A}{\rm Sn(SM)} ;
0^{+}|(S^{\dagger})^{\frac{n}{2}}| \tilde{0} \rangle |^{2}, \\
(v=2) & & & & &
|\langle ^{A}{\rm Sn(SM)} ;
J_{i}|D^{\dagger}_{JM}(S^{\dagger})^{\frac{n}{2}-1}| \tilde{0} \rangle |^{2}. 
\end{array}
\end{equation}
The vacuum state $|\tilde{0} \rangle $ is the $^{132}$Sn--core and $n$ is
the number of valence particles. These quantities tell to what extent the 
shell model states satisfy the pairing picture, or in other words, how well 
is generalized seniority conserved as a quantum number.

The squared overlaps are tabulated in Table~\ref{tab:seniority}, and vary
generally from 0.95 to 0.75. As the number of valence particles increases the 
squared overlaps are gradually decreasing. The overlaps involving the $4^{+}$
states show a fragmentation. In $^{128}$Sn, the $4^{+}_{1}$ (SM)
state is mainly a seniority $v=2$ state. As approaching the middle of the 
shell, the next state, $4^{+}_{2}$, takes more and more over the 
structure of a seniority $v=2$ state. The fragmentation of seniority over these
two states can be understood from the fact that they are rather close in 
energy and therefore may have mixed structure.

\begin{table}[htbp]
\caption{ Seniority $v=0$ overlap 
         $|\langle ^{A}Sn;0^{+}|(S^{\dagger})^{\frac{n}{2}}| 
         \tilde{0} \rangle |^{2}$ and the seniority $v=2$ 
         overlaps $|\langle ^{A}Sn ;J_{f}|
         D^{\dagger}_{JM}(S^{\dagger})^{\frac{n}{2} - 1}| 
         \tilde{0} \rangle |^{2}$ for the lowest--lying eigenstates 
         of $^{128-120}$Sn.}
\label{tab:seniority}
\begin{center}
\begin{tabular}{|cccccc|}
\hline
 & A=128 & A=126 & A=124 & A=122 & A=120 \\ 
\hline
$0^{+}_{1}$ & 0.96 & 0.92 & 0.87 & 0.83 & 0.79 \\ 
$2^{+}_{1}$ & 0.92 & 0.89 & 0.84 & 0.79 & 0.74 \\ 
$4^{+}_{1}$ & 0.73 & 0.66 & 0.44 & 0.13 & 0.00 \\ 
$4^{+}_{2}$ & 0.13 & 0.18 & 0.39 & 0.66 & 0.74 \\
$6^{+}_{1}$ & 0.81 & 0.85 & 0.83 & 0.79 & 0.64 \\
\hline
\end{tabular}
\end{center}
\end{table}

\subsection{\it Tin isotopes above $A=132$}

For tin isotopes above $A=132$ calculations were performed for isotopes with $134 \le A \le 139$
taking $Z = 50, N = 82$ as core. The active $P$-space for particles includes the
$1f_{7/2}$, $1p_{3/2}$, $2p_{1/2}$, $0h_{9/2}$, $1f_{5/2}$ and $0i_{13/2}$ particle orbits,
The single-particles energies
$\varepsilon(f_{7/2}^{-}) = 0.00$~MeV,
$\varepsilon(p_{3/2}^{-}) = 0.8537$~MeV,
$\varepsilon(p_{1/2}^{-}) = 1.5609$~MeV,
$\varepsilon(f_{5/2}^{-}) = 1.6557$~MeV
$\varepsilon(h_{9/2}^{-}) = 2.0046$~MeV and
$\varepsilon(i_{13/2}^{+}) =  2.81$~MeV
are extracted from \cite{hoff96}. 
The obtained results are presented in Table \ref{tab:even_isotopes}
for isotopes with even values of $A$ and in Table \ref{tab:odd_isotopes} for isotopes with odd $A$.

There are however no experimental data for the $A > 134$ tin isotopes available at the present time.
We therefore cannot make conclusions on how good our effective interaction is for this mass
region. A more detailed analysis of these results can be found in Ref.~\cite{maximsn134}.

\begin{table}[htbp]
\caption{Low--lying states for  $^{134}$Sn, $^{136}$Sn and $^{138}$Sn.
The energy eigenvalues are sorted according to the angular momentum assignment.}
\label{tab:even_isotopes}
\begin{center}
\begin{tabular}{|cccc|cccc|cccc|}
\hline
\multicolumn{4}{|c|}{$^{134}$Sn} & \multicolumn{4}{|c|}{ $^{136}$Sn}& \multicolumn{4}{|c|}{ $^{138}$Sn}\\
&{$J^{\pi}_i$}&Theory&&&{$J^{\pi}_i$}&Theory&&&{$J^{\pi}_i$}&Theory& \\

\hline
%   0+ states
& $0^{+}_{1}$ & 0.00   &&   &$0^{+}_{1}$ & 0.00   &&   &$0^{+}_{1}$ & 0.00   &\\
& $0^{+}_{2}$ & 2.28   &&   &$0^{+}_{2}$ & 1.86   &&   &$0^{+}_{2}$ & 1.52   &\\
& $0^{+}_{3}$ & 2.97   &&   &$0^{+}_{3}$ & 1.92   &&   &$0^{+}_{3}$ & 1.78   &\\
&             &        &&   &$0^{+}_{4}$ & 2.34   &&   &$0^{+}_{4}$ & 1.89   &\\

\hline
%   2+ states
& $2^{+}_{1}$ & 0.77   &&   &$2^{+}_{1}$ & 0.73   &&   &$2^{+}_{1}$ & 0.75   &\\
& $2^{+}_{2}$ & 1.65   &&   &$2^{+}_{2}$ & 1.45   &&   &$2^{+}_{2}$ & 1.27   &\\
& $2^{+}_{3}$ & 2.64   &&   &$2^{+}_{3}$ & 1.52   &&   &$2^{+}_{3}$ & 1.71   &\\
& $2^{+}_{4}$ & 2.73   &&   &$2^{+}_{4}$ & 1.87   &&   &$2^{+}_{4}$ & 1.87   &\\
& $2^{+}_{5}$ & 3.13   &&   &$2^{+}_{5}$ & 2.23   &&   &$2^{+}_{5}$ & 2.02   &\\
& $2^{+}_{6}$ & 3.51   &&   &$2^{+}_{6}$ & 2.29   &&   &$2^{+}_{6}$ & 2.07   &\\
&             &        &&   &            &        &&   &$2^{+}_{6}$ & 2.15   &\\

\hline
%   4+ states
& $4^{+}_{1}$ & 1.11   &&   &$4^{+}_{1}$ & 1.15   &&   &$4^{+}_{1}$ & 1.34   &\\
& $4^{+}_{2}$ & 1.94   &&   &$4^{+}_{2}$ & 1.32   &&   &$4^{+}_{2}$ & 1.50   &\\
& $4^{+}_{3}$ & 2.67   &&   &$4^{+}_{3}$ & 1.76   &&   &$4^{+}_{3}$ & 1.80   &\\
& $4^{+}_{4}$ & 2.88   &&   &$4^{+}_{4}$ & 1.98   &&   &$4^{+}_{4}$ & 1.87   &\\
& $4^{+}_{5}$ & 3.25   &&   &$4^{+}_{5}$ & 2.07   &&   &$4^{+}_{5}$ & 2.02   &\\
& $4^{+}_{6}$ & 3.74   &&   &$4^{+}_{6}$ & 2.37   &&   &$4^{+}_{6}$ & 2.07   &\\
&             &        &&   &            &        &&   &$4^{+}_{6}$ & 2.15   &\\

\hline
%   6+ states
& $6^{+}_{1}$ & 1.25   &&   &$6^{+}_{1}$ & 1.37   &&   &$6^{+}_{1}$ & 1.52   &\\
& $6^{+}_{2}$ & 2.61   &&   &$6^{+}_{2}$ & 2.15   &&   &$6^{+}_{2}$ & 2.12   &\\
& $6^{+}_{3}$ & 2.97   &&   &$6^{+}_{3}$ & 2.37   &&   &            &        &\\
& $6^{+}_{4}$ & 3.63   &&   &            &        &&   &            &        &\\

\hline
%   8+ states
& $8^{+}_{1}$ & 2.46   &&   &$8^{+}_{1}$ & 2.11   &&   &            &        &\\

\hline
%   3- states
& $3^{-}_{1}$ & 3.56   &&   &$3^{-}_{1}$ & 3.49   &&   &$3^{-}_{1}$ &  3.42  &\\

%   5- states
& $5^{-}_{1}$ & 3.91   &&   &$5^{-}_{1}$ & 3.76   &&   &$5^{-}_{1}$ &  3.53  &\\

%   7- states
& $7^{-}_{1}$ & 4.00   &&   &$7^{-}_{1}$ & 3.85   &&   &$9^{-}_{1}$ &  3.68  &\\

%   9- states
& $9^{-}_{1}$ & 4.04   &&   &$9^{-}_{1}$ & 3.89   &&   &$7^{-}_{1}$ &  3.68  &\\

%&&&&&&&&&&&\\
\hline
\end{tabular}
\end{center}
\end{table}

\begin{table}[htbp]
\caption{Low--lying states for  $^{135}$Sn and $^{137}$Sn.}
\label{tab:odd_isotopes}
\begin{center}
\begin{tabular}{|cccc|cccc|}
\hline
\multicolumn{4}{|c|}{ $^{135}$Sn} & \multicolumn{4}{|c|}{ $^{137}$Sn}\\
&{$J^{\pi}_i$}&  Theory&&    &{$J^{\pi}_i$}&   Theory& \\

\hline

&  7/2$_1^-$  &  0.00  &&    &  7/2$_1^-$  &   0.00&    \\
&  5/2$_1^-$  &  0.30  &&    &  5/2$_1^-$  &   0.29&    \\
&  3/2$_1^-$  &  0.41  &&    &  3/2$_1^-$  &   0.37&    \\
&  3/2$_2^-$  &  0.64  &&    &  3/2$_2^-$  &   0.60&    \\
& 11/2$_1^-$  &  0.74  &&    &  1/2$_1^-$  &   0.72&    \\
&  9/2$_1^-$  &  0.86  &&    & 11/2$_1^-$  &   0.75&    \\
& 15/2$_1^-$  &  1.09  &&    &  9/2$_1^-$  &   0.77&    \\
&  9/2$_2^-$  &  1.16  &&    &  5/2$_2^-$  &   0.81&    \\
&  9/2$_3^-$  &  1.19  &&    &  9/2$_2^-$  &   0.88&    \\
&  1/2$_1^-$  &  1.21  &&    &  7/2$_2^-$  &   0.95&    \\
&  7/2$_2^-$  &  1.27  &&    &  5/2$_3^-$  &   1.04&    \\
&  5/2$_2^-$  &  1.37  &&    &  1/2$_2^-$  &   1.05&    \\
&  3/2$_3^-$  &  1.40  &&    &  9/2$_3^-$  &   1.09&    \\
&  7/2$_3^-$  &  1.40  &&    &  3/2$_3^-$  &   1.10&    \\
&  5/2$_3^-$  &  1.52  &&    &  7/2$_3^-$  &   1.11&    \\
&  5/2$_4^-$  &  1.59  &&    & 15/2$_1^-$  &   1.26&    \\
& 13/2$_1^-$  &  1.70  &&    &  5/2$_4^-$  &   1.29&    \\
&  1/2$_2^-$  &  1.72  &&    &  7/2$_4^-$  &   1.30&    \\
& 11/2$_2^-$  &  1.73  &&    &  9/2$_4^-$  &   1.37&    \\
& 11/2$_3^-$  &  1.77  &&    & 13/2$_1^-$  &   1.38&    \\

\hline
\end{tabular}
\end{center}
\end{table}

\subsection{\it Light antimony isotopes}

The nucleus $^{105}$Sb was produced in the reaction $^{50}$Cr($^{58}$Ni,1p2n)
at a beam energy of 225 MeV with a 2.1 mg/cm$^2$ thick target.
The experiment was performed with the GAMMASPHERE Ge-detector array \cite{gs}
at the ATLAS accelerator
at Argonne National Laboratory. The experimental setup consisted of
78 Ge detectors, 95 CsI scintillators known as 
Microball \cite{uball} for light charged particle detection, and the newly
developed Neutron Shell. 
   \begin{figure}
   \setlength{\unitlength}{1mm}
   \begin{picture}(150,160)
   \put(0,0){\epsfxsize=16cm \epsfbox{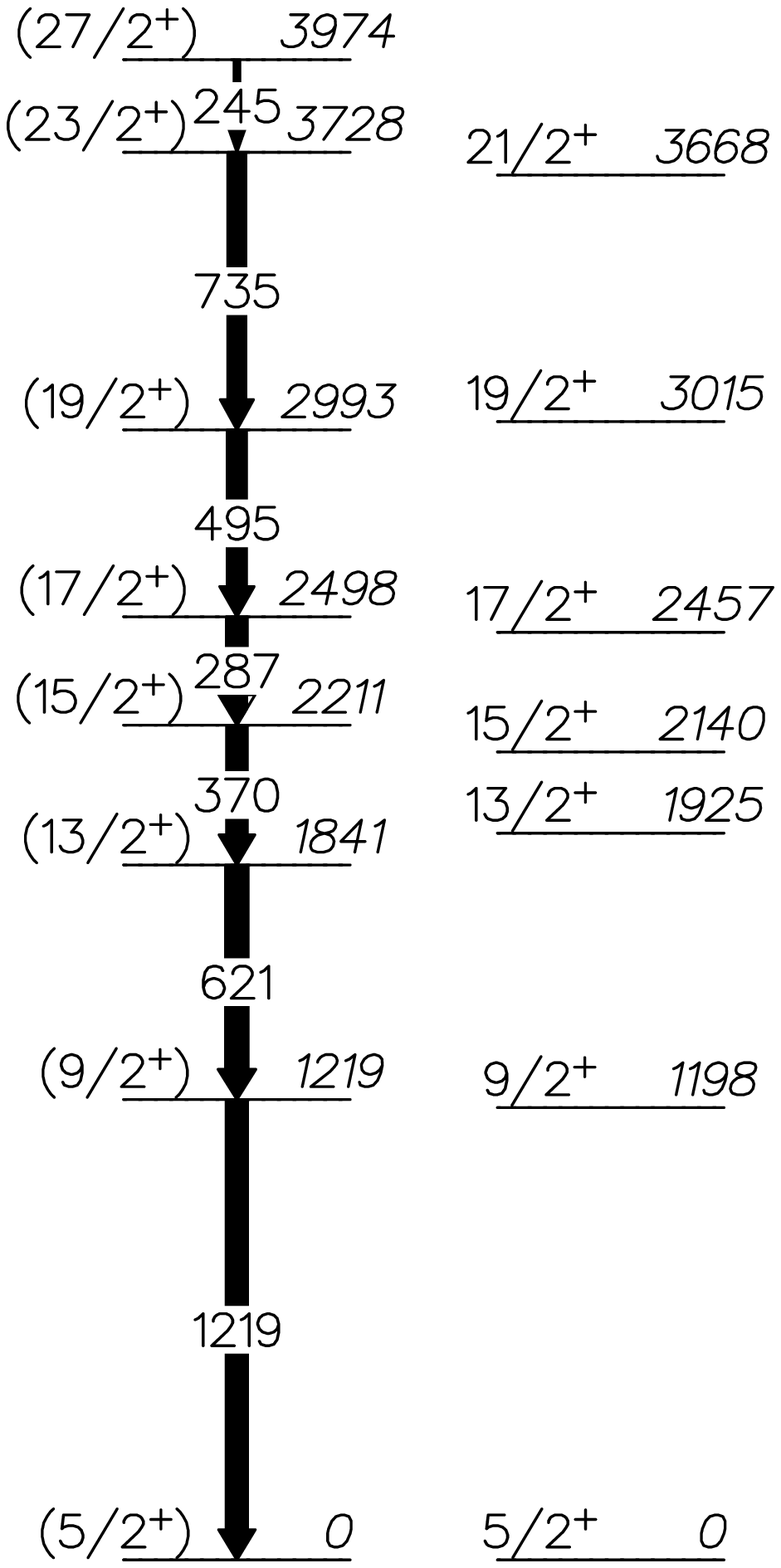}}
   \end{picture}
\caption{Proposed level scheme for $^{105}$Sb. Shell model
calculations are shown on the right hand side. The widths of
the arrows are proportional to the intensity of the transitions.\label{fig:sb105fig}}
   \end{figure}

The calculation favors a $J^{\pi}$=5/2$^+$ assignment for the ground state in
agreement with the suggestion from the proton decay data.
In this state the valence proton is mainly in the d$_{5/2}$
orbit and the two neutron pairs are almost evenly distributed
over the d$_{5/2}$ and g$_{7/2}$ neutron orbits. The situation
is very similar in the 9/2$^+$ and 13/2$^+$ states, while 
the $\nu$d$_{5/2}^3$g$_{7/2}^1$ configuration exhausts the
largest parts of the wave functions of the 15/2$^+$ and 17/2$^+$ 
states. The neutron part of the wave function of the 19/2$^+$
state is almost identical to the 17/2$^+$ state. However,
since 17/2$^+$ is the maximum spin for the 
$\pi$d$_{5/2}^1$$\nu$d$_{5/2}^1$g$_{7/2}^1$ configuration,
the odd proton resides almost exclusively in the g$_{7/2}$
orbit in the 19/2$^+$ state. The 19/2$^+$ state is therefore 
the lowest lying of the observed states, in which the $\pi$g$_{7/2}$
orbit has a significant contribution to the wave function.
This pattern repeats itself for the states with spins 21/2$^+$,
23/2$^+$, 25/2$^+$ and 27/2$^+$ in an alternating fashion with 
the state 21/2$^+$ having a proton in the single-particle orbit 
$\pi$d$_{5/2}$.
For proton degrees of freedom
the s$_{1/2}$, d$_{3/2}$ and h$_{11/2}$ single-particle 
orbits give essentially negligible
contributions to the wave functions and the energies of the excited
states, as expected. For neutrons, although
the single-particle distribution for a given state is also negligible,
these orbits are important for a good describtion of the energy spectrum,
as also demonstrated in large-scale shell-model calculations of
tin isotopes \cite{ehho98}. The effective interaction employed here
is the same as that employed in Ref.~\cite{ehho98}, with an in general
good reproduction of the data. 

We note here as well that
the agreement with the experimentally 
proposed spin assignements is very good.
The reason for such a good
agreement is most likely that the wave functions of the 
states are dominated by neutronic degrees of freedom.
The unbound proton is 
only a spectator
while the well bound neutrons change orbits and alignment
in transitions from high spin states.

The recently described spectra of $^{107}$Sb show also
low-lying $7/2^+$ and $11/2^+$ states, at approximately the 
same excitation energies as those shown in the theoretical calculation 
here. The $\gamma$-ray transitions feeding these two states in 
$^{105}$Sb are as in $^{107}$Sb expected to be much weaker than 
the main $\gamma$-ray cascade, which explains why these two states
were not identified in the experiment. They are interesting since
their wave functions contain mainly contributions from the $\pi$g$_{7/2}$
orbit that is poorly known near $^{100}$Sn, but
a more sensitive experiment
is needed for their identification.

\subsection{\it Neutron-rich nuclei in the $1s0d$-$1p0f$ shells}\label{subsec:sdpfshells}

Studies of extremely neutron-rich nuclei have revealed a number of
intriguing new phenomena.  Two sets of these nuclei that have received
particular attention are those with neutron number $N$ in the vicinity
of the $1s0d$ and $0f_{7/2}$ shell closures ($N \approx 20$ and $N
\approx 28$).  Experimental studies of neutron-rich Mg and Na isotopes
indicate the onset of deformation, as well as the modification of the
$N = 20$ shell gap for $^{32}$Mg and nearby nuclei \cite{r:motobayashi}.
Inspired by the rich set of phenomena occurring near the $N = 20$
shell closure when $N \gg Z$, attention has been directed to nuclei
near the $N = 28$ (sub)shell closure for a number of S and Ar isotopes
\cite{r:brown1,r:brown2} where similar, but less dramatic, effects
have been seen as well.
\begin{table}[hbtp]
\begin{center}
\caption{The computed and measured values of $B(E2)$ for
the nuclei in this study using $e_p=1.5$ and $e_n=0.5$.
}
\begin{tabular}{|cccc|}\hline
 & $B(E2; 0^+_{gs} \rightarrow 2^+_1)_{Expt}$ & $B(E2, total)_{SMMC}$ &
  $B(E2; 0^+_{gs} \rightarrow 2^+_1)$  \\\hline
 $^{22}$Mg & $458 \pm 183$ & $334 \pm 27 $
    & \\
 $^{30}$Ne & & $303 \pm 32$
    & 342 \cite{r:fukunishi},171 \cite{r:poves2}  \\
 $^{32}$Mg & $454 \pm 78$ \cite{r:motobayashi} & $494 \pm 44 $
   & 448 \cite{r:fukunishi},205 \cite{r:poves2} \\
 $^{36}$Ar & $296.56 \pm 28.3$ \cite{r:ensdf} & $174 \pm 48$
    & \\
 $^{40}$S & $334 \pm 36$ \cite{r:brown1} & $270 \pm 66$
    & 398 \cite{r:brown2},390 \cite{r:retamosa}  \\
 $^{42}$S & $397 \pm 63$ \cite{r:brown1} & $194 \pm 64$
   & 372 \cite{r:brown2},465 \cite{r:retamosa}  \\
 $^{42}$Si &  &  $445 \pm 62$
    & 260 \cite{r:retamosa}  \\
 $^{44}$S & $314 \pm 88$ \cite{r:brown2} & $274 \pm 68$
    & 271 \cite{r:brown2},390 \cite{r:retamosa}  \\
 $^{44}$Ti & $610 \pm 150$ \cite{r:raman} & $692 \pm 63$
    &  \\
 $^{46}$Ar & $196 \pm 39$ \cite{r:brown1} & $369 \pm 77 $
    & 460 \cite{r:brown1},455 \cite{r:retamosa}  \\\hline
\end{tabular}
\end{center}
\label{t:tab1}
\end{table}
In parallel with the experimental efforts, there have been several
theoretical studies seeking to understand and, in some cases, predict
properties of these unstable nuclei.  Both mean-field
\cite{r:werner,r:campi} and shell-model calculations
\cite{r:brown1,r:brown2,r:wbmb,r:poves1,r:fukunishi,r:retamosa,r:caurier}
have been proposed. The latter require a severe truncation to
achieve tractable model spaces, since the successful description of these
nuclei involves active nucleons in both the $1s0d$- and the $1p0f$-shells.
The natural basis for the problem is therefore the full $1s0d$-$1p0f$
space, which puts it out of reach of exact diagonalization on current
hardware\footnote{For a treatment of the CoM problem, see Ref.\
\cite{drhklz99}.}.

Shell-Model Monte Carlo (SMMC) methods
\cite{r:smmc_pr,r:smmc_ar,r:lang} offer an alternative to direct
diagonalization when the bases become very large. Though SMMC provides
limited detailed spectroscopic information, it can predict, with good
accuracy, overall nuclear properties such as masses, total strengths,
strength distributions, and deformation, precisely those quantities
probed by the recent experiments.

There is limited experimental information about the highly unstable,
neutron-rich nuclei under consideration.  In many cases only the mass,
excitation energy of the first excited state, the $B(E2)$ to that state,
and the $\beta$-decay rate is known, and not even all of this
information is available in some cases.  From the
measured $B(E2)$, an estimate of the nuclear deformation parameter,
$\beta_2$, has been obtained via the usual relation
\begin{equation}
\beta_2 = 4 \pi \sqrt{B(E2; 0^+_{gs} \rightarrow 2^+_1)}/3 Z R_0^2 e
\end{equation}
with $R_0 = 1.2 A^{1/3}$ fm and $B(E2)$ given in $e^2$fm$^4$.

Much of the interest in the region stems from the unexpectedly large
values of the deduced $\beta_2$, results which suggest the onset of
deformation and have led to speculations about the vanishing of the $N
= 20$ and $N = 28$ shell gaps.  The lowering in energy of the 2$^+_1$
state supports this interpretation.  The most thoroughly studied case,
and the one which most convincingly demonstrates these phenomena, is
$^{32}$Mg with its extremely large $B(E2) = 454 \pm 78 \, e^2$fm$^4$ and
corresponding $\beta_2 = 0.513$ \cite{r:motobayashi}; however, a word of
caution is necessary when deciding on the basis of this
limited information that we are in the presence of well-deformed
rotors: for $^{22}$Mg, we would obtain $\beta_2 = 0.67$, even more
spectacular, and for $^{12}$C, $\beta_2 = 0.8$, well above the
superdeformed bands.

Most of the measured observables can be calculated within the SMMC
framework.  It is well known that in {\it deformed} nuclei the total
$B(E2)$ strength is almost saturated by the $0^+_{gs} \rightarrow
2_1^+$ transition (typically 80\% to 90\% of the strength lies in this
transition).
Thus the total strength calculated by SMMC should only
slightly overestimate the strength of the measured transition.  In
Table 1 the SMMC computed values of $B(E2, total)$ are
compared both to the experimental $B(E2; 0^+_{gs} \rightarrow 2^+_1)$
values and to the values found in various truncated shell-model
calculations.  Reasonable agreement with experimental data across the
space is obtained when one chooses effective charges of $e_p=1.5$ and
$e_n=0.5$.
All of the theoretical
calculations require excitations to the $1p0f$-shell before reasonable
values can be obtained.  We note a general agreement among all
calculations of the $B(E2)$ for $^{46}$Ar, although they are typically
larger than experimental data would suggest. We also note a somewhat
lower value of the $B(E2)$ in this calculation as compared to
experiment and other theoretical calculations in the case of $^{42}$S.
Table 2 gives selected occupation numbers for the nuclei
considered.  We first note a difficulty in extrapolating some of the
occupations where the number of particles is nearly zero.  This leads
to a systematic error bar that we estimate at $\pm 0.2$ for all
occupations shown, while the statistical error bar is quoted in the
table. The extrapolations for occupation numbers were principally
linear. Table 2 shows that $^{22}$Mg remains as an almost
pure $sd$-shell nucleus, as expected.  We also see that the protons in
$^{30}$Ne, $^{32}$Mg, and $^{42}$Si are almost entirely confined to the
$sd$ shell.  This latter is a pleasing result in at least two regards.
First, it shows that the interaction does not mix the two shells to an
unrealistically large extent.  Second, if spurious CoM  contamination
were a severe problem, we would expect to see a larger proton
$0f_{7/2}$ population for these nuclei due to the $0d_{5/2}$-$0f_{7/2}$
``transition'' mediated by the center-of-mass creation operator.  The fact that
there is little proton $f_{7/2}$ occupation for these nuclei confirms
that the CoM contamination is under reasonable control.
See Ref.\ \cite{drhklz99} for further details.
\begin{table}[hbtp]
\begin{center}
\caption{The calculated SMMC neutron and
proton occupation numbers for the $sd$ shell, the
$0f_{7/2}$ sub-shell, and the remaining orbitals of the
$pf$ shell.  The statistical errors are given for linear
extrapolations. A systematic error of $\pm 0.2$ should also
be included. The first row represents neutron results, while the
second row represents protons.}
\begin{tabular}{|ccccc|}\hline
 & $N,Z$ & $1s0d$ & $0f_{7/2}$ & $1p0f_{5/2}$\\ \hline
$^{22}$Mg & 10,12 & $3.93 \pm 0.02$ & $0.1 \pm  0.02$ &
  $-0.05 \pm 0.01$ \\ &&  $2.04 \pm 0.02$ & $0.00 \pm 0.01$ &
  $-0.05 \pm 0.01$ \\
$^{30}$Ne & 20,10 & $9.95 \pm 0.03$ & $2.32 \pm 0.03$ &
  $-0.26 \pm 0.02$ \\ && $2.03 \pm 0.02$ & $-0.01 \pm 0.01$ &
  $-0.02 \pm 0.01$ \\
$^{32}$Mg & 20,12 & $9.84 \pm 0.03$ & $ 2.37 \pm 0.03$ &
  $-0.21 \pm 0.02$ \\ && $3.99 \pm 0.03$ & $0.05 \pm 0.02$ &
  $-0.05 \pm 0.01$ \\
$^{36}$Ar & 18,18 & $9.07 \pm 0.03$ & $1.08 \pm 0.02$ &
  $-0.15 \pm 0.02$ \\ && $9.07 \pm 0.03$ & $1.08 \pm 0.02$ &
  $-0.15 \pm 0.02$ \\
$^{40}$S & 24,16 & $11.00 \pm 0.03$ & $ 5.00 \pm 0.03 $ &
  $-0.01\pm 0.02$ \\ && $7.57 \pm 0.04$ & $0.54 \pm 0.02$ &
  $-0.12 \pm 0.02$ \\
$^{42}$Si & 28,14 & $11.77 \pm 0.02$ & $7.34 \pm 0.02$ &
  $0.90 \pm 0.03$ \\ && $5.79 \pm 0.03$ & $0.25 \pm 0.02$ &
  $-0.07 \pm 0.01$ \\
$^{42}$S & 26,16 & $11.41 \pm 0.02$ & $6.33 \pm 0.02$ &
  $0.25 \pm 0.03$ \\ && $7.49 \pm 0.03$ & $0.58 \pm 0.02$ &
  $-0.09 \pm 0.02$ \\
$^{44}$S & 28,16 & $11.74 \pm 0.02$ & $7.18 \pm 0.02$ &
  $1.06 \pm 0.03$ \\ && $7.54 \pm 0.03$ & $0.56 \pm 0.02$ &
  $-0.12 \pm 0.02$ \\
$^{44}$Ti & 22,22 & $10.42 \pm 0.03$ & $3.58 \pm 0.02$ &
  $0.00 \pm 0.02$\\ & & $10.42 \pm 0.03$ & $3.58 \pm 0.02$ &
  $0.00 \pm 0.02$ \\
$^{46}$Ar & 28,18 & $11.64 \pm 0.02$ & $7.13 \pm 0.02$ &
  $1.23 \pm 0.03$ \\ && $8.74 \pm 0.03$ & $1.34 \pm 0.02$ &
  $-0.08 \pm 0.02$ \\\hline
\end{tabular}
\end{center}
\label{t:tab3}
\end{table}
An interesting feature of Table 2 lies in the neutron
occupations of the $N = 20$ nuclei ($^{30}$Ne and $^{32}$Mg) and the $N =
28$ nuclei ($^{42}$Si, $^{44}$S, and $^{46}$Ar).  The neutron
occupations of the two $N = 20$ nuclei are quite similar, confirming
the finding of Fukunishi {\it et al.} \cite{r:fukunishi} and Poves and
Retamosa \cite{r:poves1} that the $N= 20$ shell gap is modified.  In
fact, the neutron $0f_{7/2}$ orbital contains approximately two
particles before the $N=20$ closure, thus behaving like an intruder
single-particle state.  Furthermore, we see that 2p-2h excitations
dominate although higher excitations also play some role.  We also see
that the neutrons occupying the $1p0f$-shell in $N=20$ systems are
principally confined to the $0f_{7/2}$ sub-shell.
The conclusions that follow from looking at nuclei with $N > 20$,
particularly those with $N = 28$, are that the $N = 20$ shell is nearly
completely closed at this point, and that the $N=28$ closure shell is
reasonably robust, although approximately one neutron occupies the upper
part of the $1p0f$ shell. Coupling of the protons with the low-lying
neutron excitations probably accounts for the relatively large
$B(E2)$, without the need of invoking rotational behavior.
\begin{table}[hbtp]
\begin{center}
\caption{The calculated total Gamow-Teller strength, $GT^-$,
from this study.  The results of other studies, when
available, are presented for comparison.
}
\begin{tabular}{|ccc|}\hline
 Nucleus & SMMC & Other \\
\hline
 $^{22}$Mg & $0.578 \pm  0.06$  & \\
 $^{30}$Ne & $29.41 \pm 0.25$ & \\
 $^{32}$Mg & $24.00 \pm 0.34$ & \\
 $^{36}$Ar & $2.13 \pm  0.61$ & \\
 $^{40}$S  & $22.19 \pm 0.44$ & 22.87 \cite{r:retamosa} \\
 $^{42}$S  & $28.13 \pm 0.42$ & 28.89 \cite{r:retamosa} \\
 $^{42}$Si & $40.61 \pm 0.34$ & \\
 $^{44}$S  & $34.59 \pm 0.39$ & 34.93 \cite{r:retamosa} \\
 $^{44}$Ti & $4.64 \pm  0.66$ & \\
 $^{46}$Ar & $29.07 \pm 0.44$ & 28.84 \cite{r:retamosa} \\\hline
\end{tabular}
\end{center}
\label{t:tab4}
\end{table}
In Table 3 we show the SMMC total Gamow-Teller (GT$^-$)
strength.  We compare our results to those of previous truncated
calculations, where available.  In all cases, our results are slightly
smaller than, but in good accord with, other calculations.  Since we
do not calculate the strength function, we do not compute
$\beta$-decay lifetimes.

\section{Non-perturbative resummations: parquet diagrams}
\label{sec:sec5}

\subsection{\it Inclusion of hole-hole contributions and single-particle propagators}

With the $G$-matrix defined according
to the double-partitioned scheme we can easily solve Eq.\ (\ref{eq:first12})
through matrix inversion. The number of hole-hole and particle-particle
configurations is then rather small, typically smaller than
$\sim  100$, and a matrix inversion
is then rather trivial.
Before we discuss the solution of Eq.\ (\ref{eq:first12}), it is always
instructive to consider the contributions to second order in
perturbation theory, i.e., diagrams (a) and (b) of Fig.\
\ref{fig:gamma12}. The external legs can be
particle states or hole states. Diagram (a) reads
\begin{equation}
      (a)=\frac{1}{2}\sum_{pq}V^{[12]}_{12pq J}
      \frac{1}{s-\varepsilon_p-
                \varepsilon_q} V^{[12]}_{pq34 J},
      \label{eq:secondg}
\end{equation}
and
\begin{equation}
      (b)=\frac{1}{2}\sum_{\alpha\beta}V^{[12]}_{12\alpha\beta J}
      \frac{1}{-s+\varepsilon_{\alpha}+
                \varepsilon_{\beta}} V^{[12]}_{\alpha\beta 34 J}.
\end{equation}
We note here the minus sign in the energy denominator,
since in the latter expression we are using
the hole-hole term of the propagator of
Eq.\ (\ref{eq:paulioperator12}).
If we use a double-partitioned $G$-matrix for say $^{16}$O and
are interested in an effective valence space interaction
for the $1s0d$-shell\footnote{This means that the labels $1234$ will
refer to particle states in the $1s0d$-shell.}, then typically
the single-particle orbits of the intermediate states will be represented
by states in the $1p0f$ major shell. Hole states are then defined
by single-particle states in the $0s$ and $0p$ shells.
Clearly, the number of two-body intermediate states is rather limited.
To third order we have diagrams like (c) and (d) of Fig.\
\ref{fig:gamma12}. Diagram (c) is just the third-order equivalent of
Eq.\ (\ref{eq:secondg}) and reads
\begin{equation}
      (c)=\frac{1}{4}\sum_{pqrw}V^{[12]}_{12pq J}
      \frac{1}{s-\varepsilon_p-
                \varepsilon_q}
      V^{[12]}_{pqrw J}
      \frac{1}{s-\varepsilon_r-
                \varepsilon_w}
       V^{[12]}_{rw34 J},
      \label{eq:thirdg}
\end{equation}
while diagram (d) contains both a two-particle and a two-hole
intermediate state and reads
\begin{equation}
      (d)=\frac{1}{4}\sum_{\alpha\beta pq}V^{[12]}_{12pq J}
      \frac{1}{s-\varepsilon_p-
                \varepsilon_q+\varepsilon_{\alpha}+
                \varepsilon_{\beta}}
      V^{[12]}_{pq\alpha\beta J}
      \frac{1}{s-\varepsilon_p-
                \varepsilon_q}
       V^{[12]}_{\alpha\beta 34 J}.
      \label{eq:thirdg2h}
\end{equation}
Thus, solving Eq.\ (\ref{eq:first12}) will then
yield contributions to the effective interaction
such as the above expressions.

Here we have also tacitly assumed that the energy denominators
do not diverge, i.e., we have chosen an energy $s$
so that we avoid the poles. This has always been the standard
approach in calculations of shell-model
effective interactions. To give an example, consider
now diagram (b) and suppose that we are using harmonic
oscillator wave functions. Let us also assume that the two
hole states are from the $0p$-shell and that the valence
particles are in the $1s0d$-shell. If we rescale the
energies of the valence space to zero, then the two-hole
state would yield $-28$ MeV with an oscillator parameter
$b=1.72$ fm. If $s=-28$, the denominator diverges.
In this case it is rather easy to obtain the imaginary
part, and even if we were to chose $s$ different
from $-28$ MeV, the imaginary part will influence the real
part of the effective interaction through dispersion
relations, see e.g., Refs.\ \cite{angels88,rpd89,ms92}.
It is therefore at best just a first approximation
to neglect the imaginary term.
Moreover, if we solve Dyson's equation for the
self-energy, the single-particle energies may contain
an imaginary part.
Technically it is however not difficult to deal
with imaginary contributions, one needs to
invert a complex  matrix rather than a real one.
These technicalities
will however be described elsewhere \cite{mhj99}.

Using the double-partitioned $G$-matrix, we can then rewrite
Eq.\ (\ref{eq:first12}) as
\begin{equation}
      \Gamma^{[12]}_{1234J}(s) =
      G^{[12]}_{1234J}+\frac{1}{2}
      \sum_{56}
      G^{[12]}_{1256J}\hat{{\cal G}}^{[12]}
      \Gamma^{[12]}_{5634J}(s),
      \label{eq:newfirst12}
\end{equation}
where $G^{[12]}$ is just the double-partitioned $G$-matrix
discussed above. It is also energy dependent,
in contrast to $V$.
In case we were to employ this equation for effective
interactions in the $1s0d$-shell, the intermediate two-particle
states would then come from just e.g., the $1p0f$-shell.
This equation, which now is solved within a much smaller space
than the original one spanned by the total $Q_{pp}$, allows
clearly for computationally amenable solutions. It corresponds
to the so-called {\em model-space approach} to the solution
of the Feynman-Galitskii equations.
Thus, a possible approach would consist
of the following steps
\begin{enumerate}
\item Solve the $G$-matrix equation from Eq.\ (\ref{eq:gmod})
      using the double-partitioning scheme.
\item The next step is then to solve Eq.\ (\ref{eq:first12})
      and Dyson's equation for the self-energy.
\item This scheme is iterated till self-consistency is achieved,
      see the discussion below.
\end{enumerate}

We will however not employ this {\em model-space} scheme in our
actual calculations. There are several reasons for not doing so.

Let us first assume that we omit the $[13]$ and $[14]$ channels in our
iterative scheme for Eq.\ (\ref{eq:newfirst12}).
The next iteration of Eq.\ (\ref{eq:newfirst12})
would then look like
\begin{equation}
      \Gamma^{[12]}_{(1)} =
      \Gamma_{(0)}^{[12]}+
      \Gamma_{(0)}^{[12]}
       \hat{{\cal G}}^{[12]}\Gamma^{[12]}_{(1)},
      \label{eq:second12}
\end{equation}
where the vertex function $\Gamma_{(0)}^{[12]}$ is the solution
of Eq.\ (\ref{eq:newfirst12}). However, we cannot define
the ``bare'' vertex $\Gamma_{(0)}^{[12]}$ to be the
solution  of Eq.\ (\ref{eq:newfirst12}) simply because then we
would be double-counting contributions.
Thus, $\Gamma_{(0)}^{[12]}$ has to equal the $G$-matrix.
The only change in Eq.\ (\ref{eq:second12})
arises from the solution of Dyson's equation
and thereby new single-particle energies.

Let us then for the sake of simplicity assume that the
single-particle energies are just the Hartree-Fock
solutions. The problem we are aiming at arises at the
Hartree-Fock level.
In order to obtain Hartree-Fock solutions which
are independent of the chosen harmonic oscillator
parameter $b$, we typically need to include single-particle
orbits from quite many major shells.
Typical constraints we have found when we do so-called
Brueckner-Hartree-Fock (BHF) calculations for finite nuclei is that
we need at least $2n+l \leq 10$ in order to obtain
a result which is independent of the chosen
$b$\footnote{Throughout this work our unperturbed single-particle
basis par excellence will always be that of the harmonic
oscillator.} value. The way we solve the
BHF equations is to expand the new single-particle
wave functions $\psi_{\lambda}$, with $\lambda$
representing the quantum numbers $nlj$,
in terms of harmonic oscillator wave functions,
i.e.,
\begin{equation}
     \left | \psi_{\lambda}\right\rangle=
     \sum_{\alpha =1}^{2n+l}
     C_{\alpha}^{(\lambda )}\left | \phi_{\alpha}\right\rangle
     \label{eq:selfconstbasis}
\end{equation}
where $\phi_{\alpha}$ are the harmonic oscillator wave functions
with quantum numbers $\alpha=nlj$ and $C$ are the coefficients
to be varied in the Hartree-Fock calculations.
The single-particle energies at the
Hartree-Fock level are just
\begin{equation}
  \varepsilon_{\alpha}=t_{\alpha}+
   \sum_h \bra{\alpha h} G(\varepsilon_{\alpha}+
                           \varepsilon_h)
          \ket{\alpha h},
\end{equation}
where the single-particle states are just those
of the harmonic oscillator. The $G$-matrix used in the
first iteration in the BHF calculation is the one given
by the solution of Eq.\ (\ref{eq:gmod}).
The coefficients
$C_{\alpha}$ can then be obtained by diagonalizing
a matrix of dimension $N\times N$, where $N$ is the number
of single-particle orbits with the same $lj$ values.

\subsection{\it Screening corrections and vertex renormalization, the equations
for the $[13]$ and  $[14]$ channels}
\label{subsec:sec4}

We start as in the previous section with the definition of the interaction
vertices in the $[13]$ and $[14]$ channels and the corresponding
integral equations. Thereafter, we discuss various approximations
to these equations such as the summation of TDA and RPA diagrams.
Eventually, the aim is to merge the discussion in this section and
the preceeding one into equations for a self-consistent scheme which combines
all three channels, namely the so-called set of parquet equations to
be discussed in section \ref{sec:sec5}.

\begin{figure}[hbtp]
\begin{center}
      \setlength{\unitlength}{1mm}
      \begin{picture}(100,60)
      \put(0,0){\epsfxsize=12cm \epsfbox{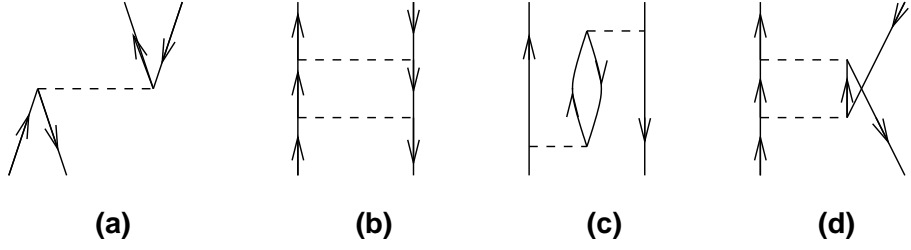}}
      \end{picture}
      \caption{Examples of irreducible and reducible diagrams in the
               $[13]$ and $[14]$ channels. See text for further details.}
      \label{fig:1314channel}
\end{center}
\end{figure}

The equations for the renormalized vertex in the $[13]$ and $[14]$
channels have the same form as Eq.\ (\ref{eq:schematic12}), namely
\begin{equation}
     \Gamma^{[13]}=V^{[13]}+V^{[13]}(gg)\Gamma^{[13]},
\end{equation}
and
\begin{equation}
     \Gamma^{[14]}=V^{[14]}+V^{[14]}(gg)\Gamma^{[14]}.
\end{equation}
The matrix elements which enter are however
defined differently and the irreducible
diagrams of $V^{[13]}$ and $V^{[14]}$ can obviously not be the same.
With irreducible in the $[13]$ channel we will mean a diagram, which by
cutting an internal particle-hole pair, cannot be separated into a piece
containing the external legs $1,3$ and another piece containing
$2,4$ as external legs. The definition for the irreducible vertex in the
$[14]$ channel is similar and we illustrate these differences in Fig.\
\ref{fig:1314channel}.
Diagram (a) is just the lowest-order
interaction in the $[13]$ channel and is therefore irreducible.
Diagram (b) is an irreducible diagram in the $[13]$ channel,
whereas it is reducible in the $[14]$ channel. Diagram (c) is in turn
irreducible in the $[14]$ channel and reducible in the $[13]$ channel.
Diagram (d) is an example of a diagram which is irreducible in both
channels. This diagram stems from the $[12]$ channel.

\begin{figure}[hbtp]
\begin{center}
      \setlength{\unitlength}{1mm}
      \begin{picture}(100,100)
      \put(0,0){\epsfxsize=10cm \epsfbox{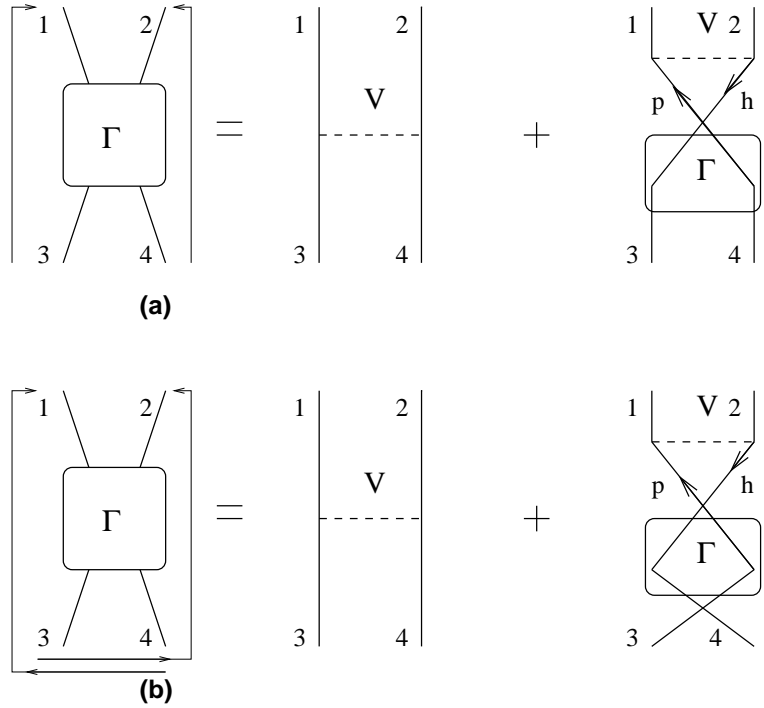}}
      \end{picture}
      \caption{(a) shows the structure of the integral equation for
               the interaction vertex in the $[13]$ channel. (b) represents
               the integral channel for the $[14]$ channel.
               The coupling order is displayed as well.}
      \label{fig:figs1314}
\end{center}
\end{figure}

The energy variables in these channels are, following
Fig.\ \ref{fig:channelsdef} and Eqs.\ (\ref{eq:13channel}) and
(\ref{eq:14channel}),
\begin{equation}
     t=\varepsilon_3-\varepsilon_1=\varepsilon_2-\varepsilon_4,
\end{equation}
for the $[13]$ channel and
\begin{equation}
     u=\varepsilon_1-\varepsilon_4=\varepsilon_3-\varepsilon_2,
\end{equation}
for the $[14]$ channel.
Defining the unperturbed particle-hole propagators
in the energy representation as 
\begin{equation}
    \hat{{\cal G}}^{[13]}=
    \frac{Q^{[13]}_{\mathrm{ph}}}{t-\varepsilon_p+\varepsilon_h+\imath \eta}
    -\frac{Q^{[13]}_{\mathrm{hp}}}{t+\varepsilon_p-\varepsilon_h-\imath \eta},
    \label{eq:paulioperator13}
\end{equation}
and
\begin{equation}
    \hat{{\cal G}}^{[14]}=
    \frac{Q^{[14]}_{\mathrm{ph}}}{u-\varepsilon_p+\varepsilon_h+\imath \eta}
    -\frac{Q^{[14]}_{\mathrm{hp}}}{u+\varepsilon_p-\varepsilon_h-\imath \eta}
    \label{eq:paulioperator14}
\end{equation}
we arrive at the following equations
for the interaction vertex in these two channels
\begin{equation}
      \Gamma^{[13]}_{1234J}(t) =
      V^{[13]}_{1234J}+
      \sum_{ph}
      V^{[13]}_{12phJ}\hat{{\cal G}}^{[13]}
      \Gamma^{[13]}_{ph34J}(t),
      \label{eq:first13}
\end{equation}
and
\begin{equation}
      \Gamma^{[14]}_{1234J}(u) =
      V^{[14]}_{1234J}-
      \sum_{ph}
      V^{[14]}_{12phJ}\hat{{\cal G}}^{[14]}
      \Gamma^{[14]}_{ph34J}(u).
      \label{eq:first14}
\end{equation}
These equations, together with Eq.\ (\ref{eq:first12}),
can then form the basis for the first iteration in a self-consistent
scheme for renormalization corrections of the parquet type.
The origin of the minus sign in Eq.\ (\ref{eq:first14}) follows from
the diagram rules \cite{kstop81} and will be examplified below.
A graphical view of these equations is given in Fig.\
\ref{fig:figs1314}.
The reader should also keep in mind the two contributions to the particle
propagators of Eqs.\ (\ref{eq:paulioperator13}) and (\ref{eq:paulioperator14}).

\subsection{\it Screened ph  and 2p2h interactions}

Here we study the screening of the particle-hole
and the 2p2h interactions given in Fig.\ \ref{fig:wavef1},
indicated by $V_{ph}$ and $V_{2p2h}$, respectively.
Before we list the final expression, it is however instructive
to consider the corrections to second order in the interaction $V$ to the ph
and 2p2h vertices.
\begin{figure}[hbtp]
\begin{center}
      \setlength{\unitlength}{1mm}
      \begin{picture}(100,60)
      \put(0,0){\epsfxsize=10cm \epsfbox{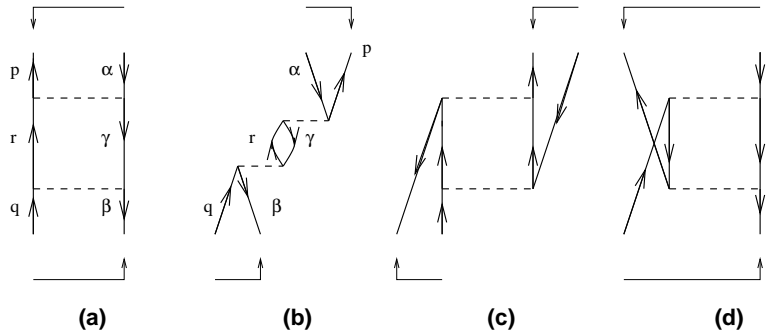}}
      \end{picture}
      \caption{Second-order perturbation theory corrections to the ph
               interaction vertex.}
      \label{fig:phvertex}
\end{center}
\end{figure}
In Fig.\ \ref{fig:phvertex} we display the second-order corrections to
the ph diagrams of Fig.\ \ref{fig:wavef1}.
Diagram (a) is the core-polarization correction term to
the particle-hole
interaction, and corresponds to a contribution from the
$[14]$ channel, as indicated by the coupling order.
The term labeled (b) corresponds to the exchange
term of (a) and is coupled in the $[13]$-order, see also
the discussion in connection with Eqs.\ (\ref{eq:ph13})
and (\ref{eq:ph14}).
The other corrections, like (c) and (d) include
particle-particle and hole-hole intermediate states, respectively.
They are irreducible in both the $[13]$ channel and the $[14]$ channel,
and can therefore enter the irreducible vertices of these two channels in
later iterations.
They are however not generated by various iterations of Eqs.\
(\ref{eq:first13})
and (\ref{eq:first14}). In fact, if we replace $V$ by $G$ in Eqs.\
(\ref{eq:first13})
and (\ref{eq:first14}), diagram (c) is already accounted for by the
$G$-matrix. It may however be included if the double-partitioned $G$-matrix
of the previous section is used.
Let us now look at the analytical expressions
in an angular momentum coupled basis for diagrams (a) and (b)
of Fig.\ \ref{fig:phvertex}. Here we just include the first term of
the propagators of Eqs.\ (\ref{eq:paulioperator13}) and
(\ref{eq:paulioperator14}). The second terms will give rise to the
2p2h contributions discussed below. In the following discussion we will
also assume that the interaction $V$ does not depend on the energy,
although
it is rather easy to generalize to an energy dependent
interaction.
Diagram (a) reads
\begin{equation}
      (a)=-\sum_{r\gamma}(-)^{j_r+j_{\gamma}-J}(-)^{2j_{\gamma}}
      V^{[14]}_{p\gamma r\alpha J}
      \frac{1}{u+\varepsilon_{\gamma}-
                \varepsilon_{r}} V^{[14]}_{r\beta q\gamma J},
       \label{eq:secordph}
\end{equation}
The factor $(-)^{2j_{\gamma}}$ stems from the opening up
and recoupling of an internal particle-hole pair \cite{kstop81}
and the phase $(-)^{j_r+j_{\gamma}-J}$ is needed in order
to rewrite the matrix elements in the coupling order of
Eq.\ (\ref{eq:13channel}).
The general structure of Eq.\ (\ref{eq:secordph}) is
just of the form $-V_{\mathrm{ph}}^{[14]}Q^{[14]}_{\mathrm{ph}}/\epsilon^{[14]}V_{\mathrm{ph}}^{[14]}$, with
$ \epsilon^{[14]}=\varepsilon_{q}+\varepsilon_{\gamma}-\varepsilon_{\beta}-
                \varepsilon_{r}=u+\varepsilon_{\gamma}-\varepsilon_{r}$
and we have defined
\begin{equation}
  u=\varepsilon_{q}-\varepsilon_{\beta}=\varepsilon_{p}-\varepsilon_{\alpha},
\end{equation}
for the on-shell energy case.
This is the equivalent of the energy variable of Eq.\ (\ref{eq:energy12}) in
the $[12]$ channel.
Diagram (b) is in turn given by
\begin{equation}
      (b)=\sum_{r\gamma}(-)^{j_r+j_{\gamma}-J}(-)^{2j_{\gamma}}
      V^{[13]}_{\gamma pr\alpha J}
      \frac{1}{\epsilon^{[13]}}V^{[13]}_{\beta rq\gamma J} ,
       \label{eq:secordphdirect}
\end{equation}
and we note that the contributions are clearly different.
The minus sign in Eq.\ (\ref{eq:secordph}) stems from the standard
diagram rules \cite{kstop81}.
In our use of the diagram rules below, we will omit the use
of the rule for the number of external valence hole lines.
In our case then, as can also be deduced from inspection of
Fig.\ \ref{fig:phvertex}, diagram (a) has  zero closed loops and
three hole lines,
giving thereby rise to a minus sign.
Diagram (b) has an additional closed
loop and thereby yielding the plus sign.
The energy denominator is in this case
\begin{equation}
      \epsilon^{[13]}=t+\varepsilon_{\gamma}-\varepsilon_{r},
\end{equation}
with
\begin{equation}
  t=\varepsilon_{q}-\varepsilon_{\beta}=\varepsilon_{p}-\varepsilon_{\alpha}.
\end{equation}
We notice, using the relations discussed
in Eqs.\ (\ref{eq:ph13})
and (\ref{eq:ph14}), that diagram (a) is simply the
exchange diagram of (b). We need however to include
both diagrams in order to obtain an
antisymmetric equation for the particle-hole channels
which exhibits the same properties as the $[12]$ channel
shown in Eq.\ (\ref{eq:symproperties}). {\em This is actually crucial
in solving the parquet equations. We wish namely that
every iteration, with a given approximation to the
vertex function $V$, preserves the antisymmetry property.}
This point cannot be emphasized enough. Let us now see what
happens to third order in the interaction.
\begin{figure}[hbtp]
\begin{center}
      \setlength{\unitlength}{1mm}
      \begin{picture}(100,60)
      \put(0,0){\epsfxsize=10cm \epsfbox{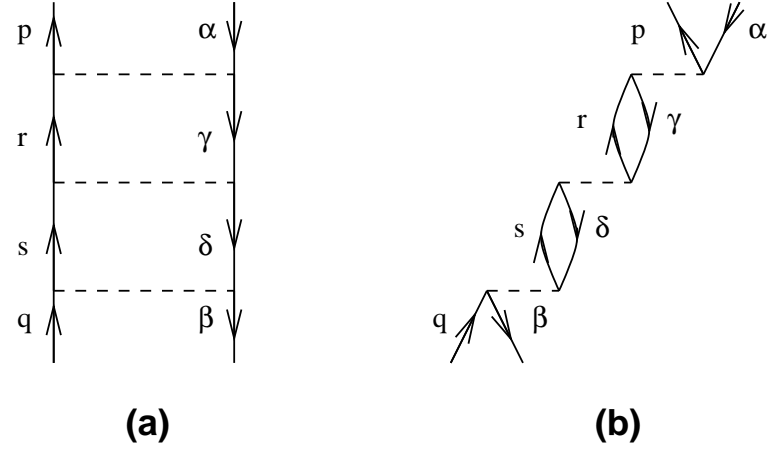}}
      \end{picture}
      \caption{Corrections beyond second order in the interaction $V$
               to the ph
               interaction vertex. (a) is in the $[14]$ channel and (b) is in
               $[13]$ channel.}
      \label{fig:phhigher}
\end{center}
\end{figure}
Third order corrections to the ph vertices (a) and (b)
 involving only ph intermediate states are
shown in (a) and (b) of Fig.\ \ref{fig:phhigher}, respectively.
The analytical expression for the third-order contribution (a) is given by
\begin{equation}
      (a)=\sum_{rs\gamma\delta}f
      V^{[14]}_{p\gamma r\alpha J}
      \frac{1}{u+\varepsilon_{\gamma}
       -\varepsilon_{r} } V^{[14]}_{r\delta s\gamma J}
      \frac{1}{u+\varepsilon_{\delta}
       -\varepsilon_{s} } V^{[14]}_{s\beta q\delta J},
       \label{eq:thirdpha}
\end{equation}
with $f=(-)^{j_r+j_s+j_{\gamma}+j_{\delta}-2J}
      (-)^{2j_{\gamma}+2j_{\delta}}$. This
equation has the general structure
\[
            V_{\mathrm{ph}}^{[14]}
            \frac{Q_{\mathrm{ph}}^{[14]}}{\epsilon^{[14]}}
             V_{\mathrm{ph}}^{[14]}
            \frac{Q_{\mathrm{ph}}^{[14]}}{\epsilon^{[14]}}
            V_{\mathrm{ph}}^{[14]}.
\]
A similar expression applies to diagram (b), whose expression
is
\begin{equation}
      (b)=\sum_{rs\gamma\delta}f
      V^{[13]}_{\gamma pr\alpha J}
      \frac{1}{t+\varepsilon_{\gamma}
        -\varepsilon_{r} } V^{[13]}_{\gamma rs\delta J}
      \frac{1}{t+\varepsilon_{\delta}
               -\varepsilon_{s} } V^{[13]}_{\beta s q\delta J}.
       \label{eq:thirdphb}
\end{equation}
It has the general structure
\[
            V_{\mathrm{ph}}^{[13]}
            \frac{Q_{\mathrm{ph}}^{[13]}}{\epsilon^{[13]}}
             V_{\mathrm{ph}}^{[13]}
            \frac{Q_{\mathrm{ph}}^{[13]}}{\epsilon^{[13]}}
            V_{\mathrm{ph}}^{[13]}.
\]
But these expressions have the same sign! Diagram (a) counts
now 4 hole lines, and (b) counts also 4 hole lines and 2 closed
loops. However, {\em there are three interaction terms $V$},
and taking
the exchange term of each of these in diagram (a) leads to
the desired results, namely $(a)=-(b)$, as it should.
Thus, to third order we keep the antisymmetry property of
$\Gamma$ in the $[13]$ and $[14]$ channels.
It is easy to see that in the $[14]$ channel we will always
have an alternating sign in front of each contribution,
since every new order in perturbation theory brings a new hole
line and no closed loop, and thus a new minus sign.
In the $[13]$ channel we have always one new hole line and
one new closed loop for every new vertex.
If we consider only the screening of the ph vertex, we can then set up
a perturbative expansion in terms of the ph vertex for the
vertex functions $\Gamma^{[13]}$ and $\Gamma^{[14]}$.
For notational economy we will skip the Pauli operators
$Q_{\mathrm{ph,hp}}^{[ij]}$ in the discussions below.
It will always be understood that the intermediate states
are two-body particle-hole states, $\left| \mathrm{ph}\right\rangle$ or
$\left| \mathrm{hp}\right\rangle$.
Consider e.g.,
$\Gamma^{[14]}$
\begin{equation}
       \Gamma^{[14]}=V^{[14]}_{\mathrm{ph}}-
        V^{[14]}_{\mathrm{ph}}
        \frac{1}{\epsilon^{[14]}}
        V_{\mathrm{ph}}^{[14]}+
        V^{[14]}_{\mathrm{ph}}
        \frac{1}{\epsilon^{[14]}}
        V^{[14]}_{\mathrm{ph}}
        \frac{1}{\epsilon^{[14]}}
        V^{[14]}_{\mathrm{ph}}-+\dots,
\end{equation}
which can be summed up to yield
\begin{equation}
  \Gamma^{[14]}=V^{[14]}_{\mathrm{ph}}-
   V^{[14]}_{\mathrm{ph}}
   \frac{1}
   {\epsilon^{[14]}-V^{[14]}_{\mathrm{ph}}}
   V^{[14]}_{\mathrm{ph}}=
   V^{[14]}_{\mathrm{ph}}-
   V^{[14]}_{\mathrm{ph}}
   \frac{1}{\epsilon^{[14]}}\Gamma^{[14]},
   \label{eq:screening1}
\end{equation}
which is the standard TDA expression for the ph term.
The corresponding expression in the $[13]$ channel
results in
\begin{equation}
  \Gamma^{[13]}=V^{[13]}_{\mathrm{ph}}+
   V^{[13]}_{\mathrm{ph}}
   \frac{1}{\epsilon^{[13]}}\Gamma^{[13]}.
\end{equation}
The signs agree with the
expressions of Blaizot and Ripka \cite{br86}, see chapter 15 and
Eq.\ (15.50).
The summations in both channels ensures that the final vertex
is antisymmetric and the combination of
the latter two equations results in the familiar
TDA equations.
We next look at the $2p2h$ matrix element and show the
corresponding corrections to second order in perturbation
theory in Fig.\ \ref{fig:pphhvertex}.
\begin{figure}[hbtp]
\begin{center}
      \setlength{\unitlength}{1mm}
      \begin{picture}(100,80)
      \put(0,0){\epsfxsize=10cm \epsfbox{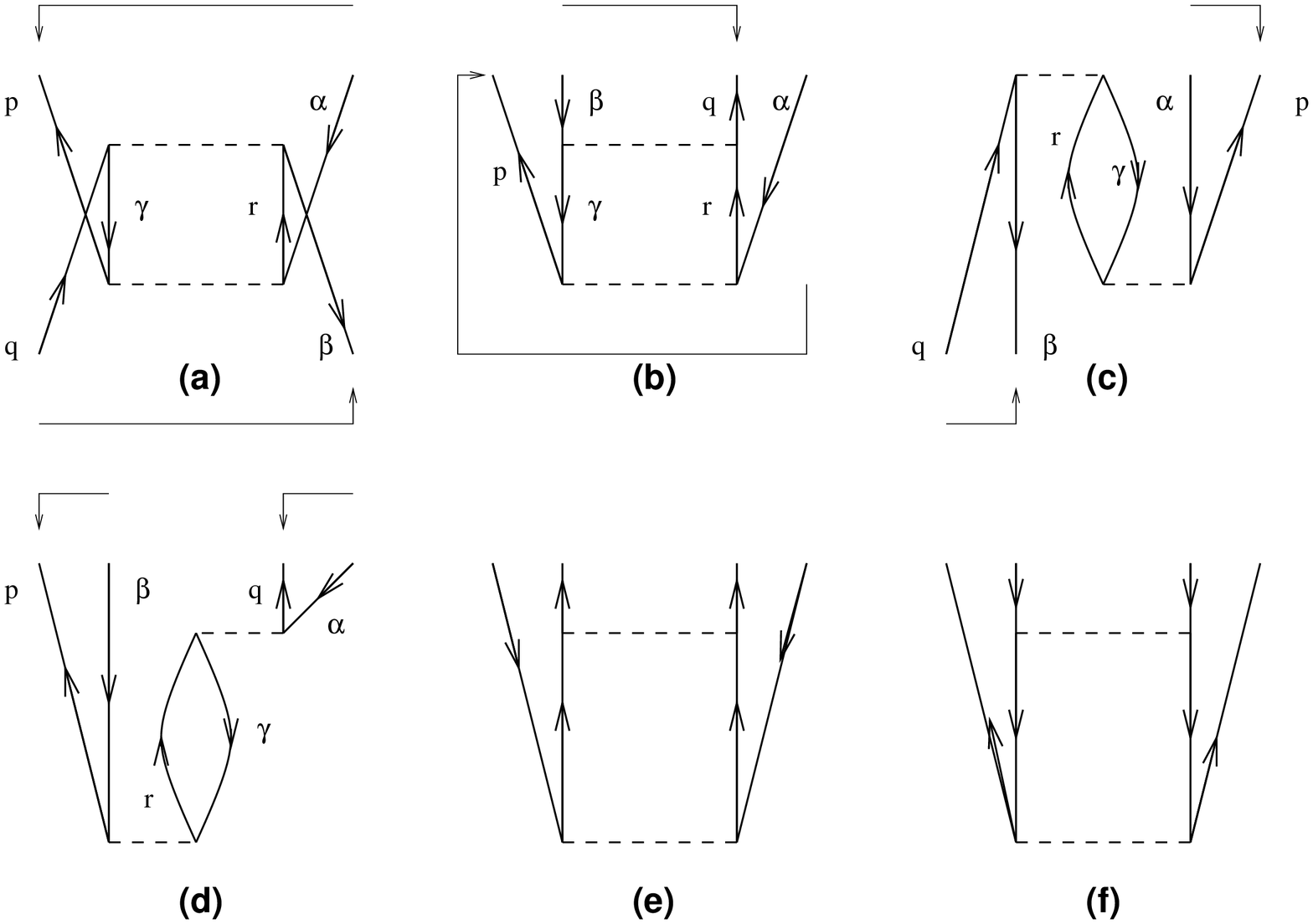}}
      \end{picture}
       \caption{Corrections to second order in $V$ of the 2p2h
               vertex.}
       \label{fig:pphhvertex}
\end{center}
\end{figure}
If we omit diagrams (e) and (f) which contain 2p and 2h
intermediate states generated by the solutions
in $[12]$ channel, we have
for diagram (a)
\begin{equation}
      (a)=-\sum_{r\gamma}(-)^{j_r+j_{\gamma}-J}
      (-)^{2j_{\gamma}}
      V^{[14]}_{\gamma\beta qr J}
      \frac{1}{-u+\varepsilon_{\gamma}-
                \varepsilon_{r}} V^{[14]}_{pr\gamma\alpha J},
       \label{eq:2p2ha}
\end{equation}
with the general structure
\begin{equation}
    -V_{\mathrm{2p2h}}^{[14]}
     \frac{1}{\epsilon^{[14]}}
     V_{\mathrm{2p2h}}^{[14]}.
\end{equation}
Note well the minus sign in front of $u$. The contribution from the
propagator can in this case be retraced to the second
term of the propagator of Eq.\ (\ref{eq:paulioperator14}).
Diagram (a) follows the coupling order of the $[14]$ channel.
It is also easy to see that diagram (b) is given by
\begin{equation}
      (b)=-\sum_{r\gamma}(-)^{j_r+j_{\gamma}-J}
      (-)^{2j_{\gamma}}
      V^{[14]}_{\gamma q \beta r J}
      \frac{1}{-u+\varepsilon_{\gamma}-
                \varepsilon_{r}} V^{[14]}_{pr\gamma\alpha J},
       \label{eq:2p2hb}
\end{equation}
and has the structure
\begin{equation}
    -V_{\mathrm{ph}}^{[14]}
     \frac{1}{\epsilon^{[14]}}
     V_{\mathrm{2p2h}}^{[14]}.
\end{equation}
Similarly, if we now move to the $[13]$ channel we have
the following expressions
\begin{equation}
      (c)=\sum_{r\gamma}(-)^{j_r+j_{\gamma}-J}
      (-)^{2j_{\gamma}}
      V^{[13]}_{\beta\gamma qr J}
      \frac{1}{-t+\varepsilon_{\gamma}-
                \varepsilon_{r}} V^{[13]}_{qp\gamma\alpha J},
       \label{eq:2p2hc}
\end{equation}
and
\begin{equation}
      (d)=\sum_{r\gamma}(-)^{j_r+j_{\gamma}-J}
      (-)^{2j_{\gamma}}
      V^{[13]}_{\gamma q r\alpha J}
      \frac{1}{-t+\varepsilon_{\gamma}-
                \varepsilon_{r}} V^{[13]}_{pr\beta\gamma J},
       \label{eq:2p2hd}
\end{equation}
with the general structure
\begin{equation}
    V_{\mathrm{2p2h}}^{[13]}
     \frac{1}{\epsilon^{[13]}}
     V_{\mathrm{2p2h}}^{[13]},
\end{equation}
and
\begin{equation}
    V_{\mathrm{ph}}^{[13]}
     \frac{1}{\epsilon^{[13]}}
     V_{\mathrm{2p2h}}^{[13]},
\end{equation}
respectively. Diagram (c) is just the exchange of (a)
and includes two 2p2h vertices, while diagram (d) is the exchange
of diagram (b) and includes a ph vertex multiplied with
a 2p2h vertex. We note again that the antisymmetry is
ensured at a given order in the interaction only
if we include the corrections at the same level in both
channels.

One can then easily sum up higher-order corrections
to the 2p2h diagrams as well in both channels.
The inclusion of the backward going particle-hole
pair in the propagators of Eqs.\ (\ref{eq:paulioperator13})
and (\ref{eq:paulioperator14}) ensures thus that we will
also sum to infinite order 2p2h corrections.
This leads ultimately to the familiar RPA equations,
see e.g., Refs.\ \cite{eo77}.

A closer inspection of Eqs.\ (\ref{eq:2p2hb}) and
(\ref{eq:2p2hd}) shows that if we only include ph vertices,
we could resum these corrections to infinite
order for the 2p2h vertex by observing that the
structure of such diagrams would be of the form
(e.g., in the $[14]$ channel )
\begin{equation}
       \Gamma_{\mathrm{2p2h}}^{[14]}=V^{[14]}_{\mathrm{2p2h}}-
        V^{[14]}_{\mathrm{ph}}
        \frac{1}{\epsilon^{[14]}}
        V_{\mathrm{2p2h}}^{[14]}+
        V^{[14]}_{\mathrm{ph}}
        \frac{1}{\epsilon^{[14]}}
        V^{[14]}_{\mathrm{ph}}
        \frac{1}{\epsilon^{[14]}}
        V^{[14]}_{\mathrm{2p2h}}.
             +\dots,
\end{equation}
which can be summed up to yield
\begin{equation}
  \Gamma^{[14]}_{\mathrm{2p2h}}=V^{[14]}_{\mathrm{2p2h}}-
   V^{[14]}_{\mathrm{ph}}
   \frac{1}
   {\epsilon^{[14]}-V^{[14]}_{\mathrm{ph}}}
   V^{[14]}_{\mathrm{2p2h}},
   \label{eq:screening2}
\end{equation}
and similarly for the $[13]$ channel, but with a plus sign.
The modification discussed in Eqs.\ (\ref{eq:screening1})
and (\ref{eq:screening2})
serve to modify the propagation of a particle-hole pair
and have normally been termed for propagator
renormalizations, as can easily be seen from
Eqs.\ (\ref{eq:screening1})
and (\ref{eq:screening2}) where the propagation of
a free particle-hole pair is modified by the presence
of the interaction $V$ in the energy denominator.
Other important processes which can affect e.g.,
various polarization terms are those
represented by so-called vertex renormalizations,
a term originally introduced by Kirson and Zamick \cite{kz70}.
These authors studied the renormalizations of the
2p1h and 2h1p vertices as well, see also Kirson \cite{kirson74}
and Ellis and Osnes \cite{eo77} for further discussions.
We will therefore end the discussion in this section
by looking at such renormalizations.

\subsection{\it Further renormalizations}

In the previous subsection we dealt mainly with what has
conventionally been labelled for propagator
renormalizations. We will therefore extend the standard TDA and RPA
scheme by looking at further ways of renormalizing
interaction vertices.
The approach discussed here follows
Kirson \cite{kirson74}. Extensions were made later
by Ellis and Goodin \cite {eg80} and
Ellis, Mavromatis and M\"uther \cite{emm91}.
We will limit the discussion here to the scheme of Kirson.
We start therefore with the contributions to second
order to the 2p1h vertex\footnote{The discussion here applies to the
other interaction vertices discussed in Fig.\ \ref{fig:wavef1}.}.
\begin{figure}[hbtp]
\begin{center}
      \setlength{\unitlength}{1mm}
      \begin{picture}(100,80)
      \put(0,0){\epsfxsize=10cm \epsfbox{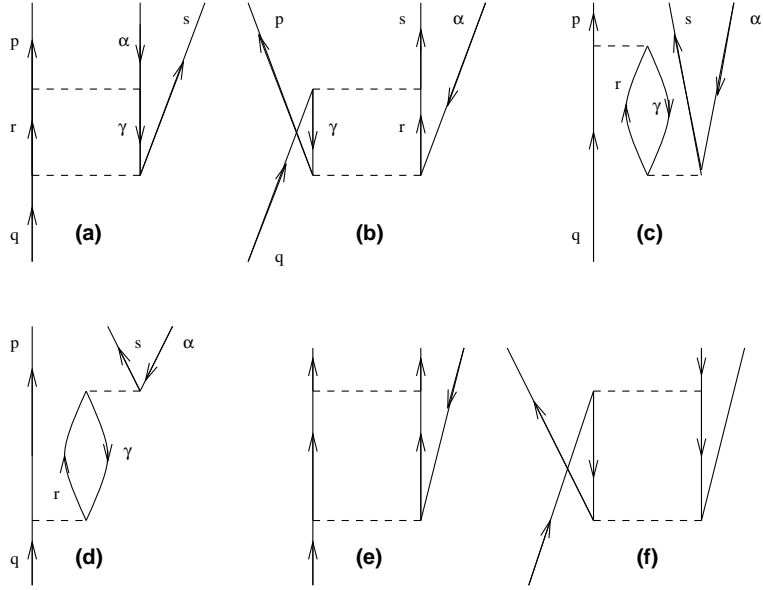}}
      \end{picture}
       \caption{The corrections to second order in $V$ of the 2p1h
               vertex.}
       \label{fig:2p1hvertex}
\end{center}
\end{figure}
These contributions are shown in Fig.\ \ref{fig:2p1hvertex}.
Diagram (a) consists of a 2p1h vertex multiplied with a
ph vertex whereas (b) stems from the multiplication
of a 2p2h vertex with a 2p1h vertex. They both contain
a particle-hole pair as an intermediate state and
follow the  coupling order of the $[14]$ channel.
The exchange diagram of (b)  is given by
(c), while that of (a) is diagram (d). Diagrams (e) and (f)
represent contributions from the $[12]$ channel
and are irreducible in both particle-hole channels.
Before we sketch the
general structure of the renormalization
procedure of Kirson, it is instructive to consider
again the equations to second order in perturbation theory,
as the general expressions can be deduced from inspection
of these diagrams.
Diagram (a) is
\begin{equation}
      (a)=\sum_{r\gamma}(-)^{j_r+j_{\gamma}-J}
      (-)^{2j_{\gamma}}
      V^{[14]}_{p\gamma r\alpha J}
      \frac{1}{u+\varepsilon_{\gamma}-
                \varepsilon_{r}} V^{[14]}_{rsq\gamma J},
       \label{eq:2p1ha}
\end{equation}
and the plus sign stems
from the diagram rules \cite{kstop81},
i.e., we
have two hole lines and no closed loop.
The propagator is that arising
from the first term in Eq.\ (\ref{eq:paulioperator14}).
The  general structure is
\begin{equation}
     V_{\mathrm{ph}}^{[14]}
     \frac{1}{\epsilon^{[14]}}
     V_{\mathrm{2p1h}}^{[14]}.
     \label{eq:2p1hseca}
\end{equation}
Diagram (b) reads
\begin{equation}
      (b)=\sum_{r\gamma}(-)^{j_r+j_{\gamma}-J}
      (-)^{2j_{\gamma}}
      V^{[14]}_{pr\gamma\alpha J}
      \frac{1}{-u+\varepsilon_{\gamma}-
                \varepsilon_r} V^{[14]}_{\gamma sqr J},
       \label{eq:2p1hb}
\end{equation}
with the following structure
\begin{equation}
     V_{\mathrm{2p1h}}^{[14]}
     \frac{1}{\epsilon^{[14]}}
     V_{\mathrm{2p2h}}^{[14]}.
     \label{eq:2p1hsecb}
\end{equation}
In this case the propagator stems from the second term
in Eq.\ (\ref{eq:paulioperator14}).
Similar equations arise for e.g.,  the 2h1p
vertices of Fig.\ \ref{fig:wavef1}.

\begin{figure}[hbtp]
\begin{center}
      \setlength{\unitlength}{1mm}
      \begin{picture}(100,80)
      \put(0,0){\epsfxsize=10cm \epsfbox{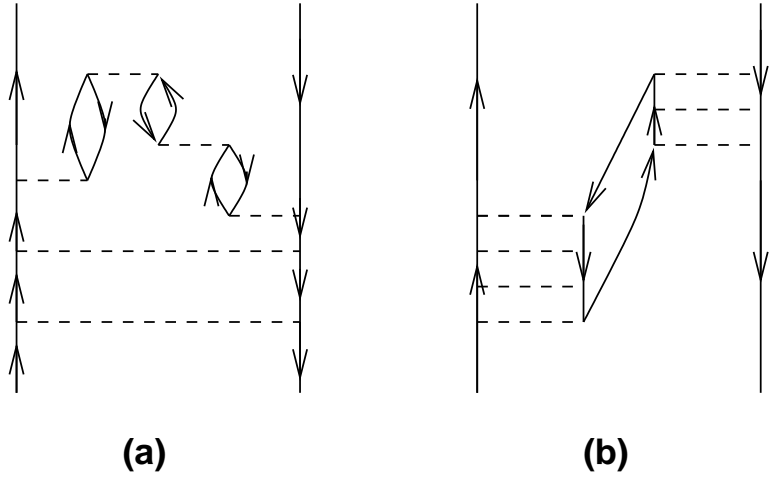}}
      \end{picture}
       \caption{Examples of diagrams which can arise from Kirson's
                self-consistent equations.}
       \label{fig:kirsoniterate}
\end{center}
\end{figure}

Before we write down the self-consistent equations
of Kirson \cite{kirson74}, let us assume that we
can approximate the 2p1h vertex in the $[14]$ channel
by the first order term and diagrams (a) and (b).
This renormalized vertex, which we here label
$\tilde{V}_{2p1h}$, is then given by
\begin{equation}
     \tilde{V}_{2p1h}\approx {V}_{2p1h}^{[14]}
     +V_{\mathrm{ph}}^{[14]}
     \frac{1}{\epsilon^{[14]}}
     V_{\mathrm{2p1h}}^{[14]}+
     V_{\mathrm{2p1h}}^{[14]}
     \frac{1}{\epsilon^{[14]}}
     V_{\mathrm{2p2h}}^{[14]}.
     \label{eq:2p1hsecondorder}
\end{equation}
If we now allow for the screening to infinite order
of the ph vertex given by Eq.\ (\ref{eq:screening1})
and replace the 2p2h vertex in the above equation
with Eq.\ (\ref{eq:screening2}) we obtain the following
renormalization of the 2p1h vertex
\begin{equation}
     \tilde{V}_{2p1h}= {V}_{2p1h}^{[14]}
     +V_{\mathrm{ph}}^{[14]}
     \frac{1}
     {\epsilon^{[14]}-V_{\mathrm{ph}}^{[14]}}
     V_{\mathrm{2p1h}}^{[14]}+
     V_{\mathrm{2p1h}}^{[14]}
     \frac{1}
     {\epsilon^{[14]}-V_{\mathrm{ph}}^{[14]}}
     V_{\mathrm{2p2h}}^{[14]}.
     \label{eq:screening3}
\end{equation}
A similar equation applies to the 2h1p vertex of
Fig.\ \ref{fig:wavef1} and
for the $[13]$ channel. Eqs.\ (\ref{eq:screening1}),
(\ref{eq:screening2}) and (\ref{eq:screening3}) form then
the starting point for the approach of Kirson \cite{kirson74}.
Examples of diagrams which can be obtained through the
iterative solution of Eqs.\ (\ref{eq:screening1}),
(\ref{eq:screening2}) and (\ref{eq:screening3})
are given in Fig.\ \ref{fig:kirsoniterate}.

The question now is however how to relate
Eqs.\ (\ref{eq:screening1}),
(\ref{eq:screening2}) and (\ref{eq:screening3}) with those
from Eqs.\ (\ref{eq:first13}) and
(\ref{eq:first14}).
This is rather trivial if we recall that the labels
$1234$ can, as was also discussed in section \ref{sec:sec2},
represent whatever single-particle states, either holes
or particles. Thus, $V_{1234}$ can represent
a 2p1h, 2h1p, 2p2h, 2p, 2h or a ph vertex.
This means that, due to the choice of propagators
in Eqs.\ (\ref{eq:paulioperator13}) and
(\ref{eq:paulioperator14}) equations like Eq.\
(\ref{eq:screening3}) are already inherent in Eqs.\
(\ref{eq:first13}) and
(\ref{eq:first14}). If we e.g., approximate Eq.\ (\ref{eq:first14})
to second order in the interaction $V$ and let the
single-particle labels $1234$ represent a 2p1h interaction
vertex, we immediately reobtain Eq.\
(\ref{eq:2p1hsecondorder}). If we let $1234$ represent
a 2p2h vertex, we find to second order diagrams
(a)-(d) of Fig.\ \ref{fig:pphhvertex}.

Till now we have however refrained from discussing the contributions
from the $[12]$ channel, examples were only shown in (c) and (d) of
Fig.\ \ref{fig:phvertex}, (e) and (f) of Figs.\ \ref{fig:pphhvertex}
and \ref{fig:2p1hvertex}.
These diagrams cannot be generated by simply iterating
the equations for the $[13]$ and $[14]$ channels, but they could enter
as contributions to the irreducible vertex
in the $[13]$ and $[14]$ channels from the first iteration
in the $[12]$-channel.
We see thus the emerging contour of an iterative scheme.
The crucial point is however how to perform the next iteration of say
Eqs.\ (\ref{eq:screening1}),
(\ref{eq:screening2}) and (\ref{eq:screening3}) or
Eqs.\ (\ref{eq:first13}),
(\ref{eq:first14})  and (\ref{eq:first12}) from the
$[12]$ channel
The question is how do we include the results
from the first iteration into the next one, i.e.,
how to modify the bare vertices $V^{[13]}$ and $V^{[14]}$
in e.g., Eqs.\ (\ref{eq:first13}) and
(\ref{eq:first14}) in order to obtain
an effective interaction for the shell model.
We have also not addressed how to deal
with the solution of Dyson's equation for the one-body
Green's function.
We mention also that Ellis and Goodin \cite{eg80}
included pp correlations, i.e., terms from the $[12]$ channel
such as diagram (e) of Fig.\ \ref{fig:2p1hvertex},
as well when they considered the screening of the 2p1h and 2h1p vertices.
Furthermore, as already mentioned in the introduction
the authors of Refs.\ \cite{emm91,hmm95}
extended the pp RPA to include the particle-hole (ph) RPA, though
screening of the 2p1h and 2h1p vertices was not included. In Ref.\
\cite{hmm95} however, a study with
self-consistent  single-particle energies was also performed.
These works represent thus a first serious step towards the solution
of the parquet equations, i.e., a many-body scheme which solves
self-consistently the equations in the $[12]$, $[13]$ and $[14]$
channels, with the addition of the self-consistent evaluation
of the self-energy. It ought also to be mentioned
that one of the really
first applications for nuclear systems was performed
in a series of papers by Dickhoff and M\"uther and co-workers
\cite{nuclearmatter}  for nuclear matter. These authors
actually performed the first iteration of the three
channels.

\subsection{\it Parquet diagrams}

The equations we discussed in the two previous sections can
be generalized
in matrix form as
\begin{equation}
      \Gamma= \Gamma^{[ij]}+\Gamma^{[ij]}\hat{{\cal G}}^{[ij]}\Gamma,
      \label{eq:generalchannel}
\end{equation}
where obviously $[ij]$ represents a given channel,
$\hat{{\cal G}}^{[ij]}$ is the particle-particle, or hole-hole or particle-hole
propagator.
The propagator is a product of two single-particle propagators $g$
which we do not specify any further here. They are
defined by the solution of Dyson's equation in Eq.\ (\ref{eq:dyson12}).
The irreducible vertices must appear in the solution of
the self-energy, and conversely, the self-energy must appear in all
single-particle propagators within the expressions for the three
channels $[12]$, $[13]$ and $[14]$.

For all of our practical purposes, the irreducible vertex used in all
channels is the $G$-matrix defined by the large model space of 
Fig.~\ref{fig:paulioperator}. 

Let us now define the contribution from the $[12]$, $[13]$ and
$[14]$ channels following Ref.\ \cite{jls82}.
Eq.\ (\ref{eq:first12}) is then rewritten as
\begin{equation}
    L=\Gamma^{[12]}-G,
\end{equation}
where obviously $L$ stands for ladder.
The ladder term can then be rewritten as
\begin{equation}
    L=G\hat{{\cal G}}^{[12]}G+G\hat{{\cal G}}^{[12]}L.
    \label{eq:ladder}
\end{equation}
In a similar way we can define the diagrams beyond
first order in the particle-hole channel as
\begin{equation}
    R^{[13]}=\Gamma^{[13]}-G,
\end{equation}
and
\begin{equation}
    R^{[14]}=\Gamma^{[14]}-G,
\end{equation}
where $G$ now is coupled in either the $[13]$ or $[14]$ way.
Rewriting $R^{[ij]}$, where $R$ refers to ring diagrams,
we obtain
\begin{equation}
    R^{[ij]}=G\hat{{\cal G}}^{[ij]}G+G\hat{{\cal G}}^{[ij]}R^{[ij]},
    \label{eq:ring}
\end{equation}
where $ij$ stands for either $[13]$ or $[14]$.
The equation for the vertex function $\Gamma$ in Eq.\
(\ref{eq:generalchannel}) becomes then
\begin{equation}
    \Gamma=G+L+R^{[13]}+R^{[14]}.
     \label{eq:gammap}
\end{equation}
{\em Note here that the vertex $\Gamma$ can be represented in the coupling
order of any of the above channels}. Our convention is that of the
$[12]$ channel.

If Eqs.\ (\ref{eq:ladder}) and (\ref{eq:ring}) define the first iteration,
it should be fairly obvious to see that the next iteration would be
\begin{equation}
    L=\left(G+R^{[13]}+R^{[14]}\right)\hat{{\cal G}}^{[12]}\left(G+R^{[13]}+R^{[14]}\right)
      +\left(G+R^{[13]}+R^{[14]}\right)\hat{{\cal G}}^{[12]}L,
    \label{eq:laddernext}
\end{equation}
where the contributions $R^{[13]}+R^{[14]} $ are recoupled according to the coupling order
of the $12$-channel. These contributions are irreducible in the $[12]$ channel.
Similarly, for the rings we have
\begin{equation}
    R^{[13]}=\left(G+L+R^{[14]}\right)\hat{{\cal G}}^{[13]}
             \left(G+L+R^{[14]}\right) +
             \left(G+L+R^{[14]}\right)\hat{{\cal G}}^{[13]}R^{[14]},
    \label{eq:ring13next}
\end{equation}
and
\begin{equation}
    R^{[14]}=\left(G+L+R^{[13]}\right)\hat{{\cal G}}^{[14]}
             \left(G+L+R^{[13]}\right) +
             \left(G+L+R^{[13]}\right)\hat{{\cal G}}^{[14]}R^{[14]}.
    \label{eq:ring14next}
\end{equation}

Our scheme for calculating $\Gamma$ will be an iterative one based on
Eqs.\ (\ref{eq:gammap})-(\ref{eq:ring14next})
and the solution of Dyson's equation
for the single-particle propagator. This set of equations will then yield
the two-body parquet diagrams.
Relating the above equations to the discussions above it is rather easy to see
that the $G$-matrix, TDA, RPA and Kirson's
screening scheme are contained in Eqs.\
(\ref{eq:gammap})-(\ref{eq:ring14next}).
Applications of this method will be presented elsewhere \cite{mhj99}. However, we present
here a numerically viable approach
to the parquet equations.
Here
we will limit ourself to just sketch 
the structure of the solution, more technical details 
can be found in Ref.~\cite{mhj99}.

The iterative scheme starts with the solution of Eq.\ (\ref{eq:first12}).
We define then $Q^{[12]}=Q_{pp}+Q_{hh}$ which results in 
\begin{equation}
   \Gamma^{[12]}_{(0)}=G+G\left(\frac{Q^{[12]}}{s-H_0}\right)\Gamma^{[12]}_{(0)}.
   \label{eq:approx12channel}
\end{equation}
The subscript $(0)$ means that this is just the first iteration.
The single-particle energies are the unperturbed
harmonic oscillator energies.  

The first step in our calculations is to evaluate $G$.  It is 
calculated following the discussion of section \ref{sec:sec3}, employing a
large model space.
The exclusion operator for this $G$-matrix was defined in
Fig.~\ref{fig:paulioperator} and is typically defined for $6-10$ major oscillator shells,
as discussed in connection with Fig.~\ref{fig:mbptox}. There, 8 major oscillator shells
were needed in order to obtain a converged result. Similar results are obtained
in the coupled cluster calculations discussed in the next section.

The second step is to solve Eq.\ (\ref{eq:approx12channel}),
which is now a complex equation. 
The hole states and particle states are then defined according to Fig.~\ref{fig:finalp}.
These single-particle states define then the projection operator  $Q^{[12]}=Q_{pp}+Q_{hh}$.
This equation is then solved by matrix diagonalization, yielding a Green's function and the 
pertinent new vertex.

These two steps lead then to the first iteration of
the ladders, i.e.,
\begin{equation}
     L_{(0)}=\Gamma^{[12]}_{(0)}-G.
\end{equation} 
It contains both hole-hole and particle-particle intermediate
states and is a complex matrix. The external single-particle
legs can be particles or holes. Only unperturbed single-particle 
energies enter the definition of the two-body propagators.

The third step is to calculate the first iteration for the  
rings, namely
\begin{equation}
    R_{(0)}^{[13]}=\left(G+L_{(0)}\right)\hat{{\cal G}}^{[13]}
             \left(G+L_{(0)}\right) + 
             \left(G+L_{(0)}\right)\hat{{\cal G}}^{[13]}R_{(0)}^{[13]},
    \label{eq:ring13first}
\end{equation}
and
\begin{equation}
    R_{(0)}^{[14]}=\left(G+L_{(0)}\right)\hat{{\cal G}}^{[14]}
             \left(G+L_{(0)}\right) + 
             \left(G+L_{(0)}\right)\hat{{\cal G}}^{[14]}R_{(0)}^{[14]}.
    \label{eq:ring14first}
\end{equation}
The equations for $L$ and $R$ are all defined within a
truncated Hilbert space. They can therefore be recast into
matrix equations of finite dimensionality.
Recall also  that we need to recouple the contribution
from the $[12]$ into the relevant ones for the $[13]$ and $[14]$
channels. This is done employing Eqs.\ (\ref{eq:13channel}) and
(\ref{eq:14channel}). 
With these contributions, we can now obtain the vertex function
$\Gamma$ after the first interaction
\begin{equation}
    \Gamma_{(0)}=G+L_{(0)}+R_{(0)}^{[13]}+R_{(0)}^{[14]}.
\end{equation}
The fourth step is to compute the self-energy and thereby obtain
new single-particle energies. In so doing, care has to be exercised
in order to avoid double-counting problems. A thourough discussion
of this topic can be found in Ref.\ \cite{jls82}. More details
will also be presented in Ref.\ \cite{mhj99}.
The new single-particle wave functions are obtained
by diagonalizing a matrix of dimension 
$n_{\alpha}\times n_{\alpha}$, $n_{\alpha}$ the quantum
number $n$ of the single-particle state $\alpha$. 

The fifth step is to repeat steps 1-4 with the new single-particle
energies till a predetermined self-consistency is obtained. 
But now the rings have to be included in all equations, i.e.,
we solve Eqs.\ (\ref{eq:gammap})-(\ref{eq:ring14next}). 

The final vertex $\Gamma$ can then be used to define a 
new effective interaction to be applied in shell model studies,
where many more diagrams are considered than in present 
state of the art calculations, see e.g., Fig.\ 8 of Ref.\ \cite{jls82}
for a list of diagrams to sixth order entering the definition
of  the irreducible vertex $\Gamma$. 

The same technique employing a large model-space is used in our 
coupled cluster calculations discussed in the next section.

\section{Non-perturbative resummations: Coupled cluster theory}\label{sec:sec6}

As we have seen from the above discussions, nuclear many-body theory
often begins with a $G$-matrix interaction which is derived from 
an underlying bare nucleon-nucleon interaction. In this section we limit
the discussion to the no-core $G$-matrix so that all particles are active
within our chosen model space.  Using a given basis expansion
of the many-body wave function we could then solve the nuclear problem by
diagonalization as has been pursued by the No-Core shell model collaboration
\cite{bruce1,bruce2,bruce3,petr_erich2002}. 
In fact, the current and most advanced no-core oscillator expansion
techniques have approached $^{12}$C, with nearly
converged solutions \cite{hayes03}.

It should be evident, however, that diagonalization procedures scale 
almost combinatorally with the number of particles in a given number of 
single-particle orbitals. Because of this scaling, diagonalization simply
becomes untenable at some point. The efforts to 
expand diagonalization into $p$-shell nuclei with all 
nucleons active, an effort that
spans over ten years, illustrates the problem. The
computational complexity of the nucleus grows dramatically as the size
of the nucleus increases. As a simple example consider 
oscillator single-particle states,
and single-particle spaces consisting of 4 and 7 major
oscillator shells, and compare the number of uncoupled many-body basis states
there are for 4,8,12, and 16 particles. From table \ref{tab:table_1}
we see an enormous growth of the standard shell-model diagonalization
problem within the space. We calculated the number of $M=0$ states for
He and B
within the model space consisting of 4 major shells
and estimated the number of basis states for C and O. Also
indicated are similar estimates for seven major oscillator
shells. The important lesson to learn from these numbers is
that the model-space expansion becomes astronomical quite quickly.
\begin{table}[th]
\caption{Dimensions of the shell-model problem in four major oscillator
shells  and 7 major oscillator
shells with $M=0$.}
\label{tab:table_1}
\begin{center}
\begin{tabular}{|ccc|}
\hline
System &   4 shells & 7 shells \cr
\hline
$^{4}$He & 4E4  &  9E6 \cr
$^{8}$B  & 4E8  &  5E13 \cr
$^{12}$C & 6E11 &  4E19 \cr
$^{16}$O & 3E14 &  9E24  \cr
\hline
\end{tabular}
\end{center}
\end{table}

Yet, because of the advent of radioactive nuclear beam
accelerators, such as the proposed Rare Isotope Accelerator (RIA) in the
U.S., we face the daunting task of moving beyond
$p$-shell nuclei in {\em ab initio} calculations. We should therefore
investigate several ways of approaching the nuclear many-body 
problem in order to successfully make the move into the RIA era.

One motivation for developing 
Auxiliary Field Monte Carlo (see section \ref{sec:sec4} above) for the shell model was 
to overcome the scaling problem. Other many-body 
methods that resum major classes of many-body diagrams have simply
been neglected in nuclear science to this point. 
In this Section we will discuss the coupled-cluster 
technique which can be used to pursue nuclear many-body calculations to 
heavier systems beyond the $p$-shell. 

Coupled cluster theory originated in nuclear physics
\cite{coester58,coester60} around 1960.  Early studies in the
seventies \cite{klz78} probed ground-state properties in limited
spaces with free nucleon-nucleon interactions available at the
time. The subject was revisited
only recently by Bishop {\it et al.}
\cite{ticcm}, for further theoretical development, and by Mihaila and
Heisenberg \cite{hm99}, for coupled cluster calculations
using realistic
two- and three-nucleon
bare interactions
and expansions in the
inverse particle-hole energy spacings.
However, much of
the impressive development in
coupled cluster theory made in quantum chemistry in
the last 15-20 years
\cite{comp_chem_rev00,Bartlett95,Paldus99,Piecuch02a,Piecuch02b}
still awaits applications to the nuclear many-body problem.

Many solid theoretical reasons exist that motivate a pursuit of
coupled-cluster methods. First of all, the method is fully
microscopic and is capable of systematic and hierarchical improvements.
Indeed, when one expands the cluster operator in coupled-cluster theory
to all $A$ particles in the system, one exactly produces the fully-correlated
many-body wave function of the system. The only input that the method
requires is the nucleon-nucleon interaction. 
The method may also be extended
to higher-order interactions such as the three-nucleon interaction.
Second, the method is size extensive which means that only linked
diagrams appear in the computation of the  
energy (the expectation value of the Hamiltonian) and amplitude equations.
As discussed in Ref.~\cite{comp_chem_rev00} all shell model calculations
that use particle-hole truncation schemes
actually suffer from the inclusion of unconnected diagrams
in computations of the energy.
Third, coupled-cluster theory is also size
consistent which means that the energy of two non-interacting fragments
computed separately is the same as that computed for both fragments
simultaneously. In chemistry, where the study of reactions
is quite important, this is a crucial property not available
in the interacting shell model (named configuration interaction in
chemistry).
Fourth, while the theory
is not variational,
the energy behaves as a variational quantity in most instances.
Finally, from a
computational point of view, the practical implementation of coupled
cluster theory is amenable to parallel computing.

We are in the process of applying quantum chemistry inspired coupled cluster
methods
\cite{comp_chem_rev00,cizek66,Bartlett95,Paldus99,Piecuch02a,Piecuch02b,cizek69,Stanton:1993,Piecuch99} to
finite nuclei \cite{dean03,kowalski03n}. We show one result
from our current studies, namely the convergence of $^{16}$O
as a function of the model space in which we perform the calculations.

The basic idea of coupled-cluster theory is that the correlated many-body
wave function $\mid \Psi\rangle$ 
may be obtained by application of a correlation operator, 
$T$, such that
\begin{equation}
\mid\Psi \rangle =\exp\left(-T\right)\mid\Phi\rangle\;,
\end{equation}
where $\Phi$ is a reference Slater determinant chosen as a convenient starting
point.  For example, we use the filled $0s$ state as the reference 
determinant for $^4$He.

The correlation operator $T$ is given by
\begin{equation}
T=T_1 + T_2 + \cdots T_A\;,
\end{equation}
and represent various
$n$-particle-$n$-hole ($n$p-$n$h) excitation amplitudes such as
\begin{eqnarray}
T_1 &=& \sum_{a\langle\varepsilon_f, i\rangle\varepsilon_f}t^a_i a^\dagger_a a_i\;, \\
T_2 &=& \frac{1}{4}\sum_{i,j\langle\varepsilon_f; ab \rangle \varepsilon_f}t^{ab}_{ij}
a^\dagger_a a^\dagger_b a_j a_i\;,
\end{eqnarray}
and higher-order terms for $T_3$ to $T_A$.  
We are currently exploring
the coupled-cluster method at the $T_1$ and $T_2$ level. This is 
commonly referred to in the literature as Coupled-Cluster Singles and
Doubles (CCSD). 

We compute the expectation of the ground-state energy from
\begin{equation}
E_{\rm g.s.}=\langle\Psi_0\mid \exp\left(-T\right) H \exp\left(T\right)
\mid\Psi_0\rangle\;. 
\end{equation}
The Baker-Hausdorf relation may be used to rewrite the similarity
transformation as
\begin{equation}
\exp\left(-T\right) H \exp\left(T\right) =
H+\left[H,T_1\right]+\left[H,T_2\right]
+\frac{1}{2}\left[\left[H,T_1\right],T_1\right]
+\frac{1}{2}\left[\left[H,T_2\right],T_2\right]
+\left[\left[H,T_1\right],T_2\right]+\cdots \;.
\end{equation}
The expansion terminates exactly at four nested commutators  when
the Hamiltonian contains, at most, two-body terms, and at six-nested 
commutators when  three-body potentials are present. 
We stress that
this termination is exact, thus allowing for a derivation of exact
expressions for the $T_1$ ($1p$-$1h$) and 
$T_2$ ($2p$-$2h$) amplitudes. 
The equations for amplitudes are found by left projection of
excited Slater determinants
so that
\begin{eqnarray}
0 &=& \langle\Phi_i^a\mid 
\exp\left(-T\right) H \exp\left(T\right) \mid \Phi\rangle\;,  \nonumber \\ 
0 &=& \langle\Phi_{ij}^{ab}\mid 
\exp\left(-T\right) H \exp\left(T\right) \mid \Phi\rangle \;.
\label{project_eqns}
\end{eqnarray}
The commutators also generate nonlinear terms within these expressions. 
To derive these equations
is straightforward, but tedious, work \cite{comp_chem_rev00}.
While the resulting equations for the single and double excitation
amplitudes  appear quite lengthy, they are 
solvable through iterative techniques. 

The Hamiltonian may be written in a slightly more convenient
form by explicitly calculating the expectation of the Hamiltonian in the
reference state $\mid\Phi\rangle$,
$E_0=\langle\Phi\mid H \mid \Phi\rangle$. This reference state is a single
Slater determinant and represents, in this work, a doubly closed shell system.
In this case, the Hamiltonian becomes
\begin{equation}
H=\sum_{pq}f_{pq}\left\{a^\dagger_p a_q\right\} +
\frac{1}{4}\langle pq \mid G\mid rs \rangle \left\{a^\dagger_p a^\dagger_q
a_s a_r\right\} + E_0 \;,
\end{equation}
where the $\{\}$ indicates normal ordering relative to the Fermi vacuum.
The Fock operator is given by
\begin{equation}
f_{pq}=\langle p\mid K\mid q\rangle + 
\sum_{i}\langle pi \mid G\mid qi \rangle\;.
\end{equation}
Using this form, and solving for the 
amplitudes, the energy of the system may then be calculated. The CCSD
energy is 
\begin{equation}
\langle H \rangle = E_{CCSD}=\sum_{ia}f_{ia}t^a_i 
+\frac{1}{4}\sum_{aibj}\langle ij \mid G\mid ab\rangle t^{ab}_{ij}
+\frac{1}{2}\sum_{aibj}\langle ij \mid G\mid ab\rangle t^a_i t^b_j +E_0\;.
\label{energy_eqn}
\end{equation}
This equation is not restricted to the CCSD approximation. Since 
higher-order excitation operators such as $T_3$ and $T_4$ cannot 
produce fully contracted terms with the two-body Hamiltonian, 
their contribution
to the energy equation is zero. Higher-order excitation clusters can 
contribute indirectly to the energy through the equations used to 
determine the amplitudes. 

Because of the nonlinearity of the equations,
one must have a good first guess for the $np$-$nh$ amplitudes. 
For closed-shell nuclei, we
use a M\o ller-Plesset-like approach to generate the first guess 
for the iteration: 
\begin{eqnarray}
t^a_i(1) &=& \frac{f_{ai}}{D_{ai}} \;, \nonumber \\ 
t^{ab}_{ij}(1) &=& \frac{<ab\mid G \mid ij>}{D_{ijab}} \;,
\label{amp_start}
\end{eqnarray}
where $D_{ia}=f_{ii}-f_{aa}$, and $D_{ijab}=f_{ii}+f_{jj}-f_{aa}-f_{bb}$
are single-particle energy denominators. 
Using this as a first guess, we can iterate the projection
equations 
(\ref{project_eqns})
until we find the converged amplitudes. Finally, we list the expressions for the 
$T_1$ and $T_2$ amplitudes.
The $T_1$ amplitude equations are given by
\begin{eqnarray}
0 & = & f_{ai} + \sum_cf_{ac}t^c_i - \sum_k f_{ki}t^a_k +  
\sum_{kc}\langle ka\mid\mid ci\rangle t^c_k + \sum_{kc}f_{kc}t^{ac}_{ik}
+\frac{1}{2}\sum_{kcd}\langle ka\mid\mid cd\rangle t^{cd}_{ki} \nonumber \\
  & - & \frac{1}{2}\sum_{klc}\langle kl\mid\mid ci \rangle t^{ca}_{kl}
-\sum_{kc}f_{kc}t^c_it^a_k -\sum_{klc}\langle kl\mid\mid ci\rangle t^c_k t^a_l
+\sum_{kcd} \langle ka\mid\mid cd \rangle t^c_k t^d_i  \nonumber \\
  & - & \sum_{klcd}\langle kl\mid\mid cd\rangle t^c_kt^d_it^a_l
+ \sum_{klcd}\langle kl\mid\mid cd\rangle t^c_kt^{da}_{li}
- \frac{1}{2} \sum_{klcd}\langle kl\mid\mid cd\rangle t^{cd}_{ki}t^{a}_{l}
- \frac{1}{2} \sum_{klcd}\langle kl\mid\mid cd\rangle t^{ca}_{kl}t^{d}_{i}\;.
\label{t1_eqn}
\end{eqnarray}
This equation is non-linear in the $T_1$ amplitudes, and linear in
the $T_2$ amplitudes. 

The $T_2$ amplitude equations are given by 
\begin{eqnarray}
0 & = & \langle ab \mid\mid ij \rangle 
+ \sum_c\left(f_{bc}t^{ac}_{ij} - f_{ac}t^{bc}_{ij}\right)
- \sum_k\left(f_{kj}t^{ab}_{ik} - f_{ki}t^{ab}_{jk}\right)  \nonumber \\
  & + & \frac{1}{2}\sum_{kl}\langle kl \mid \mid ij\rangle t_{kl}^{ab}
+ \frac{1}{2}\sum_{cd}\langle ab \mid \mid cd\rangle t_{cd}^{ij}
+ P(ij)P(ab)\sum_{kc}\langle kb \mid \mid cj\rangle t_{ac}^{ik} \nonumber \\
  & + & P(ij)\sum_c \langle ab \mid\mid cj\rangle t^c_i
- P(ab)\sum_k \langle kb \mid\mid ij\rangle t^a_k \nonumber \\
  & + & \frac{1}{2}P(ij)P(ab)\sum_{klcd} \langle kl\mid\mid cd \rangle 
        t_{ik}^{ac}t_{lj}^{db} 
+ \frac{1}{4}\sum_{klcd} \langle kl\mid\mid cd \rangle 
        t_{ij}^{cd}t_{kl}^{ab} \nonumber \\
&-& \frac{1}{2}P(ab)\sum_{klcd} \langle kl\mid\mid cd \rangle 
        t_{ij}^{ac}t_{kl}^{bd} 
- \frac{1}{2}P(ij)\sum_{klcd} \langle kl\mid\mid cd \rangle 
        t_{ik}^{ab}t_{jl}^{cd}  \nonumber \\
  & + & \frac{1}{2}P(ab)\sum_{kl}\langle kl \mid\mid ij\rangle t^a_kt^b_l
+ \frac{1}{2}P(ij)\sum_{cd}\langle ab\mid\mid cd\rangle t^c_i t^d_j
- P(ij)P(ab)\sum_{kc}\langle kb\mid\mid ic\rangle t^a_k t^c_j \nonumber \\
  & + & P(ab) \sum_{kc} f_{kc} t^a_k t^{ab}_{ij} 
+  P(ij) \sum_{kc} f_{kc} t^c_i t^{ab}_{jk}  \nonumber \\
  & - & P(ij)\sum_{klc} \langle kl\mid\mid ci\rangle t^c_k t^{ab}_{lj}
+  P(ab)\sum_{kcd} \langle ka\mid\mid cd\rangle t^c_k t^{db}_{ij}
+  P(ij)P(ab)\sum_{kcd} \langle ak\mid\mid dc\rangle t^d_i t^{bc}_{jk}
   \nonumber \\
  & + & P(ij)P(ab)\sum_{klc} \langle kl\mid\mid ic\rangle t^a_l t^{bc}_{jk}
+ \frac{1}{2} P(ij)\sum_{klc} \langle kl\mid\mid ck\rangle t^c_i t^{ab}_{kl}
- \frac{1}{2} P(ab)\sum_{kcd} \langle kb\mid\mid cd\rangle t^a_k t^{cd}_{ij}
  \nonumber \\
  & - & \frac{1}{2} P(ij)P(ab)\sum_{kcd}\langle kb\mid\mid cd \rangle 
         t^c_it^a_kt^d_j
+ \frac{1}{2} P(ij)P(ab)\sum_{klc}\langle kl\mid\mid cj \rangle 
         t^c_it^a_kt^b_l \nonumber \\
  & - & P(ij)\sum_{klcd} \langle kl \mid\mid cd \rangle t^c_k t^d_i t^{ab}_{lj}
- P(ab)\sum_{klcd} \langle kl \mid\mid cd \rangle t^c_k t^a_l t^{db}_{ij}
+ \frac{1}{4}P(ij)\sum_{klcd} \langle kl \mid\mid cd \rangle 
    t^c_i t^d_j t^{ab}_{kl} \nonumber \\
  & + & \frac{1}{4}P(ab)\sum_{klcd} \langle kl \mid\mid cd \rangle 
    t^a_k t^b_l t^{cd}_{ij} 
+ P(ij)P(ab)\sum_{klcd} \langle kl \mid\mid cd \rangle 
t^c_i t^b_l t^{ad}_{kj}\nonumber \\
&+& \frac{1}{4}P(ij)P(ab)\sum_{klcd} \langle kl \mid\mid cd \rangle 
  t^c_i t^a_k t^d_j t^b_l \;.
\label{t2_eqn}
\end{eqnarray}
The permutation operator $P$ yields
\begin{equation}
P(ij)f(ij) = f(ij) - f(ji)
\end{equation}
The equations of the $t_2$ amplitudes are nonlinear in both $t_1$ and
$T_2$ terms. While these equations appear quite lengthy, they are 
solvable through iterative techniques that we will discuss below. 
We note that the amplitude equations include terms that allow for 
4p-4h excitations. Indeed, while we speak of doubles in terms of 
amplitudes, the class of diagrams involved in the theory include 
fourth-order terms. This is a very important difference 
and distinction between the shell model with up to 
$2p$-$2h$ excitations and CCSD. Furthermore, when
the energy is computed in CCSD, all terms are linked and connected. 
For further details, see Ref.~\cite{dean03}.
\begin{figure}[ht]
\begin{center}
\includegraphics[angle=270, scale=0.5]{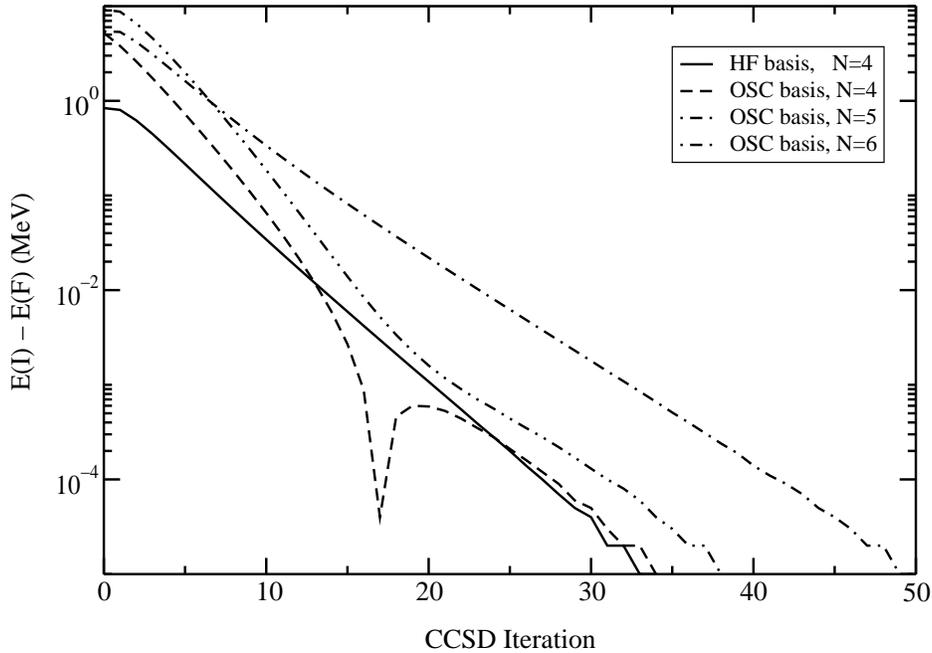}
\caption{The convergence of the ground-state energy
as a function of the CCSD iterations in the $^{16}$O system 
for various shell-model spaces.}
\label{converge_fig}
\end{center}
\end{figure}

Shown in Fig.~\ref{converge_fig} is the difference between the 
energy $E(I)$ at a given iteration $I$ and the final energy $E_F$
as a function of iteration number for a calculation 
of $^{16}$O in different oscillator spaces. 
Here we  achieve convergence at the $10^{-5}$ level
by 50 iterations in a model-space that includes seven major oscillator
shells. 

Eq.~(\ref{amp_start}) represent the terms
that one uses to compute the energy
in second-order perturbation theory of the
M{\o}ller-Plesset type \cite{mp34}. Shown in Table~\ref{e_tab} are
the energies obtained from the $0^{th}$ $E(0)$ iteration and the final
iteration $E(F)$ as functions of increasing oscillator levels in the $^{16}$O
system. The difference between the converged CCSD energies
and the initial $0^{th}$ order energies increases as the basis
space increases. The converged summation of the CCSD
equations yields approximately
10 MeV (or 0.6~MeV per particle) in extra binding at the $\hbar\omega$
minima.
These findings are corroborated by those from many-body 
perturbation theory.  It is therefore worth comparing 
these results with those from second-order and third-order
many-body perturbation theory as well. These are labeled
$E^{\mathrm{2nd}}_{\mathrm{MBPT}}$ and $E^{\mathrm{3rd}}_{\mathrm{MBPT}}$
in the same table. The reader should notice that the zeroth iterations of the
coupled-cluster schemes already includes corrections to the one-body amplitudes
$t_1$. However, the energy 
denominators used in the computation of the second-order
diagrams of Fig.~\ref{fig:diagrams} (diagrams 2 and 3)
have hole states determined by 
\begin{equation}
 \varepsilon_i = 
\left\langle i\right | \frac{p^2}{2m} \left | i \right \rangle 
+\sum_{j \leq F}  \left\langle ij\right | 
G(\omega=\varepsilon_i+\varepsilon_j) \left | ij\right\rangle,
\end{equation}
where $F$ stands for the Fermi energy. 
We do not perform a self-consistent Brueckner-Hartree-Fock
calculation however, as done by e.g., 
Gad and M\"uther \cite{herbert02}. 
The agreement with the zeroth order iteration
and second-order perturbation theory is very good, especially for
five and six major shells, as can be seen from Table \ref{e_tab}.
\begin{table}
\begin{center}
\caption{Comparisons of the 0$^{th}$ order energy $E(0)$
and the converged CCSD results $E(F)$ for
$^{16}$O as a function of increasing model model space.
The results are also compared with many-body perturbation theory to second and third order, $E^{\mathrm{2nd}}_{\mathrm{MBPT}}$ and $E^{\mathrm{3rd}}_{\mathrm{MBPT}}$, respectively. All energies are in MeV.}
\begin{tabular}[t]{|ccccc|}
\hline
N ($\hbar\omega)$ & $E(0)$  & $E^{\mathrm{2nd}}_{\mathrm{MBPT}}$& $E^{\mathrm{3rd}}_{\mathrm{MBPT}}$&$E(F)$ \cr
\hline
4 14  & -135.12  &  -132.06 & -129.92 &-140.47 \cr
5 14  & -124.79  &  -124.84 & -121.52 & -127.79 \cr
6 14   & -121.36  &  -121.48& -118.23 & -119.73 \cr
\hline
\end{tabular}
\end{center}
\label{e_tab}
\end{table}
However, for third-order perturbation theory one clearly sees fairly large
differences compared with the coupled-cluster results.
Typically, the relation between first- and second-order in perturbation theory
for  $^{16}$O is given by a factor of $\sim 5-6$. For e.g., $N=5$ and
$\hbar\omega=14$ MeV, we have $-329.12$ MeV from first order and $-47.72$
from second order. To third order we obtain a repulsive contribution of
$3.32$ MeV, to be contrasted with the almost $3$ MeV of attraction given
by higher-order terms in the coupled-cluster expansion. This indicates that
many-body perturbation theory to third order is
most likely not a converged result. An interesting feature to be noted from
many-body perturbation theory calculations is that higher terms loose their
importance as the size of the system is increased. For
$^{4}$He the relation between first-order and second-order perturbation theory
is given by a factor of $\sim 3-4$, depending on the value of $\hbar\omega$.
Calculations for $^{40}$Ca not reported here indicate a relation of
$\sim 7-9$ between first-order and second-order perturbation theory.
This is somewhat expected since the $G$-matrix is smaller for larger
systems, although the energy denominators become smaller.

\begin{figure}
\begin{center}
\includegraphics[angle=270, scale=0.35]{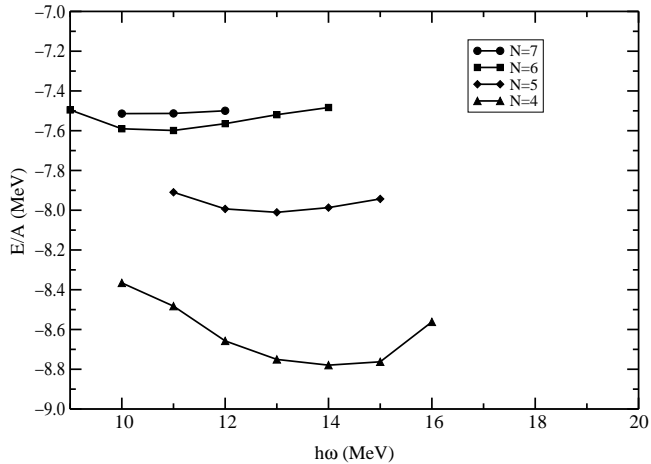}
\caption{Dependence of the ground-state energy of $^{16}$O  on $\hbar\omega$
as a function of increasing model space.}
\label{fig_ox_hw}
\end{center}
\end{figure}

In the initial coupled-cluster study, we performed calculations of 
the $^{16}$O ground state for up to seven major oscillator 
shells as a function of $\hbar\omega$. 
Fig.~\ref{fig_ox_hw} indicates the level of convergence
of the energy per particle for $N=4,5,6,7$ shells. The experimental value
resides at 7.98~MeV per particle.  This calculation is practically converged.
By seven oscillator shells, the $\hbar\omega$ dependence becomes rather
minimal and we find a ground-state binding energy of 7.52 MeV per particle in
oxygen using the Idaho-A potential. Since the Coulomb interaction should give
approximately 0.7 MeV/A of repulsion, and is not included in this
calculation, we actually obtain approximately 6.90 MeV of nuclear binding
in the 7 major shell calculation which is somewhat above the experimental
value. We note that the entire procedure ($G$-matrix plus CCSD) tends to
approach from below converged solutions.

We also considered chemistry inspired 
noniterative triples corrections to the ground
state energy. We performed this study in the model space consisting
of four major oscillator shells. 
Table~\ref{table_ox16_gs} shows the total ground-state energy values
obtained with the CCSD and one of the
triples-correction approaches (labeled CR-CCSD(T) 
\cite{Piecuch02a,Piecuch02b,Kowalski00,Kowalski03}
in the table). Slightly
differing triples-corrections yield similar corrections to the
CCSD energy.
The coupled cluster methods recover the bulk of the correlation
effects, producing the results of the SM-SDTQ, or better, quality.
SM-SDTQ stands for the expensive shell-model (SM) diagonalization in
a huge space spanned by the reference and all
singly (S), doubly (D), triply (T), and
quadruply (Q) excited determinants.
To understand this result, we note that
the CCSD $T_1$ and $T_2$ amplitudes are similar in order of magnitude. (For
an oscillator basis, both $T_1$ and $T_2$ contribute to the first-order
MBPT wave function.)
Thus, the $T_1 T_2$ {\it disconnected} triples are large, much larger than
the $T_3$ {\it connected} triples, and the difference
between the SM-SDT (SM singles, doubles, and triples)
and SM-SD energies is mostly due to $T_1 T_2$.The small $T_3$
effects, as estimated by CR-CCSD(T), are consistent
with the SM diagonalization calculations. If the $T_3$ corrections
were large, we would observe a significant lowering of the
CCSD energy, far below the SM-SDTQ result.
Moreover, the CCSD and CR-CCSD(T) methods
bring the nonnegligible higher-than-quadruple excitations,
such as $T_1^3 T_2$, $T_1 T_2^2$, and $T_{2}^{3}$, which are
not present in SM-SDTQ. It is, therefore, quite likely that the
CR-CCSD(T) results are very close to the results of the exact
diagonalization, which cannot be performed.
\begin{table}[ht]
\caption{The ground-state energy of $^{16}$O
calculated using various coupled cluster methods
and oscillator basis states.  }
\begin{center}
\begin{tabular}{|cc|}
\hline
Method & Energy \cr
\hline
CCSD                       & -139.310 \cr
CR-CCSD(T)                 & -139.467 \cr
SM-SD                        & -131.887 \cr
SM-SDT                       & -135.489 \cr
SM-SDTQ                      & -138.387 \cr
\hline
\end{tabular}
\end{center}
\label{table_ox16_gs}
\end{table}

These results indicate that the bulk of the correlation energy within
a nucleus can be obtained by solving the CCSD equations. This gives us
confidence that we should pursue this method in opened shell systems
and to excited states. We have recently 
\cite{kowalski03n} performed excited state calculations on $^{4}$He
using the EOMCCSD (equation of motion CCSD) method.
For the excited
states $|\Psi_{K}\rangle$ and energies $E_{K}^{\rm (CCSD)}$ ($K > 0$),
we apply the EOMCCSD (``equation of motion CCSD'') approximation
\cite{Stanton:1993,Piecuch99} (equivalent to the 
response CCSD method \cite{Monkhorst:1977}),
in which
\begin{equation}  
|\Psi_{K}\rangle=R_{K}^{\rm (CCSD)} \exp(T^{\rm (CCSD)}) |\Phi\rangle .  
\label{eomfun}  
\end{equation}
Here $R_{K}^{\rm (CCSD)} = R_{0}+ R_{1} + R_{2}$ is a sum of the
reference ($R_{0}$), one-body ($R_{1}$), and two-body ($R_{2}$)
components
obtained by diagonalizing
$\bar{H}^{{\rm (CCSD)}}$
in the same space of singly and doubly excited determinants
$|\Phi_{i}^{a}\rangle$ and $|\Phi_{ij}^{ab}\rangle$ as used in the
ground-state CCSD calculations. These calculations may also be 
corrected in a non-iterative fashion using the completely renormalized
theory for excited states 
\cite{Piecuch02a,Piecuch02b,Kowalski00,Kowalski03,Kowalski01}.  
The low-lying
$J=1$ state most likely results from the center-of-mass contamination
which we have removed only from the ground state.  The $J=0$ and $J=2$
states calculated using EOMCCSD and CR-CCSD(T) are in excellent
agreement with the exact results. 
\begin{table}[ht]
\caption{The excitation energies of $^4$He   
calculated using the  
oscillator basis states (in MeV).  
}  
\begin{center}  
\begin{tabular}{|ccccc|}  
\hline
State & EOMCCSD & CR-CCSD(T) & CISD & Exact \cr
\hline
J=1   &  11.791 & 12.044 & 17.515    & 11.465 \cr
J=0   &  21.203 & 21.489 & 24.969    & 21.569 \cr
J=2   &  22.435 & 22.650 & 24.966    & 22.697 \cr
\hline
\end{tabular}
\end{center}
\label{table_2}
\end{table}

Our experience thus far with the 
quantum chemistry inspired coupled cluster
approximations to calculate the ground and excited states of the
$^{4}$He and $^{16}$O nuclei indicates that this will be a promising
method for nuclear physics.  By comparing coupled cluster results
with the exact results obtained by diagonalizing the Hamiltonian in
the same model space, we demonstrated that relatively inexpensive
coupled cluster approximations recover the bulk of the nucleon
correlation effects in ground- and excited-state nuclei. These results
are a strong motivation to further develop coupled cluster methods for
the nuclear many-body problem, so that accurate {\it ab initio}
calculations for small- and medium-size nuclei become as routine as in
molecular electronic structure calculations.

\section{Three-body forces in shell-model studies}\label{sec:sec7}

An important feature of large scale shell-model calculations
is that they allow one to probe the underlying many-body
physics in a hitherto unprecedented way.
As we have seen, the crucial starting point in all such shell-model 
calculations is
the derivation of an effective interaction, be it
either an approach based on a microscopic theory
starting from the free $NN$ interaction or a more 
phenomenologically determined interaction. 
In shell-model studies of e.g., the Sn isotopes, one may have
up to 31 valence particles or holes interacting via e.g.,
an effective two-body interaction. The results of such 
calculations can therefore yield, when compared with 
the availiable body of experimental data, critical
inputs to the underlying theory of the effective interaction.
Thus, by going to the
tin isotopes, in which the major neutron shell between neutron numbers 50 and
82 is being filled beyond the $^{100}$Sn closed shell core, we have the opportunity
of testing the potential of large-scale 
shell-model calculations as well as the realiability of
realistic effective interactions in systems with many valence particles. It should
be noted that in many current shell-model calculations the effective interaction
is frequently either parametrized or adjusted in order to optimize the fit to the
data. As a matter of principle we shall refrain from making any such
adjustments and stick to the interaction obtained by a 
rigorous calculation consistent with the many-body scheme chosen. 
Only then may one 
be able to assess the quality and reliability of the interaction 
obtained and the possible
needs for improvement. 
It is our firm belief that 
one of the important  aims  
behind many-body based derivations of effective
interactions for the shell model is namely
to provide a link between e.g., the free
nucleon-nucleon interaction and properties of
finite nuclei. 
Recalling our discussion in section \ref{sec:sec1}, there are indications that
three-body 
interactions, both real and effective ones, may be of
importance. The Green's function Monte Carlo calculations of the Argonne-Urbana
group \cite{bob1,bob2,bob3} and the recent no-core shell-model calculations
\cite{petr_erich2002,petr_erich2003} clearly indicate the need for
three-body forces. Similar arguments, based on an analysis of $0p$, $1s0d$ and $1p0f$
nuclei by Zuker \cite{zuker1} lend support to this picture for heavier nuclei as well.
Real three-body forces are necessary
in order to reproduce the saturation properties of nuclear matter as well, see for example
Ref.~\cite{apr98}.

Thus, with many valence nucleons present, such
large-scale shell-model calculations may
tell us how well e.g., an effective interaction
which only includes two-body terms does in
reproducing properties such as excitation spectra and
binding energies. 

In general, 
excitation spectra for the chain of both odd and
even isotopes from  
$^{116}$Sn to $^{130}$Sn exhibit an excellent agreement
with the data, see for example our discussion in section \ref{sec:sec4}.
All these results are
based on shell-model calculations starting with
a two-body effective interaction determined from the 
recent charge dependent $NN$ potentials of Machleidt
and co-workers \cite{cdbonn,cdbonn2000}. These $NN$-interactions
were renormalized for the given nuclear medium, using for example $^{132}$Sn as closed
shell core, through
the introduction of the so-called reaction matrix $G$,
which corresponds to solving the Lippmann-Schwinger
equation for a finite nucleus. The $G$-matrix formed then
the basis for a perturbative calculation of more complicated
Feynman-Goldstone diagrams.
However, if one studies Table \ref{tab:tablesnbe},
\begin{table}[htbp]
     \caption{Binding energies for Sn isotopes.}
     \label{tab:tablesnbe}
     \begin{center}
\begin{tabular}{|l|ccccccc|} \hline
& $^{130}$Sn& $^{129}$Sn& $^{128}$Sn& $^{126}$Sn& $^{124}$Sn& $^{122}$Sn& $^{116}$Sn\\
\hline
Exp& -2.09& -1.70& -3.64& -4.79&-5.47&-5.64&-2.61\\
$V^{(2)}$& -2.24& -2.63& -4.60& -6.99&-9.39&-11.77& -18.58\\
Mod.\ Shell Model &-2.09&-2.78& -3.72&-4.81&-5.32& -5.22& -1.12\\
$+V^{(3)}$&  & +0.009& +0.029& +0.096& +0.213& +0.394&+0.998\\ 
Trp. contr.& & +0.009& +0.034& +0.170& +0.504&+1.08&+4.760\\\hline
      \end{tabular}
     \end{center}
\end{table}
one sees that the binding energy relative to $^{131}$Sn  
is clearly at askance with the data. 
The binding energy is defined as
\begin{equation}
      BE[^{132 - n}Sn)] = BE[^{132 - n}Sn] - BE[^{132}Sn] 
      - n  \left (BE[^{131}Sn] -  BE([^{132}Sn] \right ).
\end{equation}
Experiment indicates a minimum around $^{124}$Sn-$^{122}$Sn and consequently
a shell closure around $^{116}$Sn whereas theoretical 
binding energies increase linearly all the way 
down to $^{116}$Sn. 

The above is just an example of one of the problems which beset the theory
of effective interactions for the shell model. In this case, as also
pointed out by Zuker and co-workers \cite{zuker1},
one is not able to obtain simultaneously a good reproduction
of both the excitation spectra and the binding energy. Similar problems
have also been discussed in connection with large scale shell-model
calculations of $1f0p$ shell nuclei. To give an example, effective
interactions derived from two-body $NN$ interactions which fit 
nucleon-nucleon scattering data, are not able to reproduce the 
well-known shell clousure in $^{48}$Ca or the excitation spectra 
of $^{47}$Ca and $^{49}$Ca, see the discussion in Refs.\ \cite{alex,hko95,zuker1}.

In this work we wish to address the discrepancy between theory and experiment
shown in Table \ref{tab:tablesnbe} by including effective three-body
forces in our shell-model calculations. The reason for this follows
from the observation that the introduction of a global monopole correction
of the form $Wn(n-1)/2$, with $n$ being the number of valence particles and
$W$ a quantity which is related to the energy centroids, 
can, when added to our theoretical binding energy, partly cure the 
discrepancy seen in Table \ref{tab:tablesnbe}. Three-body forces
typically yield repulsive corrections which scale as $n(n-1)$.  
In Table \ref{tab:tablesnbe} we adjusted the term $W$ by simply requiring
it to equal the difference in binding energy between the calculated and
experimental values for $^{130}$Sn. This resulted in $W=0.15$ MeV.
Adding such a global correction to even nuclei shown in Table  
\ref{tab:tablesnbe} results in the column labelled Mod.\ shell model.
Clearly this improves the theoretical results in the correct direction
and such a global modification of the matrix elements has no
effect on the excitation spectra. 

\begin{figure}[hbtp]
\begin{center}
      \setlength{\unitlength}{1mm}
      \begin{picture}(100,80)
      \put(0,0){\epsfxsize=10cm \epsfbox{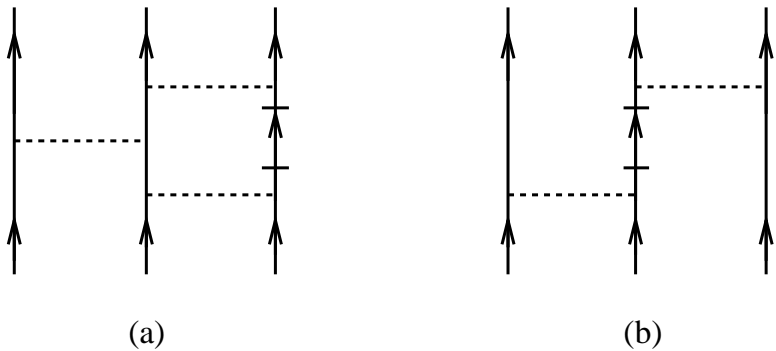}}
      \end{picture}
       \caption{Examples of unfolded three-body diagrams.}
       \label{fig:threebody}
\end{center}
\end{figure}

These observations form thus the starting point for our investigation
of three-body effective interactions. 
More explicitely, we aim at seeing whether three-body interactions
may yield a more microscopic understanding of the above problem. 
The results presented employ effective three-body diagrams to second order in the
$G$-matrix (see Polls {\em et al.} \cite{polls81} for details)
and have been shown at various conferences \cite{eho}. Examples of three-body diagrams
are shown in Fig.~\ref{fig:threebody}. Diagram (a) is a third-order contribution not included
in the $\hat{Q}$-box. Diagram (b) is included.
The inclusion of such effective three-body diagrams do add a needed repulsion as seen
from the column labelled $V^{(3)}$ in Table \ref{tab:tablesnbe}. However
there is still large discrepancy between theory and experiment for the binding energy
and even 
multiplying the number of triplets 
\[
\mbox{number of triplets} = \frac{n!}{3! (n-3)!}
\]
with the three-body contribution from $^{129}$Sn yields too little attraction.
Most likely the lack of repulsion can be ascribed to real three-body forces.
Effective three-body forces for identical particles tend to be small due to the absence
of the strong tensor channels in the intermediate 
states. This is also in line with 
recent works on the energy of pure neutron 
drops, where three-body clusters are included \cite{ndrops97}.
The reader should however note that 
our inclusion of three-body effective diagrams is incomplete. The summation of folded diagrams
is based on only a two-body and three-body $\hat{Q}$-box. We have not included a one-body $\hat{Q}$-box.
The latter introduces unlinked diagrams which need to be subtracted correctly when summing folded
diagrams.
To investigate whether the discrepancies seen for the binding energies are due to real three-body 
forces or not, we plan to include these in future studies, together with a three-body 
$G$-matrix and properly determined $\hat{Q}$-box and folded diagrams.
One needs thus to carefully
distinguish between three-body forces and effective three-body
interactions.

\section{Perspectives}\label{sec:sec8}

The study of exotic nuclei opens new 
challenges to nuclear physics. 
The challenges and the excitement arise 
because exotic nuclei will present new and 
radically different manifestations of nucleonic matter 
that occur near the bounds of nuclear existence, 
where the special features of weakly bound, quantal systems 
come into prominence. Furthermore, many of these
nuclei are key to understanding matter production in the universe.
Given that present and future nuclear structure research facilities
will open significant
territory into regions of medium-mass and heavier nuclei,
it becomes important to investigate theoretical methods that will allow
for a description of medium-mass nuclear systems. 
Such systems pose significant
challenges to existing nuclear structure models, especially since many of
these nuclei will be unstable and short-lived. How to deal with weakly
bound systems and coupling to resonant states is an unsettled problem in
nuclear spectroscopy. 

The aim of this work has been to expose both 
the weakness and strength of three many-body 
methods for studying nuclei heavier than $^{16}$O. 
For lighter nuclei, both Green's function 
Monte Carlo \cite{vijay,bob1,bob2,bob3} and no-core shell-model calculations 
\cite{bruce1,bruce2,bruce3,petr_erich2002,petr_erich2003}
offer true benchmarks for nuclear structure studies.
The nuclear shell-model, combined with microscopically or phenomenologically 
derived effective interactions, has been extremely succesful in reproducing 
properties of nuclei 
employing $0\hbar\omega$ model spaces such as the $1s0d$ \cite{alex},
the $1p0f$ shell \cite{taka1,etienne6} and, as discussed in 
this work, medium heavy systems
in the region of $A=100$ to $A=132$. However, 
if one wishes to study weakly bound systems
or resonant states within this framework, the dimensionality 
of the problem soon exceeds
present and most likely future hardware capabilities. 

For the heavier systems which are expected to be 
studied with facilities like RIA, many-body methods like the 
coupled cluster or parquet diagrams offer possibilities for extending
microscopic ab-initio calculations to nuclei like $^{40}$Ca.
Especially the coupled-cluster methods are
very promising, since they allow one to study 
ground- and excited-state properties of nuclei
with dimensionalities beyond the capability of present shell-model
approaches. As demonstrated in Ref.~\cite{kowalski03n} we 
show for the first time how to calculate 
excited states for a nucleus using coupled cluster 
methods from quantum chemistry.
For the weakly bound nuclei to be produced by future low-energy 
nuclear structure facilities
it is almost imperative to increase the
degrees of freedom under study in order to reproduce
basic properties of these systems. 
We are presently working on deriving effective interactions 
for weakly bound systems to be used in coupled cluster 
studies of these weakly bound nuclei. An extension of this project is to extract effective 
two-body interactions for open shell systems. These will in turn  
be used in shell calculations. 

We have based most of our analysis on effective interactions
using two-body nucleon-nucleon interactions only. 
We feel this is important since
techniques like the coupled cluster methods or 
parquet diagrams allow one to include a much larger
class of many-body terms than done earlier. Eventual discrepancies 
with experiment 
such as the missing reproduction of e.g., the first 
excited $2^+$ state in a $1p0f$ calculation
of $^{48}$Ca, can then be ascribed to eventual 
missing three-body forces, as indicated by the studies
in Refs.~\cite{bob1,bob2,bob3,apr98,petr_erich2002,petr_erich2003} 
for light nuclei. 
The inclusion of real three-body interactions belongs 
to our future plans both for effective 
interactions for nuclei with $A\sim 100$ and larger 
as well as within the coupled cluster method and
parquet diagrams for nuclei like $^{40}$Ca.

\section*{Acknowledgments}

We thank Alex Brown, Hubert Grawe, Karol Kowalski, Karlheinz Langanke, Matej Lipoglav\v{s}ek, 
Herbert M\"uther, Thomas Papenbrock, Piotr Piecuch, Artur Polls and Andres Zuker 
for many stimulating discussions on nuclear structure, 
many-body physics and the shell model.
Research sponsored by the Laboratory Directed Research and Development
Program of Oak Ridge National Laboratory (ORNL), managed by
UT-Battelle, LLC for the U. S.  Department of Energy under
Contract No. DE-AC05-00OR22725 and the Research Council of Norway.


\begin{thebibliography}{200}
\bibitem{bruce1} P.~Navr\'atil and B.R.~Barrett, Phys. Rev. {\bf C57} (1998) 562.
\bibitem{bruce2} P.~Navr\'atil, J.P.~Vary, and B.R.~Barrett
Phys.~Rev.~Lett.~{\bf 84} (2000) 5728.
\bibitem{bruce3} P.~Navr\'atil, J.P.~Vary, and B.R.~Barrett, Phys. Rev. {\bf C62 }(2000) 054311.
\bibitem{bruce4} D.C.~Zheng, B.R.~Barrett, L.~Jaqua, J.P.~Vary, and R.J.~McCarthy, 
Phys.~Rev.~{\bf C48} (1993) 1083.
\bibitem{alex} B.A.~Brown, 
Prog.~Part.~Nucl.~Part.~{\bf 47}  (2001) 517, and references therein.
\bibitem{taka1} T.~Otsuka, M.~Homna, T.~Mizusaki, N.~Shimizu, and Y.~Utsuno,
Prog.~Part.~Nucl.~Part.~{\bf 47}  (2001) 319, and references therein.
\bibitem{r:smmc_pr} S.E.\ Koonin, D.J.\ Dean and K.\ Langanke,
Phys.\ Rep.\ {\bf 278} (1997) 1.
\bibitem{vijay} B.S.\ Pudliner, V.R.\ Pandharipande, J.\ Carlson, S.C.\ Pieper and R.B.\
Wiringa, Phys.\ Rev.\ {\bf C56} (1997) 1720.
\bibitem{ndrops97} B.S.\ Pudliner, A.\ Smerzi, J.\ Carlson,
                   V.R.\ Pandharipande, S.C.\ Pieper  and
                   D.G.\ Ravenhall, Phys.\ Rev.\ Lett.\
                   {\bf 76}  (1996) 2416.
\bibitem{bob1} S.C.~Pieper, V.R.~Pandharipande, R.B.~Wiringa, and J.~Carlson, Phys.~Rev.~{\bf C64}
(2001) 014001
\bibitem{bob2} S.C.~Pieper, K.~Varga, and R.B.~Wiringa, Phys.~Rev.~{\bf C66}
(2002) 0044310
\bibitem{bob3}  R.B.~Wiringa and S.C.~Pieper, Phys.~Rev.~Lett.~{\bf 89}
(2002) 182501
\bibitem{so95} K.\ Suzuki, R.\ Okamoto and H.\ Kumagai, 
Nucl.\ Phys.\ {\bf A580} (1994) 213;
K.\ Suzuki and R.\ Okamoto, 
Prog.\ Theor.\ Phys.\ {\bf 92} (1994) 1045;
ibid.\  {\bf 93} (1995) 905
\bibitem{brandow67} B.H.\ Brandow, 
Rev.\ Mod.\ Phys.\  {\bf 39} (1967) 771.
\bibitem{ko90} T.T.S.\ Kuo and E.\ Osnes, {\em Folded-Diagram Theory
of the Effective Interaction in Atomic Nuclei}, Springer Lecture
Notes in Physics, (Springer, Berlin, 1990) Vol.\ {\bf 364}; 
T.T.S.\ Kuo, Lecture Notes in
Physics; Topics in Nuclear Physics, eds.\ T.T.S.\ Kuo and S.S.M.\
Wong, (Springer, Berlin, 1981) Vol.\ {\bf 144}, p.\ 248.
\bibitem{hko95} M.\ Hjorth-Jensen, T.T.S.\ Kuo and E.\ Osnes,
Phys.\ Rep.\ {\bf 261} (1995) 125.
\bibitem{lm85}  I.\ Lindgren and J.\ Morrison, 
{\em Atomic Many-Body Theory},
(Springer, Berlin, 1985); I.\ Lindgren, J.\ Phys.\ B: At.\ Mol.\ Opt.\ Phys.\ {\bf 24} (1991) 1143.
\bibitem{so84} K.\ Suzuki, Prog.\ Theor.\ Phys.\ {\bf 68} (1982) 1627; K.\ Suzuki and 
R.\ Okamoto, Prog.\ Theor.\ Phys.\ {\bf 75} (1986) 1388; 
ibid. {\bf 76} (1986) 127.
\bibitem{coester58} F.~Coester, Nucl. Phys. {\bf 7} (1958) 421.
\bibitem{coester60} F.~Coester and H.~K\"ummel, Nucl. Phys.~{\bf 17} (1960) 477.
\bibitem{bartlett81} R.J.~Bartlett, Ann. Rev. Phys. Chem. {\bf 32} (1981) 359.
\bibitem{comp_chem_rev00} T.D.~Crawford and H.F.~Schaefer III,  Rev.~Comp.~Chem.~{\bf 14} (2000) 33.
\bibitem{mp34}C.~M{\o}ller and M.~Plesset, Phys. Rev. {\bf 46} (1934) 618. 
\bibitem{harris92} F.E.~Harris, H.J.~Monkhorst and D.L.~Freeman, 
{\em Algebraic and Diagrammatic Methods in Many-Fermion Theory}, (Oxford, New York, 1992).
\bibitem{piotr1} P.~Piecuch, P.-D.~Fan, K.~Jedziniak, and K.~Kowalski, Phys.~Rev.~Lett.~{\bf 90}
(2003) 113001.
\bibitem{helgaker} T.~Helgaker, P.~J{\o}rgensen, and J.~Olsen, 
 {\em Molecular Electronic Structure Theory. Energy and Wave Functions}, (Wiley, Chichester, 2000).
\bibitem{arponen97} J.\ Arponen, 
Phys.\ Rev.\ {\bf A55} (1997) 2686.
\bibitem{lk72a} H.\ K\"{u}mmel and K.H.\ L\"{u}hrmann, Nucl.\ Phys.\ {\bf A191} (1972) 525.
\bibitem{lk72b} H.\ K\"{u}mmel and K.H.\ L\"{u}hrmann, 
Nucl.\ Phys.\ {\bf A194} (1972) 225.
\bibitem{zabolitzky74} J.G.\ Zabolitzky, 
Nucl.\ Phys.\ {\bf A228} (1974) 285.
\bibitem{klz78} H.\ K\"{u}mmel, K.H.\ L\"{u}hrmann and J.G.\ Zabolitzky, Phys.\ Rep.\
{\bf 36} (1977) 1.
\bibitem{ticcm} R.F.\ Bishop, E.\ Buendia, M.F.\ Flynn and R.\ Guardiola, J.\ Phys.\ G:
Nucl.\ Part.\ Phys.\ {\bf 17} (1991) 857; 
ibid.\ {\bf 18} (1992) 1157; 
ibid.\ {\bf 19} (1993) 1663; 
R.\ Guardiola, P.I.\ Moliner, J.\ Navarro, R.F.\ Bishop, A.\ Puente and
N.R.\ Walet, Nucl.\ Phys.\ {\bf A609} (1996) 218; R.F.\ Bishop and
R.\ Guardiola.
\bibitem{mh00a} B.~Mihaila and J.H.~Heisenberg, Phys. Rev. Lett.~{\bf 84} (2000) 1403.
\bibitem{mh00b} B.~Mihaila and J.H.~Heisenberg, Phys.~Rev.~{\bf C61} (2000) 054309.
\bibitem{mh99} B.~Mihaila and J.H.~Heisenberg, Phys. Rev. {\bf C60} (1999) 054303.
\bibitem{hm99} J.H.~Heisenberg, and B.~Mihaila, Phys. Rev. {\bf C59} (1999) 1440.
\bibitem{dm64} C.\ de Dominicis and P.C.\ Martin, J.\ Math.\ Phys.\
{\bf 5} (1964) 14.
\bibitem{nozieres} B.\ Roulet, F.\ Gavoret and P.\ Nozieres,
Phys.\ Rev.\ {\bf 178} (1969) 1072.
\bibitem{babu} S.\ Babu and G.E.\ Brown, 
Ann.\ Phys.\ {\bf 78} (1973) 1.
\bibitem{jls82} A.D.\ Jackson, A.\ Lande and R.A.\ Smith, 
Phys.\ Rep.\  {\bf 86} (1982) 55; A.\ Lande and R.A.\ Smith, Phys.\ Rev.\ 
{\bf A45}
(1992) 913 and references therein.
\bibitem{br86} J.P.\ Blaizot and G.\ Ripka, {\em Quantum theory of finite systems},
(MIT press, Cambridge, USA, 1986), chapter 15.
\bibitem{scalapino} 
N.E.\ Bickers and D.J.\ Scalapino, Ann.\ Phys.\ 
{\bf 193} (1989) 206;
N.E.\ Bickers and S.R.\ White, Phys.\ Rev.\ 
{\bf B43} (1991) 8044; 
N.E.\ Bickers and D.J.\ Scalapino, Phys.\ Rev.\
{\bf B46} (1992) 8050.
\bibitem{ym96} J.\ Yeo and M.A. Moore, Phys.\ Rev.\ 
{\bf B54} (1996) 4218.
\bibitem{dya97} A.T.\ Zheleznyak, V.M.\ Yakovenko and I.E.\
Dzyaloshinskii, Phys.\ Rev.\ {\bf B55} (1997) 3200.
\bibitem{adelchi98} A.\ Fabrocini, F.\ Arias de Savedra, G.\ C\'o and P.\ Folgarait,
Phys.\ Rev.\ {\bf C57} (1998) 1668.
\bibitem{apr98} A.\ Akmal, V.R.\ Pandharipande and D.G.\
                Ravenhall, Phys.\ Rev.\ {\bf C58}  (1998) 1804.
\bibitem{kirson74} M.W.\ Kirson, Ann.\ Phys.\ 
{\bf 66} (1971) 624; ibid.\
{\bf 68} (1971) 556; ibid.\ {\bf 82} (1974) 345.
\bibitem{eo77} P.J.\ Ellis and E.\ Osnes, Rev.\ Mod.\ Phys.\ 
{\bf 49} (1977) 777.
\bibitem{angels88} A.\ Ramos, PhD.\ Thesis, University of Barcelona, 1988, unpublished.
\bibitem{rpd89} A.\ Ramos, A.\ Polls and W.H.\ Dickhoff, Nucl.\ Phys.\ {\bf A503} (1989) 1.
\bibitem{yhk86} S.D.\ Yang, J.\ Heyer and T.T.S.\ Kuo, Nucl.\ Phys.\ {\bf A448} (1986) 420.
\bibitem{syk87} H.Q.\ Song, S.D.\ Yang and T.T.S.\ Kuo,
Nucl.\ Phys.\  {\bf A462} (1987) 491.
\bibitem{hmtk87} H.A.\ Mavromatis, H.\ M\"{u}ther, T.\ Taigel and T.T.S.\ Kuo, Nucl.\ Phys.\ {\bf A470}
(1987) 185.
\bibitem{emm91} H.A.\ Mavromatis, P.J.\ Ellis and H.\ M\"{u}ther, Nucl.\ Phys.\ {\bf A530} 
(1991) 251.
\bibitem{hmm95} E.\ Heinz, H.\ M\"{u}ther and H.A.\ Mavromatis, Nucl.\ Phys.\ {\bf A587}
(1995) 77. 
\bibitem{db04} W.H.~Dickhoff and C.~Barbieri, Prog.~Part.~Nucl.~Phys., in press and preprint nucl-th/0402034.
\bibitem{mp00} H.~M\"uther and A.~Polls, Prog.~Part.~Nucl.~Phys.~{\bf 45} (2000) 243.
\bibitem{kstop81} T.T.S.\ Kuo, J.\ Shurpin, K.C.\ Tam, E.\ Osnes
and P.J.\ Ellis, Ann.\ Phys.\  {\bf 132} (1981) 237.
\bibitem{bbp63} H.A.\ Bethe, B.H.\ Brandow and A.G.\ Petschek,
Phys.\ Rev.\  {\bf 129} (1963) 225.
\bibitem{kkko76} E.M.\ Krenciglowa, C.L.\ Kung, T.T.S.\ Kuo and
E.\ Osnes, Ann.\ Phys.\  {\bf 101} (1976) 154.
\bibitem{ls80} S.Y.\ Lee and K.\ Suzuki, Phys.\ Lett.\  
{\bf B 91} (1980) 79;
K.\ Suzuki and S.Y.\ Lee, Prog.\ Theor.\ Phys.\
 {\bf 64} (1980) 2091.
\bibitem{des} J.\ Des Cloizeaux, Nucl.\ Phys.\  {\bf 20} (1960) 321.
\bibitem{kehlsok93} T.T.S.\ Kuo, P.J.\ Ellis, J.\ Hao, Z.\ Li,
K.\ Suzuki, R.\ Okamoto and H.\ Kumagai,
Nucl.\ Phys.\  {\bf A560} (1993); 
P.\ Navr\'atil, H.B.\ Geyer and T.T.S.\ Kuo, Phys.\ Lett.\
 {\bf B315} (1993) 1.
\bibitem{cdbonn} R.\ Machleidt, F.\ Sammarruca, and Y.\ Song,
                 Phys.\ Rev.\  {\bf C53} (1996) R1483.
\bibitem{cdbonn2000} R.~Machleidt, Phys. Rev.~{\bf C63} (2001) 024001.
\bibitem{machleidt02} D.R.~Entem and R.~and Machleidt,  Phys. Lett. {\bf B524} (2002) 93.
 \bibitem{whit77} R.R.\ Whitehead, A.\ Watt, B.J.\ Cole and I.\
                  Morrison, Adv.\ Nucl.\ Phys.\ {\bf 9} (1977) 123.
\bibitem{lawson} R.D.~Lawson, {\em Theory of the nuclear shell model},
(Clarendon Press, Oxford, 1980).
\bibitem{drhklz99} D.J.\ Dean, M.T.\ Ressell, M.\ Hjorth-Jensen, S.E.\ Koonin, 
K.\ Langanke, and A.P.\ Zuker, Phys.\ Rev.\ {\bf C59} (1999) 2474.
\bibitem{papenbrock03} T.~Papenbrock, A.~Juodagalvis, and D.J.~Dean, 
Phys. Rev. {\bf C69} (2004) 024312.
\bibitem{dean03} D.J.~Dean, and M.~Hjorth-Jensen, 
Phys. Rev. {\bf C69} (2004), in press and preprint nucl-th/0308088.
\bibitem{dean04} D.J.~Dean, and M.~Hjorth-Jensen, in preparation.
\bibitem{ols90} J. Olsen, P.~J{\o}rgensen and J.~Simons, Chem. Phys. Lett.
{\bf 169} (1990) 463
\bibitem{petr_erich2002} P.~Navr\'atil and W.E.~Ormand, Phys. Rev. Lett.~{\bf 88} (2002) 152502.
\bibitem{mihai1}   M. Horoi, J. Kaiser, and V. Zelevinsky
Phys. Rev. {\bf C67} (2003) 054309.
\bibitem{mihai2} M. Horoi, B. A. Brown, and V. Zelevinsky
Phys. Rev. {\bf C67} (2003) 034303.
\bibitem{andrius} A.~Juodagalvis, ORNL parallel Shell-model code (2003), unpublished.
\bibitem{oslo1} T.~Engeland, Oslo shell-model code (1990-2004), unpublished.
\bibitem{etienne1}P. Navr\'atil and E. Caurier
Phys. Rev. {\bf C69} (2004) 014311.
\bibitem{etienne2}E. Caurier, P. Navr\'atil, W. E. Ormand, and J. P. Vary
Phys. Rev. {\bf C66} (2002) 024314.
\bibitem{etienne3}E. Caurier, P. Navr\'atil, W. E. Ormand, and J. P. Vary
Phys. Rev. {\bf C64} (2001) 051301. 
\bibitem{etienne4}E. Caurier, G. Martinez-Pinedo, F. Nowacki, A. Poves, J. Retamosa, and A. P. Zuker
Phys. Rev. {\bf C59} (1999) 2033. 
\bibitem{etienne6}E. Caurier, A. P. Zuker, A. Poves, and G. Martinez-Pinedo
Phys. Rev. {\bf C50} (1994) 224. 
\bibitem{taka2} M.~Homna, T.~Mizusaki, and T.~Otsuka,
Phys.~Rev.~Lett..~{\bf 75}  (1995) 1284.
\bibitem{taka3} M.~Homna, T.~Mizusaki, and T.~Otsuka,
Phys.~Rev.~Lett..~{\bf 77}  (1996) 3315.
\bibitem{talmi} I.~Talmi, {\em Simple Models of Complex Nuclei},
(Harwood Academic Publishers, Chur, 1993).
\bibitem{dav89} E. R. Davidson, Comp. Phys. Comm. {\bf 53} (1989) 49;
                E. R. Davidson, Comp. in  Phys. {\bf 5} (1993) 519
\bibitem{anne} A.~Holt, T.~Engeland, M.~Hjorth-Jensen and E.~Osnes,
Phys.~Rev.~{\bf C61}, 064318 (2000).
\bibitem{leander} G. Leander {\it et al.}, Phys. Rev. {\bf C30} (1984) 416.
\bibitem{dz1999} J. Duflo and A. P. Zuker, Phys. Rev. {\bf C59}  (1999) R2347.
\bibitem{grawe}H.~Grawe {\em et al.}, Proc. 6$^{th}$ Int. Seminar Highlights of
Modern Nuclear Structure, (World Scientific, Singapore, 1999) 137.
\bibitem{in102}M. Lipoglav\v{s}ek {\it et al.}, Phys. Rev. {\bf C65} (2002) 021302.
\bibitem{matjaz}M.~Vencelj {\it et al.}, to be published.
\bibitem{alber} D. Alber {\it et al.}, Z. Phys. {\bf A332} (1989) 129.
\bibitem{sn103} C. Fahlander {\it et al.}, Phys. Rev. {\bf C63} (2001) 021307.
\bibitem{fe55} A.~R.~Poletti {\it et al.},
Phys.~Rev.~{\bf C10}  (1974) 2312.
\bibitem{co57} N.~Bendjaballah {\it et al.},
Nucl.~Phys. {\bf A280}, 228 (1977).
\bibitem{ehho98} A.\ Holt, T.\ Engeland, M.\ Hjorth-Jensen and E.\ Osnes,
Nucl.\ Phys.\ {\bf A634} (1998) 41.
\bibitem{haavard2004}H.~Gausemel, B.~Fogelberg, T.~Engeland, P.~Hoff, M.~Hjorth-Jensen, 
H.~Mach, K.A.~Mezilev, and J.P.~Omtvedt, Phys.~Rev.~{\bf C69} (2004), in press.
\bibitem{hoff96}  P. Hoff {\em et al}, Phys.~Rev.~Lett. {\bf 77} (1996) 1020.  
\bibitem{maximsn134} M.~Kartamyshev, T.~Engeland, M.~Hjorth-Jensen, and E.~Osnes,
unpublished. 
\bibitem{gs} I.~Y.~Lee, Nucl. Phys. {\bf A520}, 641c (1990).
\bibitem{uball} D.G. Sarantites {\it et al.}, Nucl. Instr. and Meth.
{\bf A381}, 418 (1996).
\bibitem{r:motobayashi} T.\ Motobayashi et al.,
Phys.\ Lett.\ B {\bf B 346} (1995) 9.
\bibitem{r:brown1} H. Scheit et al., Phys.\ Rev.\ Lett.\ {\bf 77} (1996) 3967.
\bibitem{r:brown2} T. Glasmacher et al.,  Phys.\ Lett.\ {\bf B395} (1997) 163.
\bibitem{r:werner} T.R.\ Werner et al., Nucl.\ Phys.\ {\bf A597} (1996) 327.
\bibitem{r:campi} X.\ Campi, H.\ Flocard, A.K.\ Kerman and S.E.\ Koonin,
Nucl.\ Phys.\ {\bf A251} (1975) 193.
\bibitem{r:wbmb} E.K.\ Warburton, J.A.\ Becker and B.A.\ Brown, Phys.\
 Rev.\ {\bf C41} (1990) 1147.
\bibitem{r:poves1} A.\ Poves and J.\ Retamosa, Nucl.\ Phys.\ {\bf A571}
(1994) 221.
\bibitem{r:fukunishi} N.\ Fukunishi, T.\ Otsuka and T.\ Sebe,
Phys.\ Lett.\ {\bf B296} (1992) 279.
\bibitem{r:retamosa} J.\ Retamosa, E.\ Caurier, F.\ Nowacki and A.\ Poves,
Phys.\ Rev.\ {\bf C55} (1997) 1266.
\bibitem{r:caurier} E.\ Caurier, F.\ Nowacki, A.\ Poves and J.\ Retamosa,
Phys.\ Rev.\  {\bf C58} (1998) 2033.
\bibitem{r:smmc_ar} S.E.\ Koonin, D.J.\ Dean, and K.\ Langanke,
Ann.\ Rev.\ Nucl.\ Part.\ Sci. {\bf 47} (1997) 463.
\bibitem{r:lang} G.H.\ Lang, C.W.\ Johnson, S.E.\ Koonin and W.E.\ Ormand,
Phys.\ Rev.\ {\bf C48} (1993) 1518.
\bibitem{acz} A.\ Abzouzi, E.\ Caurier, and A.P.\ Zuker, Phys.  Rev.
  Lett.\  {\bf 66} (1991) 1134.
\bibitem{r:ensdf} P.M.\ Endt, Nucl.\ Phys.\ {\bf A521} (1990) 1.
\bibitem{r:raman} S.\ Raman et al., At.\ Data and Nucl.\ Data Tab.\
{\bf 36} (1987) 1.
\bibitem{r:poves2} A.\ Poves and J.\ Retamosa, Phys.\ Lett.\ 
{\bf B184} (1987) 311.
\bibitem{ms92} C.\ Mahaux and R.\ Sartor, Phys.\ Rep.\
{\bf 211} (1992) 53. 
\bibitem{mhj99} M.\ Hjorth-Jensen, in preparation.
\bibitem{kz70} M.W.\ Kirson and L.\ Zamick, 
Ann.\ Phys.\ {\bf 60} (1970) 188.
\bibitem{eg80} F.L.\ Goodin and P.J.\ Ellis,
Nucl.\ Phys.\ {\bf A334} (1980) 229.
\bibitem{nuclearmatter} W.H.\ Dickhoff and H.\ M\"{u}ther, Nucl.\
Phys.\ {\bf A473} (1987) 394 and references therein.
\bibitem{hayes03} A.C.~Hayes, P.~Navr\'atil, and J.P.~Vary,  Phys. Rev. Lett.~{\bf 91}
(2003)  012502.
\bibitem{Bartlett95} R. J. Bartlett, ed.~D. R. Yarkony, in
 {\em Modern Electronic Structure Theory} vol.~{\bf 1},
(World Scientific, Singapore, 1995) 1047.
\bibitem{Paldus99} J. Paldus and X. Li, Adv.~Chem.~Phys.~{\bf 110} (1999) 1.
\bibitem{Piecuch02a} P.~Piecuch and K.~Kowalski and I.S.O.~Pimienta and M.J.~McGuire,
 Int. Rev. Phys. Chem. {\bf 21} (2002) 527.
\bibitem{Piecuch02b} P. Piecuch and K. Kowalski and P.-D. Fan and I.S.O. Pimienta,
eds.~J. Maruani, R. Lefebvre and E. Br{\"a}ndas,
{\em Topics in Theoretical Chemical Physics} vol.~{\bf 12},
     series     Progress in Theoretical Chemistry and Physics,
 (Kluwer, Dordrecht, 2004) 119.
\bibitem{cizek66} J. \v{C}\'{\i}\v{z}ek, J. Chem. Phys. {\bf 45} (1966) 4256.
\bibitem{cizek69} J. {\v C}{\'\i}{\v z}ek, Adv. Chem. Phys. {\bf 14} (1969) 35. 
\bibitem{Stanton:1993} J. F. Stanton and R. J. Bartlett, J. Chem. Phys.~{\bf 98} (1993) 7029.
\bibitem{Piecuch99} P. Piecuch and R. J. Bartlett, Adv. Quantum Chem.~{\bf 34} (1999) 295. 
\bibitem{kowalski03n} K.~Kowalski, D.J.~Dean, M.~Hjorth-Jensen, T.~Papenbrock, 
and P.~Piecuch, Phys. Rev. Lett.~{\bf 92} (2004) 132501.
\bibitem{Kowalski01} K. Kowalski and P. Piecuch, J. Chem. Phys.~{\bf 115} (2001) 2966.
\bibitem{Kowalski03} K. Kowalski and P. Piecuch, J. Chem. Phys.~{\bf 120} (2004) 1715.
\bibitem{Monkhorst:1977} H.~Monkhorst, Int. J. Quantum Chem. Symp.~{\bf 11} (1977) 421.
\bibitem{Kowalski00} K. Kowalski and P. Piecuch, J. Chem. Phys.~{\bf 113} (2000) 18.
% three-body stuff
\bibitem{petr_erich2003} P.~Navr\'atil and W.E.~Ormand, Phys. Rev.~{\bf C68} (2003) 034305.
\bibitem{zuker1} A.P.~Zuker, Phys.~Rev.~Lett.~{\bf 90} (2003) 042502.
\bibitem{eho} T.~Engeland and M.~Hjorth-Jensen, unpublished.
\bibitem{polls81} A.\ Polls, H.\ M\"{u}ther, A.\ Faessler,
T.T.S.\ Kuo and E.\ Osnes, Nucl.\ Phys.\
 {\bf A401}  (1983) 124;
H.\ M\"{u}ther, A.\ Polls and T.T.S.\ Kuo, ibid.\ {\bf A435} (1985) 548.
\bibitem {herbert02} Kh.~Gad and 
H.~M\"uther, Phys.~Rev.~{\bf C66} (2002) 044301.

\end{thebibliography}
\end{document}